\documentclass[12pt,english]{article} \usepackage{mathptmx}
\usepackage[T1]{fontenc} \usepackage[utf8]{inputenc}
\usepackage{color} \usepackage[normalem]{ulem} \usepackage{babel} \usepackage{verbatim}
\usepackage{booktabs} \usepackage{calc} \usepackage{amsmath}
\usepackage{amsthm} \usepackage{graphicx} \usepackage{setspace}
\usepackage[unicode=true, pdfusetitle, bookmarks=true,
bookmarksnumbered=false, bookmarksopen=false, breaklinks=true,
pdfborder={0 0 0}, pdfborderstyle={}, backref=false, colorlinks=true]
{hyperref} \hypersetup{ linkcolor=blue, urlcolor=blue, citecolor=blue,
  pdfstartview={FitH}, hyperfootnotes=false, unicode=true}
\usepackage{chngcntr}

\makeatletter

\usepackage[margin=1in]{geometry}
\usepackage{makecell}
\usepackage[font=small]{caption}

\usepackage{booktabs} 

\PassOptionsToPackage{hyphens}{url}
\usepackage{xurl}

\usepackage{placeins}

\makeatother

\usepackage{natbib}
 \bibpunct[, ]{(}{)}{,}{a}{}{,}%
\newcommand{\newgw}[1]{{#1}}
\newcommand{\newgwR}[1]{{#1}}
\newcommand{\delgw}[1]{}
\DeclareMathOperator*{\argmin}{arg\,min}

\begin{document}

\title{Bidders' Responses to Auction Format Change in Internet Display
  Advertising Auctions}

\author{Shumpei Goke, Gabriel Y. Weintraub, Ralph Mastromonaco, and
  Sam Seljan%
  \thanks{Goke
    (shgoke@protonmail.com). Weintraub: FEN U. de Chile
    (gweintra@fen.uchile.cl).  Mastromonaco
    (ralph.mastro@gmail.com). Seljan: Data Science at People, Inc (seljan@gmail.com).
     The authors thank Dmitry Arkhangelsky, Santiago Balseiro, Liran Einav, Brad Larsen, Jann Spiess, Stefan Wager, and participants in various conferences and seminars, in particular the Stanford IO Student Workshop for
  helpful discussions and suggestions. Samyak Jain provided superb research assistance.  The authors thank AppNexus/Xandr for sharing
  data. However, the opinions expressed in this paper belong to the
  authors and do not necessarily reflect Xandr's nor any other companies' views. Goke thanks Stanford Graduate Fellowship (William
  R. Hewlett Fellowship). Weintraub thanks Stanford GSB, where most of this research was conducted, and gratefully acknowledges financial support from the 2023--2024 Stanford GSB Katherine and David deWilde Faculty Scholarship.}}

\date{This version: July, 2026}

\maketitle

\vspace{-1cm}

\begin{abstract}
  We study actual bidding behavior when a new auction format gets
  introduced into the marketplace. More specifically, we investigate
  this question using a novel dataset on internet display advertising
  auctions that exploits a staggered adoption by different publishers
  (sellers) of first-price auctions (FPAs), instead of the traditional
  second-price auctions (SPAs). 
  \newgwR{We analyze the auction format change using difference-in-differences regressions and a synthetic difference-in-differences estimator, which better handles pre-trends. The results show that revenue per sold impression (price) jumps considerably for treated publishers relative to control publishers, with increases ranging from 25\% to 70\% of the pre-treatment price level of the treated group. Moreover, for later auction format changes, the increase in price levels under FPAs relative to those under SPAs tends to dissipate over time, reminiscent of the revenue equivalence theorem, although the extent of this reversion depends on the specification.
We view these results as suggestive of initially insufficient bid shading following the format change, as opposed to an immediate transition to a new Bayesian Nash equilibrium, with prices tending to decline in several specifications in a manner consistent with gradual adjustment in bidding behavior as bidders learn to shade their bids.}
 Our work constitutes one of the first field studies
  on bidders' responses to auction format changes, providing an
  important complement to theoretical model predictions. As such, it
  provides valuable information to auction designers when considering
  the implementation of different formats.
\end{abstract} \hspace{10pt}


\maketitle
 
\section{Introduction}

The auction literature has thrived in the past several decades,
starting with classical theoretical work such as \citet{Vickrey61:RET}
and followed by more recent advances in empirical studies,
particularly those using structural econometric approaches (see, e.g.,
\citealp{HendricksPorter07:Survey, AtheyHaile07:Survey}). Most of this
literature assumes that auction bidders are rational and, following the game-theoretic
tradition, play Bayesian
Nash equilibrium strategies, . The equilibrium prediction is simple with second-price
auctions (SPAs) under private values: truthful bidding is a dominant
strategy. By contrast, under first-price auctions (FPAs), Bayesian
Nash equilibrium strategies require more sophistication from bidders:
they should optimally shade their bids to balance the trade-off
between paying lower prices and decreasing their chances of winning
\citep{Vickrey61:RET}.

At the same time, \newgw{economists and scientists in other fields are
becoming increasingly aware that agents can deviate from optimal behavior. In particular, an extensive
experimental literature, surveyed in \citet{Kagel95:Survey} and
\citet{KagelLevin16:Survey}, has challenged bidders' rationality
predictions from auction theory.} For example, contrary to equilibrium
predictions, the literature has observed higher prices in first-price
auctions compared to Dutch auctions
\citep{CoxRobertsonSmith82:SingleObject}. More broadly, as researchers
came to find discrepancies between predictions by conventional auction
models and reality, they also got interested in how bidders learn or
adjust their bidding strategies over time. For instance,
\citet{KagelHarstadLevin87:APV} study bidding when experiment subjects
participate in auctions repeatedly.

Despite all these works, to the best of our knowledge, there have been
few \emph{field} studies about learning in
auctions. A notable exception is \citet{Doraszelski_etal18:Learning},
who analyze bidders' learning when a new auction market is
introduced \newgw{(we provide another reference in the context of display advertising later in this section)}. Understanding bidders' responses to a market design change
is fundamental for policymakers as well as profit-maximizing
platforms as they consider which auction format to implement among the
various alternatives. For example, in the real-time sponsored search
advertising market---one of the largest auction markets
worldwide---there was a historical dilemma between using second-price
and first-price auctions and the potential implications for bids and
revenues. A similar discussion arose in the past few years in the
display advertising industry regarding a transition from the
traditionally used second-price auctions to first-price auctions (see
Section \ref{sec:display ad}).

Thus motivated, in this work we present one of the first field studies
in the literature, investigating bidders' responses to a change from
one canonical auction format to another. Specifically, we study how
bidders learn to bid when a new auction format is introduced into the
marketplace. We investigate this question in the setting of internet
display advertising auctions.  To quantify bidders' responses to the
format change over time, we exploit staggered adoptions by different
publishers (sellers) of first-price auctions (FPAs), instead of the
second-price auctions (SPAs) traditionally used in real-time
bidding. We then address two questions: (1) How do bidders learn to bid in the new (FPA) environment, and how do prices evolve as a result? (2) How do different bidders
react to the format change?

We use daily revenue data for auctions administered by a major ad
exchange platform operated by Xandr (formerly known as AppNexus). The
dataset records, for each publisher--bidder pair and on each day, the
number of sold impressions (i.e., the number of auctions resulting in
a sale) on the platform and the aggregate revenue (i.e., the sum of
the auction clearing prices). The scale of the auctions is massive:
our data tallies hundreds of millions of auctions each day. 
In our dataset, publishers switched from SPAs to FPAs in four batches:
(i) September 2017, (ii) September 2019, (iii) April 2020, and (iv)
June 2020. \newgw{We estimate separate difference-in-differences regressions for each one of these events} by contrasting these
publishers (treatment group) with other publishers that did not switch
to FPAs on these dates (control group).

Our results show that, immediately after the format change to FPAs,
the average revenue per sold impression (the average price) jumped
considerably for the treated publishers relative to the control
publishers.  The magnitude of this jump ranges from 35\% to 75\% of
the pre-treatment price level of the treatment group.  In the last three format changes, we observe that the increase in price levels under FPAs relative to those under SPAs dissipates over the next 30 to 60 days, \newgwR{even though some of the estimates are noisy.} 

\newgw{Furthermore, to weaken concerns regarding the parallel trends assumption, we also implemented the recently developed synthetic difference-in-differences (SDID) estimator introduced in \cite{SDID}. This estimator combines a difference-in-differences approach (like the one mentioned above) with a synthetic control approach. The latter can better control for pre-treatment trends, especially with a small number of treated units (one of our events has one treated unit only).} 

\newgwR{Broadly speaking, SDID yields qualitatively similar results to the difference-in-differences approach, though with notable caveats, particularly for the three later auction format changes. The final format change does not show a statistically significant price jump across specifications. The two intermediate cases exhibit patterns that are consistent with initially insufficient bid shading followed by subsequent adjustment, although the strength of these patterns varies across specifications.}

Optimal bidding in SPAs is truthful, while optimal bidding in FPAs involves bid shading. \newgwR{We interpret our results as being consistent with initially suboptimal and insufficient bid shading relative to truthful bidding following the format change from SPA to FPA.}
 If all bidders
were behaving rationally, the average price would move from the mean
of the second-order statistic of the bidder valuations under SPA (or
the reserve price, whichever is higher), to the average highest price
under some Bayesian Nash equilibrium that involves bid shading under
FPA. Furthermore, it would stabilize at the new level immediately
after the format change. \newgwR{The transitory nature of the increase, which is more clearly observed in some specifications than in others, is consistent with initially insufficient bid shading, followed by a reduction in prices as bidders adjust their bidding behavior and start bid shading under FPAs.}

It is interesting to observe that in some of the events \newgwR{and depending on the specification,} the price levels under FPA and SPA
eventually converge. This is reminiscent of the celebrated revenue
equivalence theorem shown by \citet{Vickrey61:RET},
\citet{Myerson81:RET}, and \citet{RileySamuelson81:RET}. We think that
this result is quite remarkable in light of the fact that the
prerequisites for the standard statement of the revenue equivalence
theorem (such as bidder symmetry) generally do not hold in our setting
\citep{MaskinRiley00:Asymmetric}.\footnote{Interestingly, a recent
  study by \citet{BalseiroKroerKumar21:Budgets-RET} derives revenue
  equivalence for standard auctions (including SPAs and FPAs) in a
  setting that encompasses display ad auctions, i.e., where the
  bidders have budgets for multiple auctions and need to pace their
  bids to meet the budget constraint.}

\delgw{Furthermore, it took less time for the price levels under FPAs to go
down to the price levels under SPAs following the format change in
2019 compared to that in 2017, and it took even less time following
the format changes in 2020 than the format change in 2019. 
This pattern suggests that bidders
learn over time how to better shade their bids using a combination of
first-hand experience and industry-wide learning.}

Our results suggest that existing auction theory can fail to correctly
predict bidder behavior in the short run, which is an important fact
for market designers. 
In the short run,
bidders may have trouble bidding optimally and so it may appear that
first-price auctions are driving price increases.  As a result, it is
easy for myopic market designers to draw a wrong conclusion that a
format change increases prices.
However, over the long run, as bidders adjust to market dynamics and
learn to bid more effectively, the price increase \newgwR{may} dissipate. \newgwR{While there are compelling reasons for internet display advertising auctions to switch to FPAs, our results suggest that increasing publishers’ or ad exchanges’ revenue per sold impression may not be among the long-run effects.}

We also study the heterogeneity of the effect of the auction format
changes across bidders. Specifically, we use a \newgw{difference-in-differences} design to
estimate the impact on price, separately for the bidder representing
advertisers that use the bidding algorithm provided by the ad exchange
(``AppNexus/Xandr bidder'') and the rest of the bidders that use other
bidding algorithms. We find that, in three out of four format changes,
the latter type of bidders see a bigger increase in price than the
AppNexus/Xandr bidder. This suggests that the heterogeneity of the
bidders' sophistication impacts how they respond to the format change:
advertisers that use the ad exchange's bidding algorithm seem to be more
sophisticated in bidding, and so they shade more than other, ``naive''
bidders.

Finally, we present several alternative specifications and a
falsifying test as robustness checks. We also present evidence that ad
campaign budgets play a limited role, if any, in the main results;
thus, we believe it is reasonable to interpret our results as a result
of auction format dynamics.

Our work contributes to the growing literature on first-price auctions
in the display advertising industry, such as
\citet{BalseiroKroerKumar21:Budgets-RET} and
\citet{HanZhouWeissman20:LearningFPA}. These papers use theoretical
methods, and our work complements the literature by taking an
empirical approach. Our work also contributes to the operations
management and management science literatures that study various
market design aspects of the display advertising industry; see, e.g.,
\citet{CelisLewisMobiusNazerzadeh14:BuyItNow} on the tension between
surplus from targeting and market thickness,
\citet{GolrezaeiLobelLeme21:ROIConstrained} on financially constrained
buyers, and \citet{FridgeirsdottirNajafi18:CPM} on guaranteed delivery
contracts (a form of selling ad spaces other than by auctions).
\citet{AgarwalNajafiSmith20:DigitalAdLitReview},
\citet{Choi_etal20:DisplayAdLitReview},
\citet{Korula_etal16:OptimizingDisplayAd}, and
\citet{Muthukrishnan09:AdExchangeLitReview} provide surveys on various
issues around the display advertising industry from the operations,
information systems, economics, and computer science perspectives.
More broadly, it is also related to work in operations using
quasi-experimental data to study important changes in digital
platforms, such as \citet{LiNetessine20:MarketThickness_Matching},
\citet{FarronatoFongFradkin:DogEatDog}, and
\citet{GallinoMoreno14:BuyOnlinePickupInStore}.

\newgw{To our
knowledge, our work together with \cite{Auctionformat} are the first in the literature that use data
to study an auction format change in display ad exchanges. \cite{Auctionformat} analyzes the change from second to first-price auction in display advertising in a platform in early  2019. The authors have granular bidding data for one specific publisher from one month
before the switch to three months after the switch. Using the data and structural models, they find evidence of insufficient bid shading 
for about half of the creatives analyzed even three months after the auction format change. We think our sample is more representative with many publishers and four switches that lends itself to an event study type analysis. However, they have more granular bid-level data that lends itself to structural modeling and estimation but only for one publisher and one switch. In that way, we feel our works are complementary, and it is reassuring that we find qualitatively similar results.}

The rest of the paper is organized as follows. Section
\ref{sec:background} provides a detailed background on display
advertising in general, and the overall trend of switching from SPAs
to FPAs in particular. We then explain our dataset and the particular
format changes that we exploit in our econometric studies. Sections
\ref{sec:main-analysis}-\ref{sec:robustness} is the body of our analysis: we present
summary statistics, our regressions and SDID analysis, and an interpretation of
the results. We also conduct several robustness checks
here. Section \ref{sec:bidder-hetero} augments our main analysis by
investigating the heterogeneity of spending responses by
different types of bidders. Section \ref{sec:conclusion} concludes.
The Appendices present supplementary figures and results under
alternative specifications.

\section{Institutional Background and Data}\label{sec:background}

\subsection{The Display Advertising Industry} \label{sec:display ad}

\newgwR{Internet advertising is a huge industry, with revenues of around \$260 billion in 2024. The main formats include search, in-stream video, and display advertising. Display advertising allows website publishers to monetize the advertising space on their websites and accounts for roughly 30\% of total industry revenue. Furthermore, a majority of display advertising is sold using auctions with real-time bidding (RTB), which is the focus of this study.\footnote{\url{https://www.iab.com/wp-content/uploads/2025/04/IAB_PwC-Internet-Ad-Revenue-Report-Full-Year-2024.pdf}} }


\begin{figure}[htbp]
  {\includegraphics[width=0.95\textwidth]{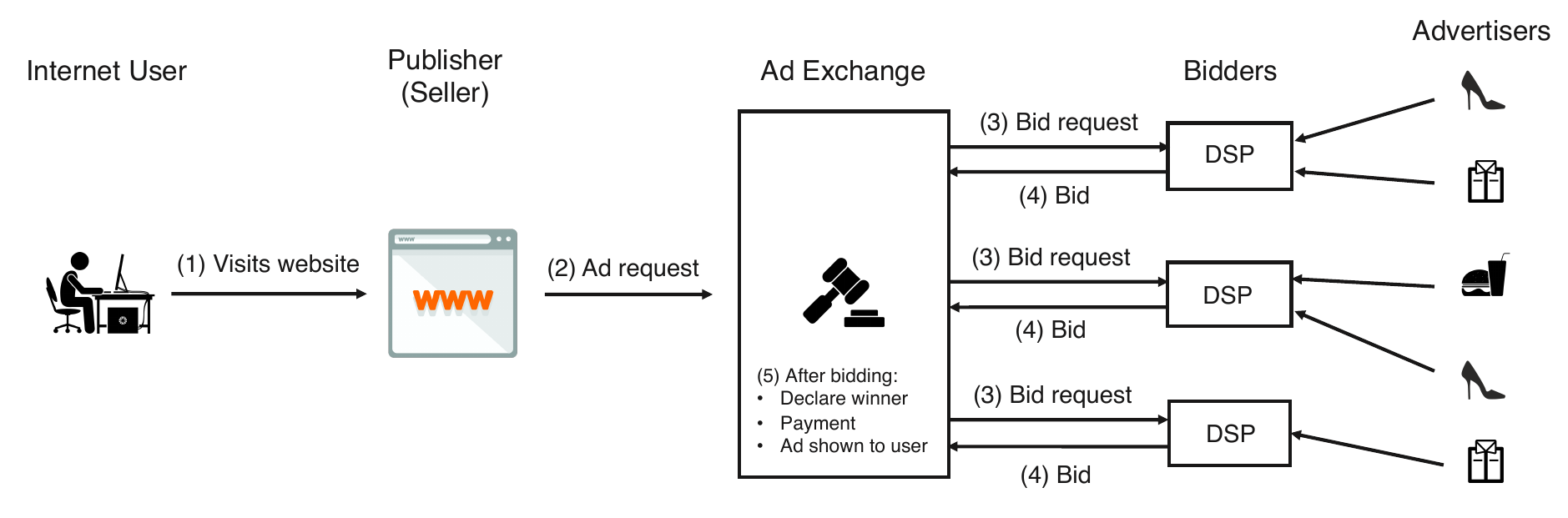}}
  \caption{Diagram of Display Advertising Auctions With Real-Time
    Bidding.  Source: \citet{Yuan_etal14:SurveyRTB}, modified by the authors.}
  \label{fig:diagram-rtb}
\end{figure}

Figure \ref{fig:diagram-rtb} is a diagram, simplified for
presentation, that explains how these auctions are run. The website
publisher prepares a web page that contains ad spaces, which are slots
on the web page dedicated for advertising contents.  The process of
RTB starts when an internet user visits that web page (1), whether on
a computer or a mobile device. The web browser loads the HTML source
code of the web page, which contains a code snippet to show the ad
content. The browser, by loading that code snippet, sends a request to
the ad exchange that an ad be served, which is called an ad
request or ad call (2). The ad exchange then uses an auction to decide
which advertiser will serve the ad. (For many digital publishers,
there are additional systems that make decisions before the ad request
reaches the ad exchange or after the ad exchange selects a winning
bid. However, for the publishers in this study, Figure
\ref{fig:diagram-rtb} is a useful depiction of the process during the
data period.)

Typically advertisers retain intermediaries, called demand-side
platforms (DSPs), that submit bids to the ad exchange on their behalf.
The ad exchange requests bids from the DSPs (3), and the DSPs submit
bids on behalf of advertisers based on the parameters that the
advertisers configure in their advertising campaigns (4). When bids
are collected, the ad exchange determines the winning bidder and the
auction price, and the winning advertiser (selected by the winning
bidder) gets to serve its advertising content to the internet user
(5).

This entire process (starting from the internet user's visit to a
webpage and ending with the ad content being served) is automated and
completed in milliseconds. The ad exchange runs hundreds of billions
of such auctions every day.  Each instance of serving an advertisement
in one ad space is called an \emph{impression}. There is, in
principle, one auction per impression. If the web page contains
multiple ad spaces (e.g., at the top of the page and in the right
column), there are multiple auctions and multiple impressions each
time a user loads that page.

Traditionally, display advertising was sold using second-price
auctions in parallel with the tradition of search advertising
\citep{EdelmanOstrovskySchwarz07:SearchAd, WangZhangYuan17:SurveyRTB}.
However, there has been a growing trend of shifting from second-price
auctions to first-price auctions to sell display ads, culminating in
Google's decision to change its auction format for Google Ad Manager
from SPA to FPA announced in March
2019.\footnote{\url{https://www.blog.google/products/admanager/simplifying-programmatic-first-price-auctions-google-ad-manager/}}
This movement started with the selling side's desire to extract
revenue above the second-highest bid: often, publishers observed a
large gap between the highest and second-highest bids, sometimes as
much as 70\% \citep{AdExchanger16:DynamicFloors}. As a result, the
selling side developed yield-enhancing technologies, such as ``hard
floors'' and ``soft floors.''\footnote{Hard floors are traditional
  reserve prices. Soft floors work as follows. If there are bids above
  the soft floor, the winner, i.e., the bidder with the highest bid,
  will pay the soft floor or the second-highest bid, whichever is
  higher (second-price auction). If all bids are below the soft floor,
  the winner will pay her own bid (first-price
  auction). \citet{Zeithammer19:SoftFloors} analyzes equilibrium
  bidding in auctions with soft floors.} One such technology, Dynamic
Price Floors, which adjusts the price floors programmatically and in
real time \citep{AdExchanger16:DynamicFloors}, was criticized as
opaque \citep{AdExchanger14:DynamicFloorsColdWar}. Advertisers were
especially concerned that the price floors were manipulated so that
they got very close to the highest bid, essentially requiring them to
pay what they bid \citep{AdExchanger15:DynamicFloorsAdFraud}: they
even had suspicions that the price floors were being set after the
bids had been submitted \citep{Digiday17:DeceptiveFloors}. (This is
exactly the incentive compatibility concern regarding SPAs raised by
\citealp{AkbarpourLi20:CredibleAuctions}). First-price auctions have
been seen as a way to solve this transparency concern, while
ostensibly solving sellers' concerns about the gap between the highest
and second-highest bid
\citep{AdExchanger17:ChangesToFirst-Price}. The advent of header bidding also strengthened the argument for the adoption of FPAs \citep{Despotakis_etal21:FPA_ad}. This work excludes data of publishers running header bid auctions.

\subsection{Data and Auction Format Change from SPAs to
  FPAs}\label{subsec:data-chage-dates}

Our goal is to investigate bidders' responses to the switch from SPAs
to FPAs by publishers. For this purpose, we use the dataset of a major
ad exchange platform operated by Xandr. The data is aggregated in the
following manner.  Publisher revenue and the number of sold
impressions are tallied for all auctions run on each day, separately
for each publisher--bidder pair. In other words, our data records that
a given publisher earned \$X by selling Y impressions to a given
bidder.\footnote{In this example, the number of impressions Y excludes
  auctions that did not result in the delivery of advertising contents
  for reasons such as server timeouts or failure of bids to meet the
  reserve price.} We do not have auction-level data such as revenue
and losing bids for each auction. We focus on real-time bidding (RTB)
auctions with no pre-negotiated deals between the publisher and any
bidders.\footnote{For example,  \citet{KimMirrokniNazerzadeh21:Deals} studies the
  impact of deals on publisher revenue.}

We use data on two sets of publishers: publishers owned by a company
that operates globally (``Global Company''), and publishers owned by
three different media companies operating in Europe (``European Media
Companies''). The Global Company has several different functionalities,
and each functionality has a website (publisher) in 
virtually every country/jurisdiction across the world. The three
European Media Companies have many websites (publishers) such as those
for newspapers and magazines: one such company, Company A, has many
publishers in different parts of Europe, while the other two
companies, Companies B and C, operate exclusively in one European
country, Country Y.

\begin{figure}[htbp]
  {\includegraphics[width=0.95\textwidth]{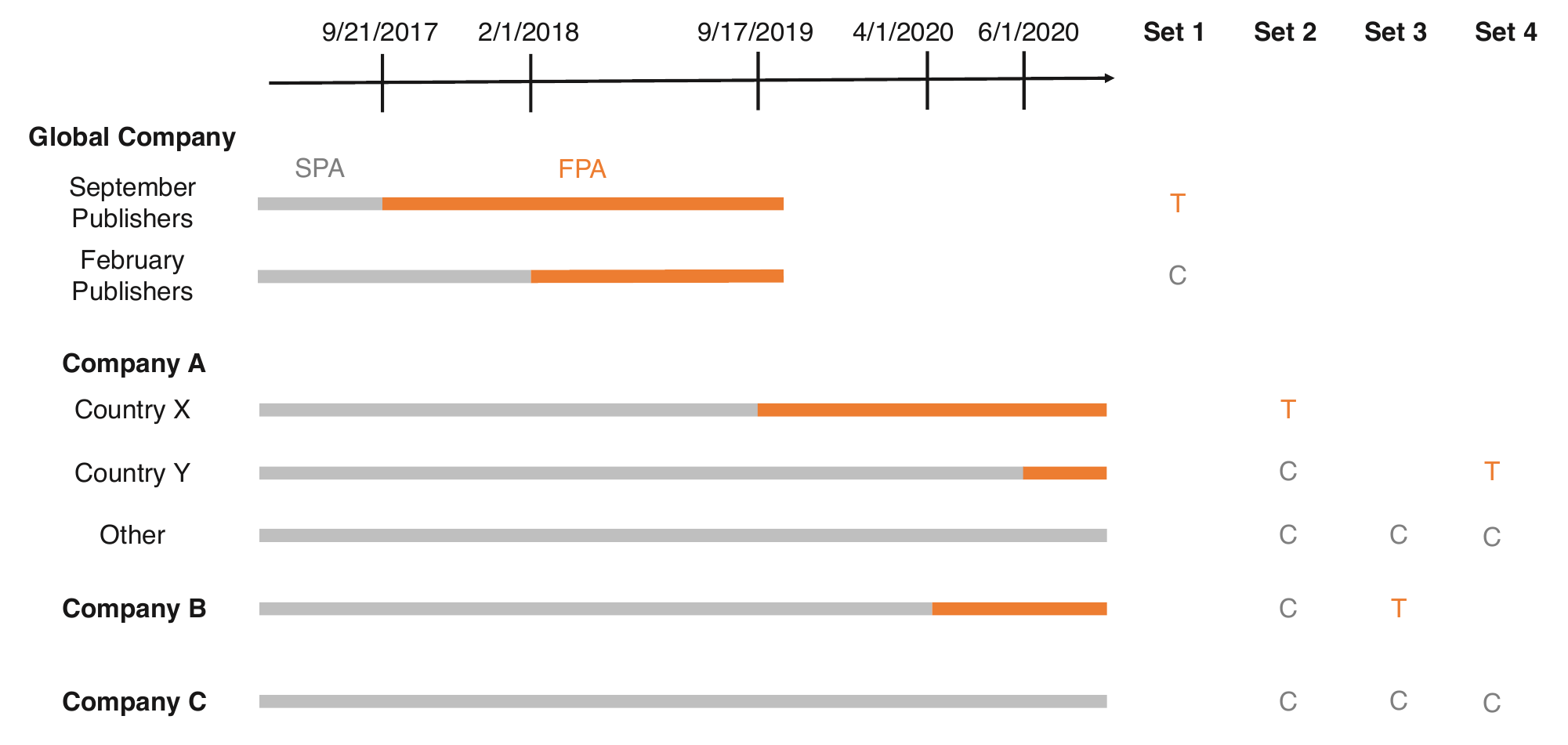}}
  \caption{Types of Publishers and Dates of Format
    Changes.  The four columns on the right indicate treatment--control pairs in
    the \newgw{difference-in-differences} regressions (``T'' indicates the treatment group
    and ``C'' indicates the control group). The gray bars represent
    periods under SPA and the orange bars represent periods under
    FPA.}
  \label{fig:diagram-changedates}
\end{figure}

We compare four sets of treatment--control pairs of publishers (Figure
\ref{fig:diagram-changedates}):

\begin{enumerate}
\item Publishers of the Global Company switched to FPA in two waves: a
  large majority of publishers on September 21, 2017, and the
  remaining publishers on February 1, 2018. The change took place at
  the country level: all publishers in smaller countries/jurisdictions
  switched in September (``September Publishers''), and all publishers
  in larger countries/jurisdictions switched in February (``February
  Publishers''). The data period is June 2011 to September 2019. We
  compare September Publishers (treatment group) to February
  Publishers (control group).
\item European Media Company A, operating internationally, changed the
  auction format for its publishers in Country X to FPA on September
  17, 2019. We compare these publishers (treatment group) to other
  publishers of European Media Companies, i.e., publishers of Company
  A outside of Country X and publishers of Companies B and C (control
  group). The data period is January 2017 to August 2020.
\item All publishers of European Media Company B, operating in Country
  Y, switched to FPA on April 1, 2020. We compare these publishers
  (treatment group) to (i) all publishers of Company A outside of
  Countries X and Y and (ii) all publishers of Company C (control
  group). The data period is January 2017 to August 2020.
\item A single publisher of Media Company A, operating in Country Y,
  switched to FPA on June 1, 2020. We compare this publisher
  (treatment group) to (i) all publishers of Company A outside of
  Countries X and Y and (ii) all publishers of Company C (control
  group). The data period is January 2017 to August 2020.
\end{enumerate}

Switching to FPAs was a big business decision. As a result, the Global
Company piloted FPAs in smaller markets (September Publishers) before
adopting them worldwide. The European Media Companies are smaller and
have less capabilities to ``test and learn'' like the Global Company
does, and so it took longer for them to embrace the change.



\section{Aggregate Response: Summary Statistics}\label{sec:main-analysis}

As a motivating fact, we compare how the average auction clearing
price changes in response to the format change from SPAs to FPAs.
Table \ref{tab:sumstats-pub-level} shows a pre--post comparison at the
treatment--control group level for each of the four format
changes. Panels A to D each correspond to one batch of the auction format
change. \newgw{For each of these changes, we report average daily sold impressions and publishers' revenue, separately for all treated
publishers (left two columns) and all control publishers (right two
columns),}\footnote{For European Media Companies, we exclude certain
  minor/inactive publishers in order to have enough observations
  around the format change date. For the format change in September
  2019, we include only publishers that sold at least 10,000
  impressions and earned \$100 USD every month from August 2017 to
  August 2019. For the two format changes in 2020, we include only
  publishers that sold at least 1,000 impressions every month from
  October 2019 to July 2020. The included publishers account for more
  than 90\% of impressions and revenue in the indicated periods. The
  number of publishers in Table \ref{tab:sumstats-pub-level} does not
  count the excluded publishers. } and separately for the 30-day
period immediately before the format change and the 30-day period
immediately after the format change. We then compute the average daily price
by dividing the revenue by the number of sold impressions. \newgw{We report standard deviations for all quantities.}

For all format changes, we observe that the average price for treated
publishers exhibits an important increase after the format change:
39\% for September Publishers (from \$0.61/1000 to \$0.85/1000), 21\%
for Company A in Country X, 21\% for Company B, and 80\% for Company A
in Country Y.  The corresponding numbers for control publishers are
smaller in magnitude and sometimes negative ($-8$\%, 5\%, $-20$\%, and
15\%, respectively). We see this increase in price across all format
changes even though the price levels differ substantially across
publisher groups.\footnote{The difference in price levels across
  publishers is due mainly to the quality of ad spaces. For instance,
  European Media Companies tend to have higher-quality ad spaces
  because they are media companies that earn an important portion of
  revenue from selling ads. By contrast, the Global Company has its
  main business that does not depend on advertising revenue.}

\begin{figure}[htbp]
  {
    \begin{minipage}{\textwidth}
      \begin{center}
        \includegraphics[height=0.27\textheight]{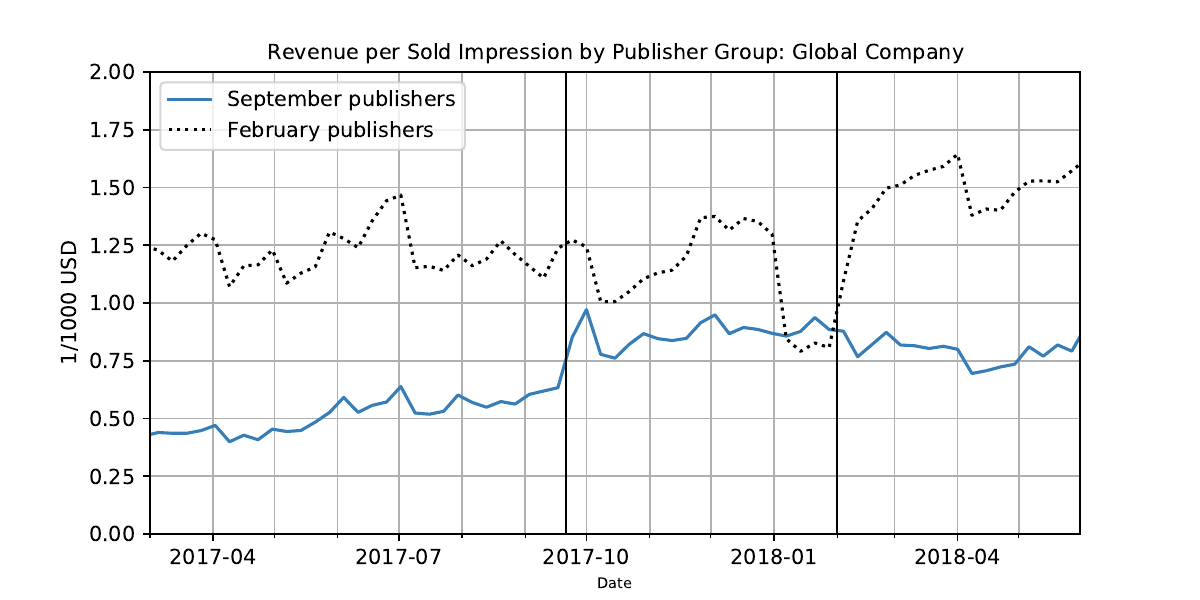} \\
        \includegraphics[height=0.27\textheight]{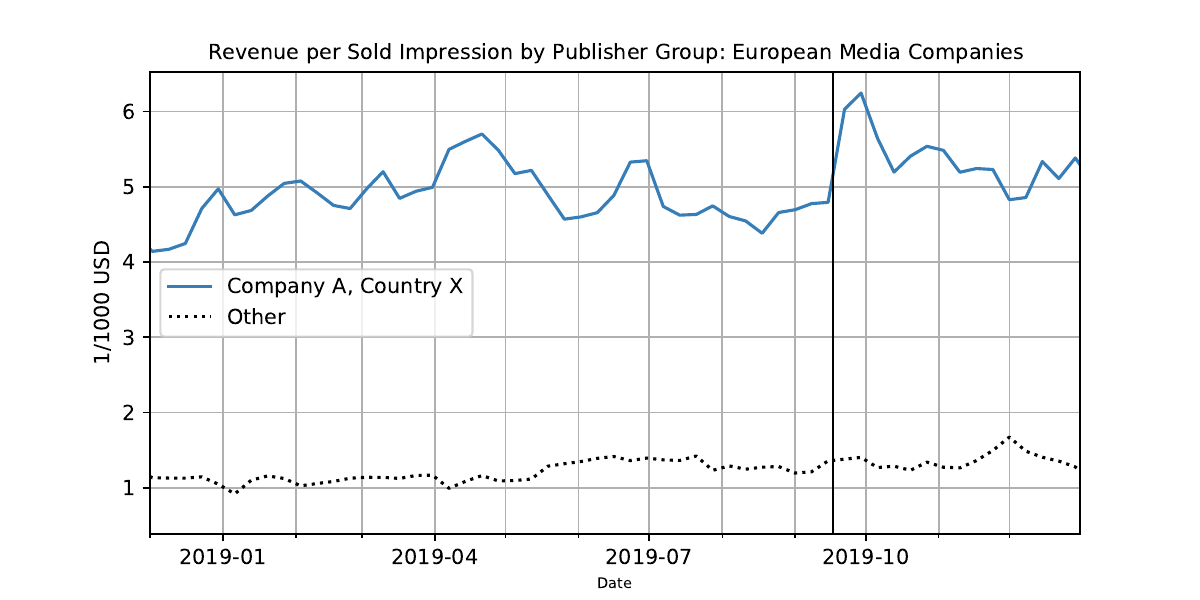} \\
        \includegraphics[height=0.27\textheight]{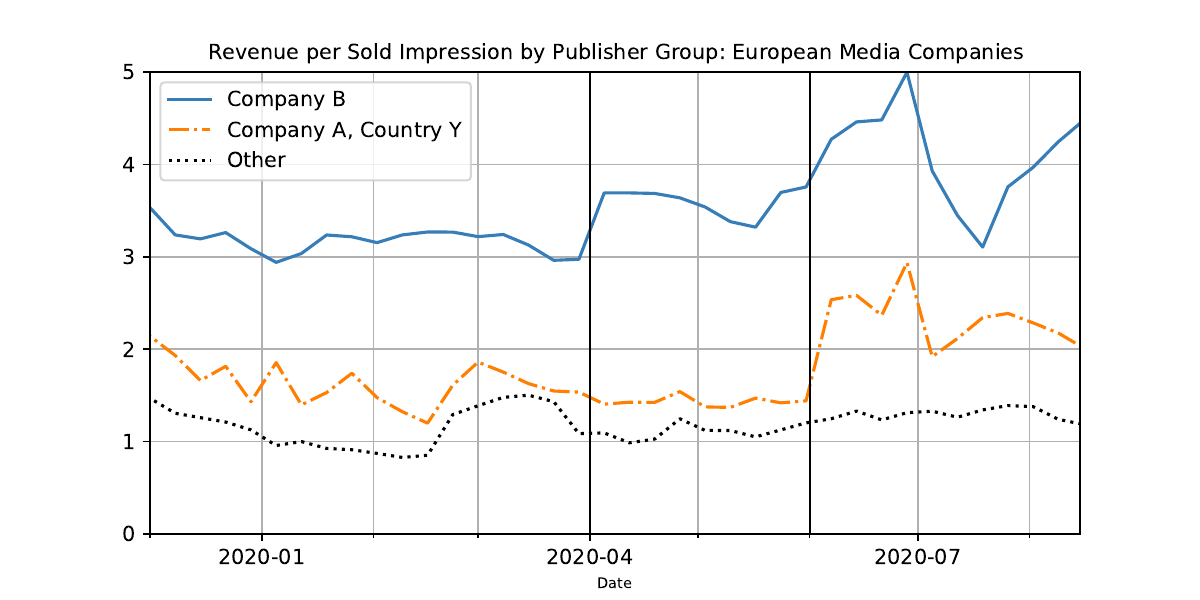}
      \end{center}
    \end{minipage}
  }
  \caption{Weekly Time Series of the Average Price by Publisher
    Group.  The publisher groups are defined as in Table
    \ref{tab:sumstats-pub-level}. The top panel plots the time series
    for September Publishers and February Publishers of the Global
    Company, and the vertical lines indicate their dates of format
    change. The middle panel plots the time series for European Media
    Companies, separately for the publishers of Company A in Country X
    and other publishers (publishers of Company A outside of Country X
    and publishers of Companies B and C). The vertical line indicates
    the date of format change by publishers of Company A in Country
    X. The bottom panel plots the time series for European Media
    Companies that did not switch to FPAs before 2020. The two
    vertical lines indicate the dates of format change by publishers
    of Company B and by the publisher of Company A in Country
    Y.}
  \label{fig:sumstats-pub-level-trend}
\end{figure}

Figure \ref{fig:sumstats-pub-level-trend} visualizes this observation
by plotting the weekly time series of the average price. In this plot,
we aggregate the revenue and the number of sold impressions for all
treated publishers and all control publishers in each week, and
compute the average price by dividing the revenue by the number of
sold impressions. The top panel shows the time series for the Global
Company, and the two vertical lines indicate format change dates for
September Publishers and February Publishers. Looking at the first
format change, we observe a spike in the average price for September
Publishers immediately after they switched to FPAs, but the trend is
stable for February Publishers. The pattern is reversed in the
February 2018 switch: the plot exhibits a spike in the average price
for February Publishers. A similar observation holds for Company A in
Country X (Figure \ref{fig:sumstats-pub-level-trend}, middle panel)
and the 2020 format changes (Figure
\ref{fig:sumstats-pub-level-trend}, bottom panel). Somewhat
surprisingly, the surge in coronavirus cases in Europe and the ensuing
social disruption starting in March 2020 did not affect the average
price for European Media Companies, at least in any obvious manner.

\begin{table}[htbp] 
    \footnotesize
    \begin{tabular}{lrrrr}
      \toprule 
      Panel A\@: Global Company & \multicolumn{2}{c}{} & \multicolumn{2}{c}{}\\
      \midrule 
                                & \multicolumn{2}{c}{September publishers} & \multicolumn{2}{c}{February publishers}\\
      Number of publishers & \multicolumn{2}{c}{160} & \multicolumn{2}{c}{38}\\
                                & 8/22--9/20/2017 & 9/21--10/20/2017 & 8/22--9/20/2017 & 9/21--10/20/2017\\
        Avg \# of sold impressions / day [000 000] & 245.32 (61.42) & 254.26 (65.05) & 423.27 (102.08) & 432.55 (103.8) \\ 
        Avg revenue / day [000 USD] & 149.23 (29.65) & 215.78 (46.16) & 502.95 (110.34) & 474.33 (114.6) \\ 
        Avg daily price [1/1000 USD] & 0.61 (0.06) & 0.85 (0.15) & 1.19 (0.10) & 1.1 (0.16) \\ 
                                &  &  &  & \\
      \multicolumn{3}{l}{Panel B\@: European Media Companies, first batch} &  & \\
      \midrule
                                & \multicolumn{2}{c}{Company A, Country X} & \multicolumn{2}{c}{Controls}\\
      Number of publishers & \multicolumn{2}{c}{44} & \multicolumn{2}{c}{32}\\
                                & 8/18--9/16/2019 & 9/17--10/16/2019 & 8/18--9/16/2019 & 9/17--10/16/2019\\
        Avg \# of sold impressions / day [000 000] & 2.98 (0.40) & 1.71 (0.31) & 17.41 (2.81) & 21.43 (2.27) \\ 
        Avg revenue / day [000 USD] & 14.04 (1.63) & 9.74 (1.52) & 22.07 (4.40) & 28.51 (3.37) \\ 
        Avg daily price [1/1000 USD] & 4.71 (0.15) & 5.7 (0.69) & 1.27 (0.08) & 1.33 (0.12) \\ 
                                &  &  &  & \\
      \multicolumn{3}{l}{Panel C\@: European Media Companies, second batch} &  & \\
      \midrule
                                & \multicolumn{2}{c}{European Media Company B} & \multicolumn{2}{c}{Controls}\\
      Number of publishers & \multicolumn{2}{c}{15} & \multicolumn{2}{c}{48}\\
                                & 3/2--3/31/2020 & 4/1--4/30/2020 & 3/2--3/31/2020 & 4/1--4/30/2020\\
        Avg \# of sold impressions / day [000 000] & 1.3 (0.14) & 0.91 (0.13) & 6.29 (1.98) & 5.55 (0.65) \\ 
        Avg revenue / day [000 USD] & 4.00 (0.49) & 3.38 (0.53) & 8.68 (3.30) & 6.15 (0.98) \\ 
        Avg daily price [1/1000 USD] & 3.07 (0.16) & 3.72 (0.29) & 1.38 (0.20) & 1.11 (0.13) \\ 
                                &  &  &  & \\
      \multicolumn{3}{l}{Panel D\@: European Media Companies, third batch} &  & \\
      \midrule
                                & \multicolumn{2}{c}{Company A, Country Y} & \multicolumn{2}{c}{Controls}\\
      Number of publishers & \multicolumn{2}{c}{1} & \multicolumn{2}{c}{48}\\
                                & 5/2--5/31/2020 & 6/1--6/30/2020 & 5/2--5/31/2020 & 6/1--6/30/2020\\
        Avg \# of sold impressions / day [000 000] & 0.25 (0.04) & 0.20 (0.04) & 7.31 (2.29) & 8.65 (0.88) \\ 
        Avg revenue / day [000 USD] & 0.36 (0.05) & 0.52 (0.12) & 8.21 (2.72) & 11.15 (1.19) \\ 
        Avg daily price [1/1000 USD] & 1.42 (0.07) & 2.55 (0.42) & 1.12 (0.09) & 1.29 (0.09) \\ 
      \bottomrule
    \end{tabular}
  
  \caption{\newgw{Summary Statistics Before and After Format Changes. Daily Averages (Standard Deviations).}.  \newgw{Comparison of daily number of sold impressions, revenue, and average
    price (revenue divided by the number of sold impressions) during the 30-day period before and after the auction format
    change. The left two columns summarize data for publishers that
    switched from SPAs to FPAs on September 21, 2017 (Panel A),
    September 17, 2019 (Panel B), April 1, 2020 (Panel C), and June 1,
    2020 (Panel D) (treatment group), and the right two columns
    summarize data for publishers that did not switch at these times
    (control group). The number of impressions and the revenue are
    aggregated to the daily level for both groups of publishers, and the resulting 30 day time series are used to calculate averages and standard deviations (in parenthesis). }}
  \label{tab:sumstats-pub-level}
\end{table} 

\section{Aggregate Response: Difference-in-Differences  Regressions} \label{sec:DID}

\subsection{Specification}

As Table \ref{tab:sumstats-pub-level} and Figure
\ref{fig:sumstats-pub-level-trend} indicate, the publishers that
switched to FPAs experienced a surge in average prices. In this
section, we formalize this observation through a rigorous \newgw{difference-in-differences (DID) design} 
that controls for publisher and time-fixed effects, as
well as seasonality, which is known to be prevalent in the industry.

\newgw{First, we create a dataset for each treatment--control group pair in Figure
\ref{fig:diagram-changedates}; note that there is a single treatment date within
each dataset. We truncate the data period before the control
publishers (e.g., Global Company February Publishers) switched to
FPAs, so that the data period is June 2011 to February 2018 for Set 1,
January 2017 to February 2020 for Set 2, and January 2017 to August
2020 for Sets 3 and 4.}

\newgw{Second, we aggregate data to the publisher--day level. That is for every publisher $p$ and day $t$, we tally the
revenue and the number of impressions sold by publisher $p$ in day $t$, and by dividing those two numbers, we obtain the average price, $y_{pt}$.  
The data is winsorized by capping the values of
$y_{pt}$ at the 0.1 percentile from below and at the 99.9 percentile from
above. }

We then estimate
the following regression equation, separately for each
treatment--control group pair:
\begin{equation}
  y_{pt} = \alpha_{p} + \sum_{\underline{k}\leq k\leq\overline{k},\,k\neq-1}\beta_{k}D_{p}\cdot\boldsymbol{1}(K_{t}=k)
  + \gamma_{t}
  + \gamma_{p,\mathrm{dow}(t)} + \gamma_{p,\mathrm{dom}(t)} + \gamma_{p,\mathrm{month}(t)} + \gamma_{p,\mathrm{eoq}(t)}
  + \varepsilon_{pt},
  \label{eq:1step-reg}
\end{equation}
where $p$ is a publisher, $t$ is a day, $y_{pt}$ is average price,
$\alpha_{p}$ is publisher fixed effect, and $\gamma_{t}$ is time (day)
fixed effect. We also include publisher-specific seasonal fixed
effects $\gamma_{p,\mathrm{dow}(t)}$, $\gamma_{p,\mathrm{dom}(t)}$,
$\gamma_{p,\mathrm{month}(t)}$, and $\gamma_{p,\mathrm{eoq}(t)}$. In
other words, we have the following fixed effects, separately for each
publisher $p$: (i) day of week (7 fixed effects per publisher before
perfect multicollinearity is removed), (ii) day of month (30 fixed
effects per publisher),\footnote{We use the same fixed effect for the
  30th and 31st days of the month, as there are fewer observations on
  the 31st day.}  (iii) month (12 fixed effects per publisher), and
(iv) end of quarter (2 fixed effects per publisher, one for the last
14 days of every March, June, September, and December combined, and
another for days other than at the end of the quarter).

The coefficients of interests are $\beta_{k}$. The variable $D_{p}$ is
the treatment indicator, which takes a value of 1 if publisher $p$ is
in the treatment group. This is interacted with dummy variables for
$K_{t}$ (number of days from the date of format change till $t$, which
is censored at a negative number $\underline{k}$ from below and a
positive number $\overline{k}$ from above). In the estimation, we take
$\underline{k}=-65$, $\overline{k}=65$ and plot estimates for
$-60\leq k\leq60$. We omit the parameter for $k=-1$, and hence all
estimates are with respect to the day before the auction format
change.

The regressions are weighted by the number of impressions sold
by publisher $p$ on day $t$ so that larger publishers have more
influence on the estimates than smaller publishers. In other words,
$y_{pt}$ is an average of ``grouped data''
\citep[][p.~92]{AngristPischke09:MHE}. The standard errors are
clustered at the publisher level \citep{Bertrand_etal04:DIDClusterSE}.

It is worth mentioning how our regression specification stands in
relation to the recent literature on event study regressions with
two-way fixed effects. First,
\citet{DeChaisemartinD'Haultfoeuille20:TwowayFE} and
\citet{BorusyakJaravelSpiess21:EventStudy} show that (the probability
limit of) the estimated coefficient in a static
DID regression (i.e., only one coefficient
for the treatment effect) is a weighted average of unit- and
time-specific treatment effects, where the weights may be negative. As
a result, it is possible to obtain an estimate with the wrong sign
(e.g., the estimated coefficient may be negative even if all true
treatment effects are positive). Second,
\citet{Goodman-Bacon21:DID-DeltaATT} shows that the estimated
coefficient in a static DID regression may be biased, because it may
be confounded by the time-series change in the treatment effect of the
treated units. Both these problems arise only if there are units with
different event dates, i.e., different cohorts. Thirdly, \citet{SunAbraham21:Event_DynamicEffects} consider a dynamic
specification like ours, i.e., a specification that uses a separate
dummy variable for each relative period from the event date. They show
that, when there are multiple cohorts, $\beta_k$ (to use our notation)
is a weighted average not only of conditional average treatment
effects $k$ periods after treatment, but also of conditional average
treatment effects $k'$ periods after treatment for all possible $k'$
corresponding to other cohorts. We note that none of these problems are present in
our regressions, as we have a single cohort/event date within each of
the four treatment--control comparisons  (DID regressions).

\subsection{Identifying Assumptions} \label{sec:assDID}

The main identifying assumptions of \newgw{DID} regressions are (i)
the exogeneity of treatment assignment and (ii) the common trend
(parallel trend) assumption
\citep{DeChaisemartinD'Haultfoeuille20:TwowayFE}.  As for (i), Section
\ref{subsec:data-chage-dates} explains the high-level motivation for
deciding the format change dates. Display advertising was adopting FPAs
as an industry trend, and companies that have the capability to
``test and learn'' are doing so first. When deciding on particular
dates (i.e., why those dates rather than one week earlier), there are
various factors to consider, such as staff availability to
support/supervise the process of format change. Overall, the
publishers set dates when there are unlikely to be any factors that
confound the impact of format change on prices so that their analysts
can investigate the impact of the format change.  To investigate
whether the parallel trend assumption holds (ii), we will discuss the
pre-trend (estimates of $\beta_{k}$ for $k<0$) in Section
\ref{subsec:publevel-results}, as in \citet{Autor03:GrangerCausality}
and \citet{AngristPischke09:MHE}. \newgw{In addition, in Section \ref{sec:SDID} we will introduce a synthetic difference-in-differences approach that alleviates potential issues with pre-treatment trends.}

On a different note, our estimates may potentially reflect
market equilibrium effects: if, for instance, the average price for
treated publishers goes up, bidders may substitute away from treated
publishers to control publishers.\footnote{Technically, this is a
  violation of the stable unit treatment value assumption (SUTVA;
  \citealp[][p.~10]{ImbensRudin15:Book}).} 
We believe that
such concerns are limited in our case, however. For the Global
Company, September Publishers and February Publishers serve distinct
geographical markets: they serve internet users of different countries
and jurisdictions, often using different languages. As for the format
change by European Media Company A in 2019, no control publishers
operate in Country X (because there were few suitable candidates on
the ad exchange)\@. In 2020, the control group does include publishers
operating in Country Y, but they occupy only 3.3\% of the impressions
sold and 12.6\% of the revenue earned by all the control
publishers. See also Section \ref{subsec:alt-control-grp} for a
robustness check, where we rerun the regressions only using
publishers outside of Country Y as the control group.

\subsection{Estimation Results} \label{subsec:publevel-results}

\begin{figure}[htbp]
  {
    \begin{minipage}{\textwidth}
      \begin{center}
        \includegraphics[width=0.48\textwidth]{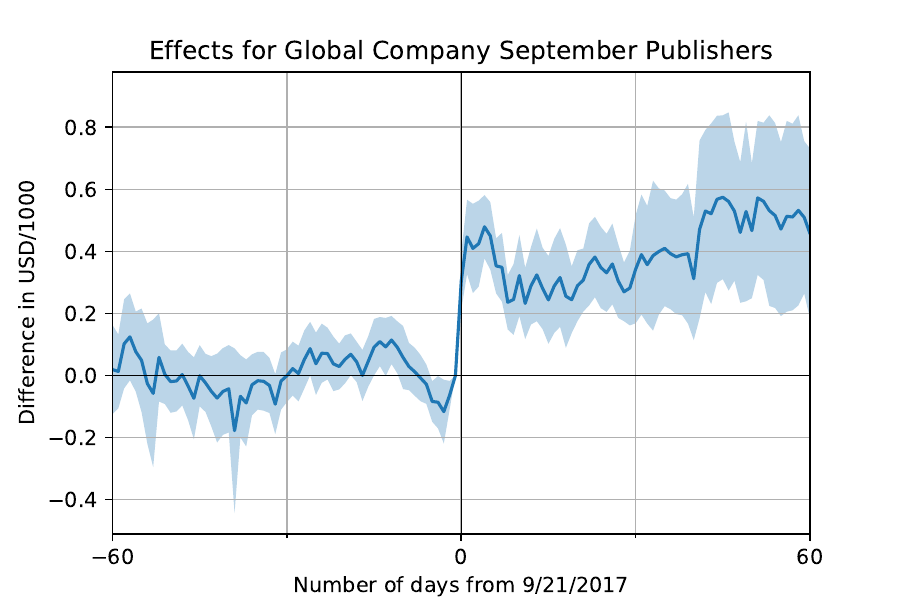}
        \includegraphics[width=0.48\textwidth]{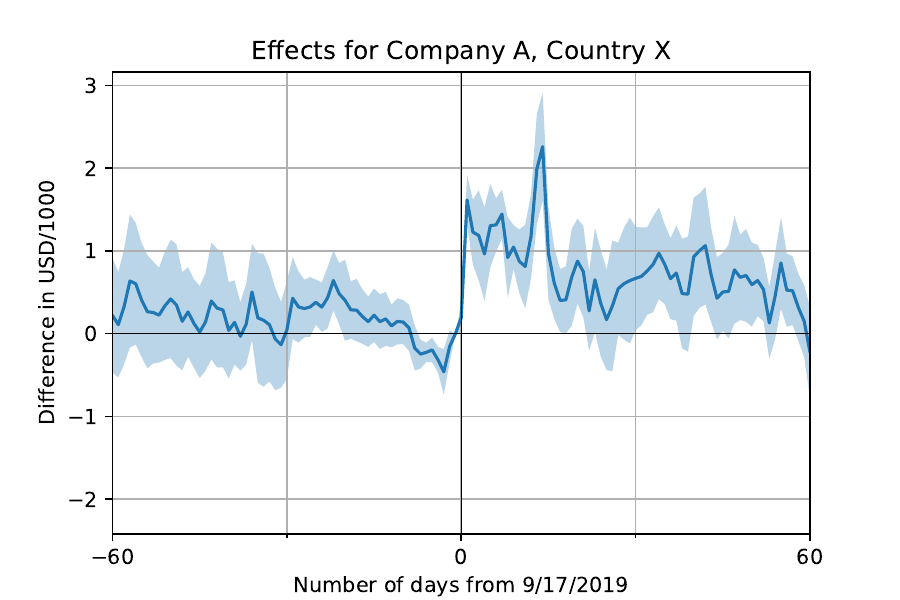} \\
        \includegraphics[width=0.48\textwidth]{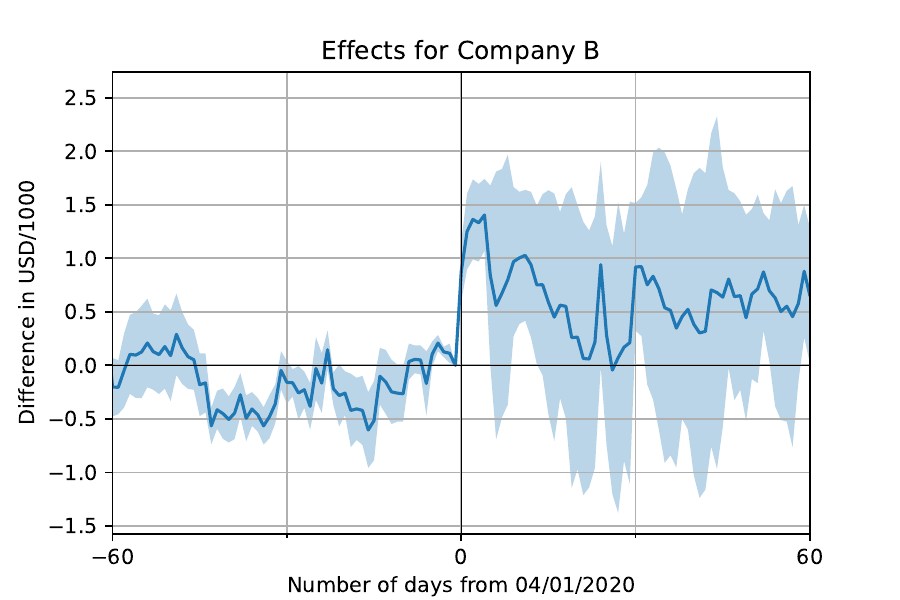}
        \includegraphics[width=0.48\textwidth]{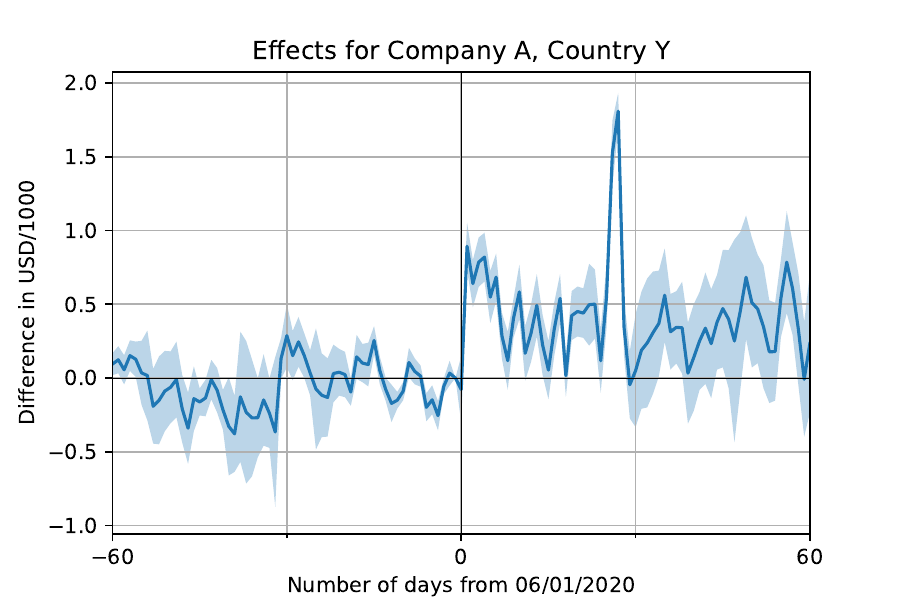}
      \end{center}
    \end{minipage}
  }
  \caption{Estimated Effects of Format Change on Average
    Price.  The solid line indicates point estimates of $\beta_{k}$, and the
    band indicates 95\% confidence intervals.}
  \label{fig:event-estimates}
\end{figure}

Figure \ref{fig:event-estimates} depicts the point estimates and 95\%
confidence intervals for $\beta_{k}$'s for different pairs of
treatment and control groups. In each of the four treatment--control
pairs, we observe an immediate jump in the average price for the
treated publishers. The pre-trends are either statistically
insignificant or, even if they are significant, much smaller in
magnitude than the estimated treatment effects.

The top left panel shows estimates for Global Company's September
Publishers. The format change increased the average price on the day
following the format change ($\beta_{1}$) by 0.45/1000 USD, relative
to the counterfactual price level that would have been obtained if
these publishers had continued to run SPAs. This increase in price is
substantial: it is 73\% of the average price for September Publishers
during the 30-day period immediately before the change (shown in Table
\ref{tab:sumstats-pub-level}).\footnote{\newgw{There is evidence
that due to targeting, auctions in display advertising may be ‘thin’ and the difference between
the highest and second highest bid may be large \cite{CelisLewisMobiusNazerzadeh14:BuyItNow}. This together with insufficient bid shading as discussed below may explain these large differences.}} The average price under FPAs continues
to be higher than SPAs until $k = 60$.

The top right panel shows estimates for the publishers of Company A in
Country X with respect to the September 2019 format
change. $\beta_{1}$ is estimated to be 1.62/1000 USD, which is 34\% of
the average price for the publishers of Company A in Country X during
the 30-day period immediately before the format change. This time, the
increase in average price is transitory: the estimated effect
diminishes over time and becomes statistically insignificant as $k$ approaches 60.
The bottom two panels show estimates for the two batches of format
changes in 2020, with the left panel showing the effects for the
publishers of Company B after April 1 and the right panel showing the
effects for the publisher of Company A in Country Y after June 1. The
estimates for $\beta_{1}$ are 1.25/1000 USD for publishers of Company
B and 0.89/1000 USD for Company A's publisher in Country Y (41\% and
63\% of their respective average price levels during the immediately
preceding 30-day period). Again, the increase in the average price
under FPAs diminishes over time and becomes statistically insignificant as $k$
approaches 30.\footnote{\newgw{When considering longer panels we obtain similar results. The exception is Company B, for which the impact on the average price of the auction format change becomes significant after two months and insignificant again after three months. We obtain somewhat of a similar pattern with synthetic differences-in-differences below.}}

\subsection{Interpretation of Results}\label{subsec:Pub-Level-Discussion}

\newgwR{The 
results suggest that at least a subset of bidders responded
suboptimally to the format change.} Imagine that all bidders were
rational and assume, for simplicity, that they have private values:
under the SPA, they bid their valuation of each impression, and so the
average revenue is the mean of the maximum of the second-highest
valuation by bidders and the reserve price. After the format change,
the bidders would shade their bids relative to their valuations
according to some Bayesian Nash equilibrium. The average price would
stabilize, immediately after the format change, at a level sustained
by the equilibrium. Contrary to these predictions, we observe that the
average price levels initially went up after each of the format
changes compared to the average price levels under SPAs, and that the
increase \newgwR{seems to dissipate} over time for {\em the three format changes in 2019 and
2020.}  

We interpret this \emph{transitory} increase in prices as
evidence that (i) some bidders shaded their bids insufficiently under
the new regime of FPAs relative to their rational, best-response
strategy, and (ii) these bidders gradually learned to shade their bids
to a level sustained by a rational strategy. It is important to note
that the transition to FPA for each publisher was both transparent at
the auction level---the auction type was sent in the bid
request---and communicated proactively by publishers to demand-side
platforms.

Incidentally, the average price level under FPAs eventually falls \newgwR{to levels that are statistically close to, and generally statistically indistinguishable from, those under SPAs.}
This is an interesting observation reminiscent of the celebrated revenue
equivalence theorem shown by \citet{Vickrey61:RET},
\citet{Myerson81:RET}, and \citet{RileySamuelson81:RET}. We believe
that this result is noteworthy and intriguing in light of the fact
that the prerequisites for the classic revenue equivalence theorem
(such as bidder symmetry) generally do not hold in our setting
\citep{MaskinRiley00:Asymmetric}.\footnote{We note another difference between our result and the classic revenue equivalence theorem. We compare revenue per \emph{sold} impression, i.e., revenue per auction where the highest bid exceeds the reserve price, under SPAs and FPAs. On the other hand, the revenue equivalence theorem concerns revenue per \emph{available} impression, i.e., revenue per auction considering the possibility of not selling the impression (if all bids are below the reserve price), in which case the auctioneer receives its opportunity cost. Unfortunately, due to data limitations, we have not been able to test the equivalence of revenue per available impression under SPAs and FPAs. We do not observe in our data the number of available but unsold impressions nor relevant reserve prices (price floors).}

\newgwR{Having said all of that, we interpret the DID regression results cautiously, given evidence of pre-trends in some specifications and the imprecision of certain estimates (e.g., for Company B). To address these concerns—particularly pre-trends—we introduce a synthetic difference-in-differences approach in the next section and revisit the results in light of this analysis.}

\delgw{Second, we note that it took less time for the average price levels
under FPAs to go down to the average price levels under SPAs following
the format changes in 2020 compared to the format change in 2019. Note that in the 2017 format change, the average price
was still not going down 60 days after the initial jump. By contrast,
in the 2019 format change, the average price levels under FPAs went
down to the average price levels under SPAs within 60 days. This
period until the price decrease was reduced to 30 days in the 2020
format changes.  We interpret this as evidence of long-term learning
whereby bidders got better and faster in adjusting their bids, whether
through their first-hand experience of format changes, industry-wide
learning, or a combination of the two.}
 
\section{Aggregate Response: Synthetic Difference-in-Differences} \label{sec:SDID}

\subsection{Specification}

\newgw{As discussed in Section \ref{sec:assDID}, a key assumption in a DID regression is the parallel trends assumption. In a recent paper, \cite{SDID} introduces synthetic difference-in-differences (SDID) estimators, which combines the strengths of DID with synthetic controls. Synthetic controls, as pioneered by \cite{Abadie2003}, address potential violations of parallel trends by re-weighting units, and have been particularly used in settings with a small number of treated units. SDID merges these two approaches, offering better controls and alleviating pre-treatment trend issues.}

\newgw{We provide a brief description of SDID, following the presentation of \cite{SDID} (more technical details can be found in that paper). Suppose we index the $P$ units (publishers) so that the first $P_{co}$ (control) units are not treated. The rest of the units are treated. Let $y_{pt}$ be the outcome variable (e.g., average price) for unit $p$ and time period $t$, where $T$ is the total number of periods and $T_{pre}<T$ is the number of pre-treatment periods. The first step of SDID, as in synthetic control methods, is to find weights, $\hat\omega_p$, to match control outcomes with treatment outcomes in pre-treatment periods:} $$\sum_{p=1}^{P_{co}}\hat \omega_p y_{pt}\approx \frac{1}{P-P_{co}}\sum_{p=P_{co}+1}^{P}y_{pt}\ ,\forall t=1,..,T_{pre}.$$ SDID also considers time-weights, $\hat\lambda_t$, \newgwR{to align average pre and post-treatment outcomes for control units:}
$$ \sum_{t=1}^{T_{pre}}\hat \lambda_t y_{pt}\approx \frac{1}{T-T_{pre}}\sum_{t=T_{pre}+1}^{T}y_{pt}\ ,\forall p=1,..,P_{co}.$$

 \newgw{Then, these weights are used in a DID regression, similar to the one in Section \ref{sec:DID}.} \newgwR{The use of the weights in the SDID estimator place greater emphasis on units whose pre-treatment outcomes are more similar to those of the treated units, as well as on time periods that are more comparable to the treated periods.}
To be more specific, the SDID estimator solves:
\begin{equation} \label{eq:SDID} (\hat{\tau}^{SDID}, \hat{\mu}, \hat{\alpha}, \hat{\beta}) = \argmin_{\tau,\mu,\alpha,\beta} \left\{ \sum_{p=1}^P \sum_{t=1}^T \left(y_{pt} - \mu - \alpha_p - \gamma_t - D_{pt} \tau \right)^2 \hat{\omega}_p \hat{\lambda}_t \right\} \ ,\end{equation}
\newgw{where as before $\alpha_p$ are unit fixed effects, $\gamma_t$ are time fixed-effects, and $D_{pt}$ is a treatment indicator if unit $p$ has been treated in period $t$. Note that 
SDID weighs more heavily units that are more similar on average in the past to treated units as well as pre-treatment periods that are more similar on average to treated periods. We run four separate SDID regressions, one per treatment--control group pair in Figure
\ref{fig:diagram-changedates}. We use the implementation provided by the authors of \cite{SDID}.}\footnote{https://github.com/synth-inference/synthdid}

\newgw{To prepare the data, we follow a similar procedure to Section \ref{sec:DID} with a few changes to accommodate SDID as we describe now. First, to reduce the dimensionality we aggregate data to weeks. We considered 20 weeks prior to the treatment week in our data sets, which seems like a reasonable time period to match synthetic controls (we also tried 10 weeks without major changes in the results). In addition, the data is winsorized by capping the values of
$y_{pt}$ at the 0.1 percentile from below and at the 99.9 percentile from
above.\footnote{To align with the implementation of SDID we did not weight impressions in the regressions as before. However, we re-run the regressions, removing publishers with an oversized number of impressions without observing major  changes to the results.}}

\newgw{In Section \ref{sec:DID}, we considered average price as the outcome variable. In Section \ref{sec:robust} we also consider a specification in which the outcome variable is the log of average price. While the results are similar, we decided to keep the average price in our main specification of DID because the log specification exhibits more pronounced pre-treatment trends. Since one of the purposes of SDID is to correct for these pre-treatment trends, we consider the log of average price here since percentage changes may be appropriate when considering price changes (the results without logs are qualitatively similar).}

\newgw{To correct for seasonality in a way consistent with the implementation we use for SDID, we implement a `two-step' method, in which in the first step we `de-season' the data and in the second step we run SDID. In the first de-season step, using the entire time series of weekly data before the format change we regress, for each publisher $p$:\footnote{We weigh the observations
  by the number of impressions.}}
  \[ log ~ y_{pt} = \gamma_{p,\mathrm{wom}(t)} +
    \gamma_{p,\mathrm{month}(t)} + \gamma_{p,\mathrm{eoq}(t)}+\delta_{pt}, \]
\newgw{ where $\gamma_{p,\mathrm{wom}(t)}$ is a week of month fixed effect (1,2,3,4,5), $\gamma_{p,\mathrm{month}(t)}$ is a month fixed effect (1 to 12) and $\gamma_{p,\mathrm{eoq}}$ is an end of quarter fixed effect (takes value of 1 if week corresponds to the last two in a quarter); all of these are publisher specific. Then, we compute the residuals of these regressions (that have the seasonality component removed through the fitted values), and these residuals are the outcome variables used for SDID, $\tilde{y}_{pt}$.}

\newgw{Once we have obtained the de-seasonalized time series $\tilde{y}_{pt}$, we produce SDID estimates separately for each week up to 10 weeks after the auction format change date. 
The standard SDID implementation does not estimate dynamic treatment effects, that is a different treatment effect for each week after the auction format change, as we intend to do. To accommodate this dynamic specification, we implement the following procedure. Let us define the week in which the auction format changes for treated publishers is week 0. All other weeks are assigned a numeric label equal to the number of weeks from the auction format change (e.g., the time period 3 weeks prior to treatment is assigned -3, the time period 3 weeks after treatment is assigned 3).}

\newgw{Then, for each post-treatment week $w \in [1, 10]$, we 
create a subset of the full dataset with weeks $\{[-20, 0]\} \cup \{w\}$. Then, we calculate an SDID estimate $\tau_w$ with log average price as the outcome, weeks $[-20, 0]$ as pre-treatment periods, and week $w$ as the single post-treatment period using equation \eqref{eq:SDID}. $\tau_w$ is the treatment effect of the auction format change on log average price of week $w$.\footnote{We thank Dmitry Arkhangelsky for suggesting this implementation of dynamic treatment effects.} We use the implementations of \cite{SDID} to compute standard errors. For three of the events we use the bootstrap method. For the last event we only have one treated unit, so we use the placebo method.}

\subsection{\newgw{Estimation Results}}

\newgw{In Figure \ref{fig:parallel-trends}  we present treated and synthetic control trajectories produced by SDID for the four events that we study. Note that the pre-treatment trajectories do not change as we iterate through treated week $w$ in the procedure described at the end of the last subsection. The figures show how SDID generates trajectories that look reasonably parallel. The treatment effects and the time weights are presented for a single treated week, $w=1$.}

\begin{figure}[htbp]
  {
    \begin{minipage}{\textwidth}
      \begin{center}
        \includegraphics[width=0.48\textwidth]{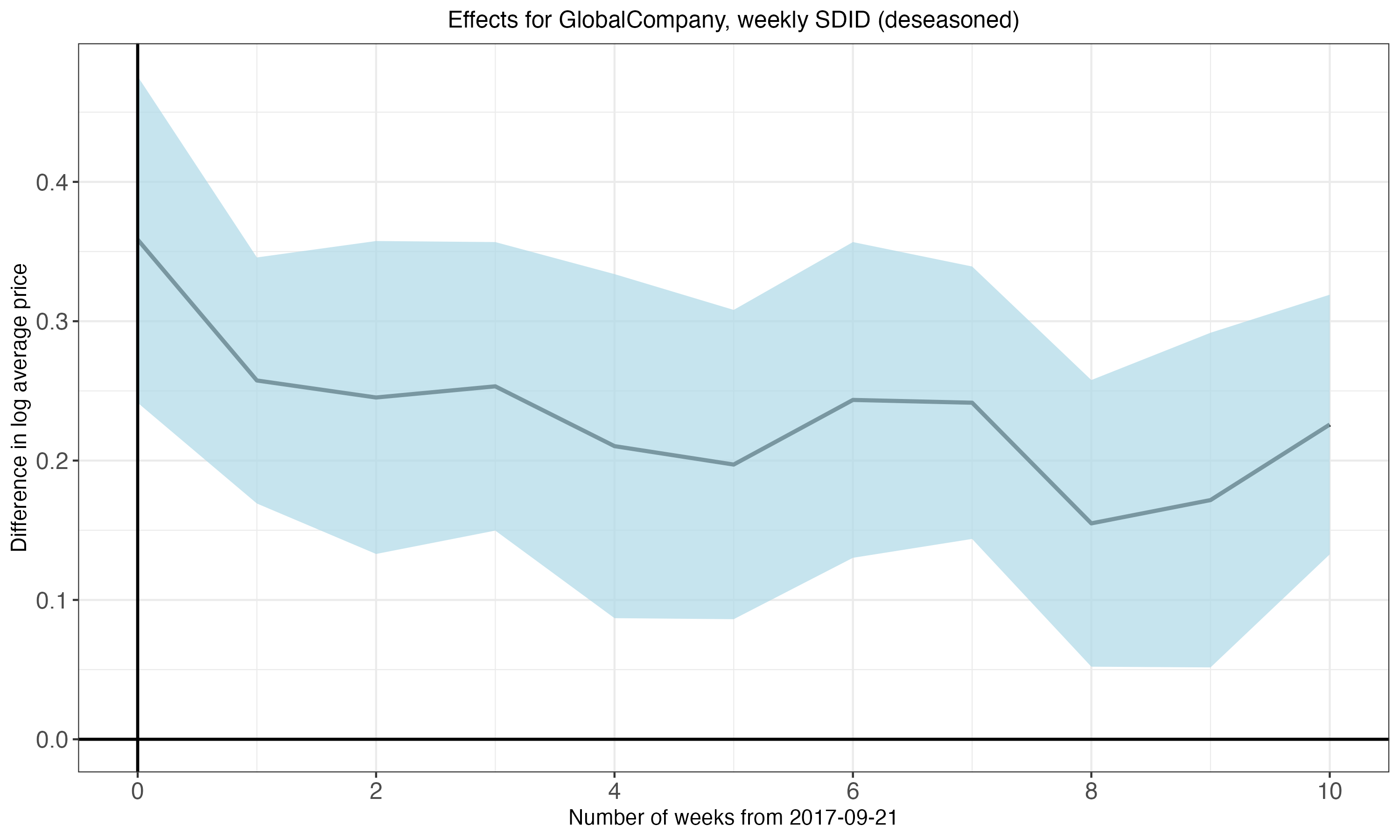}
        \includegraphics[width=0.48\textwidth]{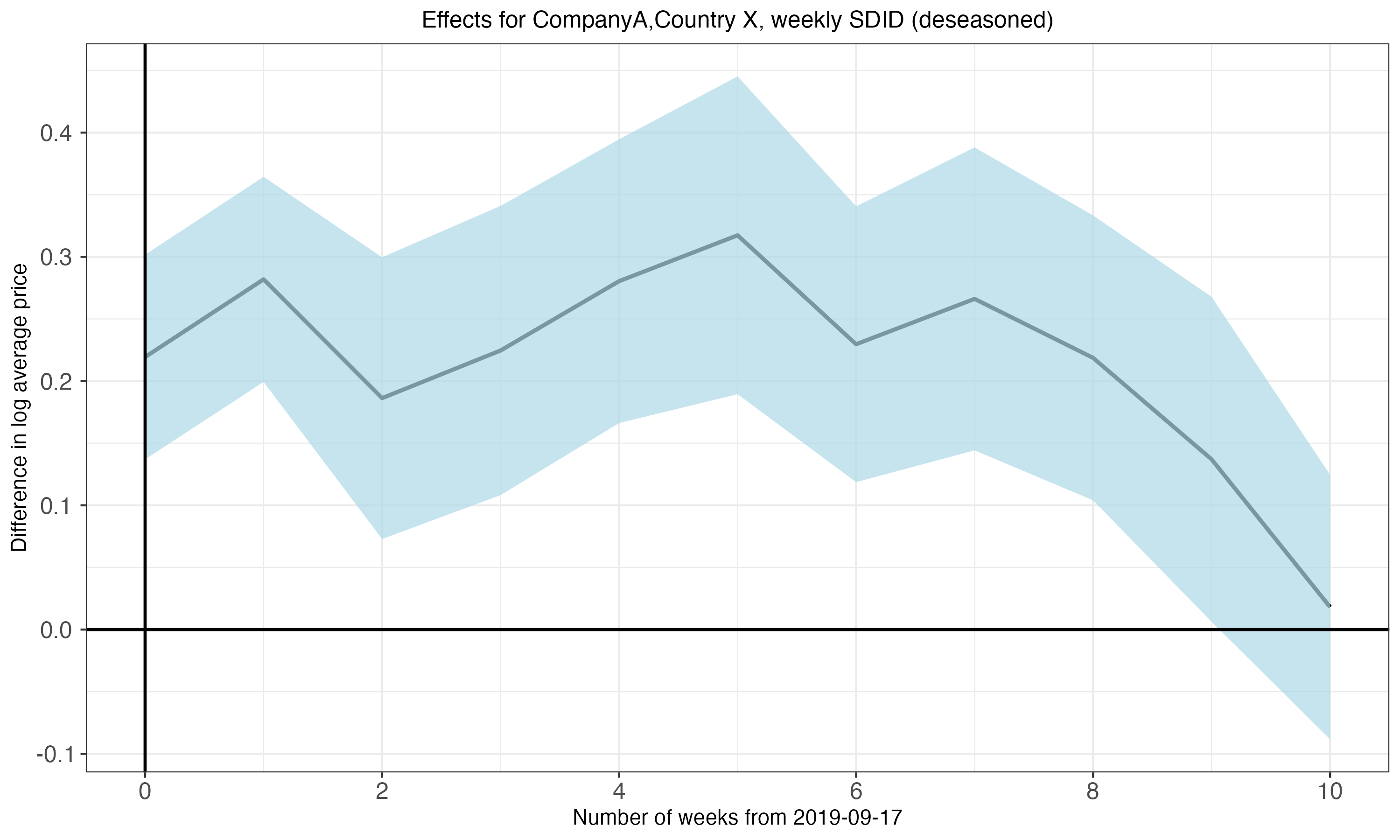} \\
        \includegraphics[width=0.48\textwidth]{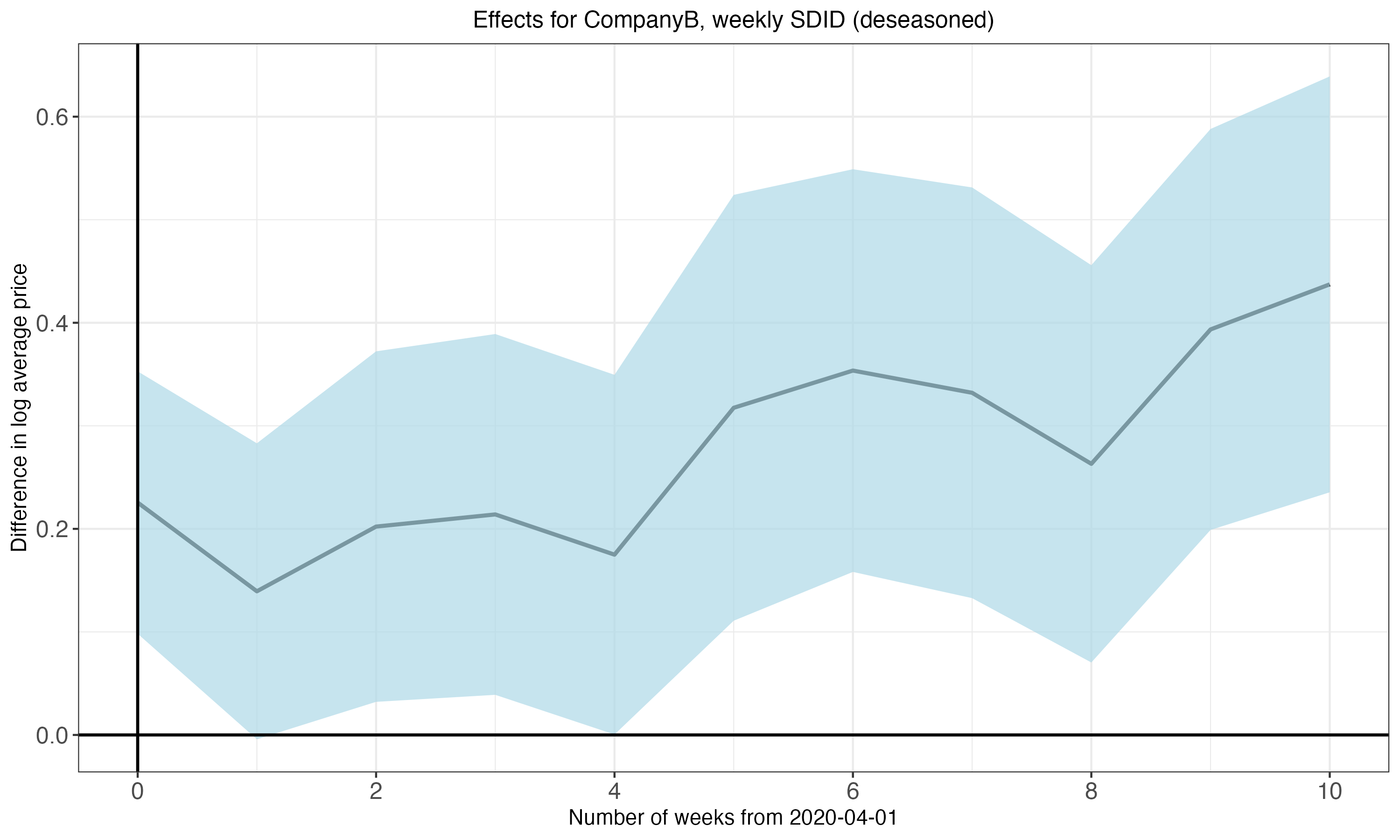}
        \includegraphics[width=0.48\textwidth]{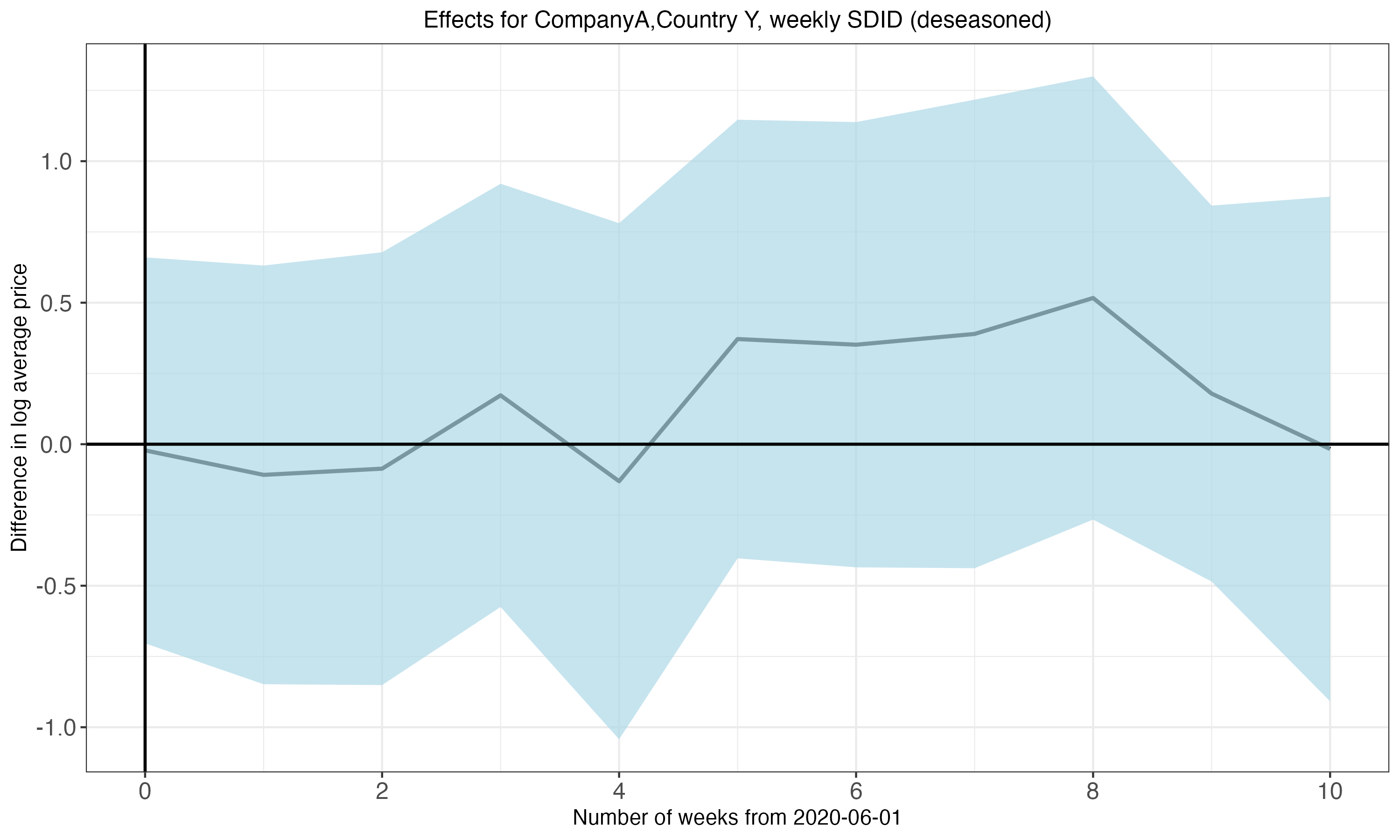}
      \end{center}
    \end{minipage}
  }
  \caption{SDID Estimated Effects of Format Change on Average
    Price (Logs).  The solid line indicates point estimates of $\tau_{w}$, and the band indicates 95\% confidence intervals.}
  \label{fig:SDID-estimates}
\end{figure}

\newgw{In Figure \ref{fig:SDID-estimates} we show the estimated SDID treatment effects with 95\% confidence intervals for each of the pairs of treatment and control groups discussed earlier. The top left panel shows the SDID estimates for the Global Company's September Publishers. In the immediate week following the auction format change, log average price increases by 0.36, indicating a 43\% increase in average price relative to the pre-treatment level. While the percentage increase is not as pronounced as in the DID estimates, qualitatively, the results are similar. In fact, even though we see a slight decrease over time, the average price does not return to pre-treatment levels before the control publishers also switch to first-price auctions. }


\begin{figure}[htbp]
  {
    \begin{minipage}{\textwidth}
      \begin{center}
        \includegraphics[width=0.48\textwidth]{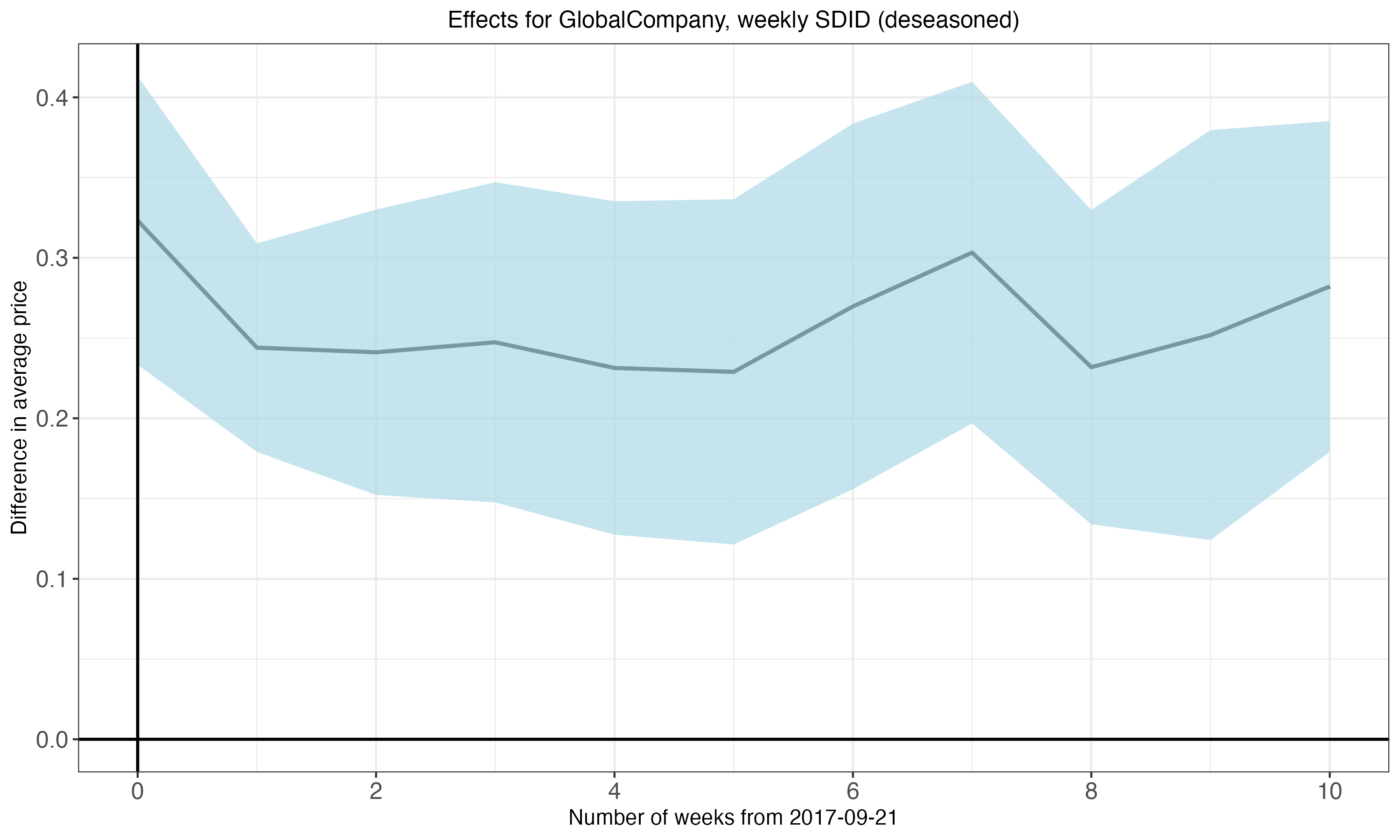}
        \includegraphics[width=0.48\textwidth]{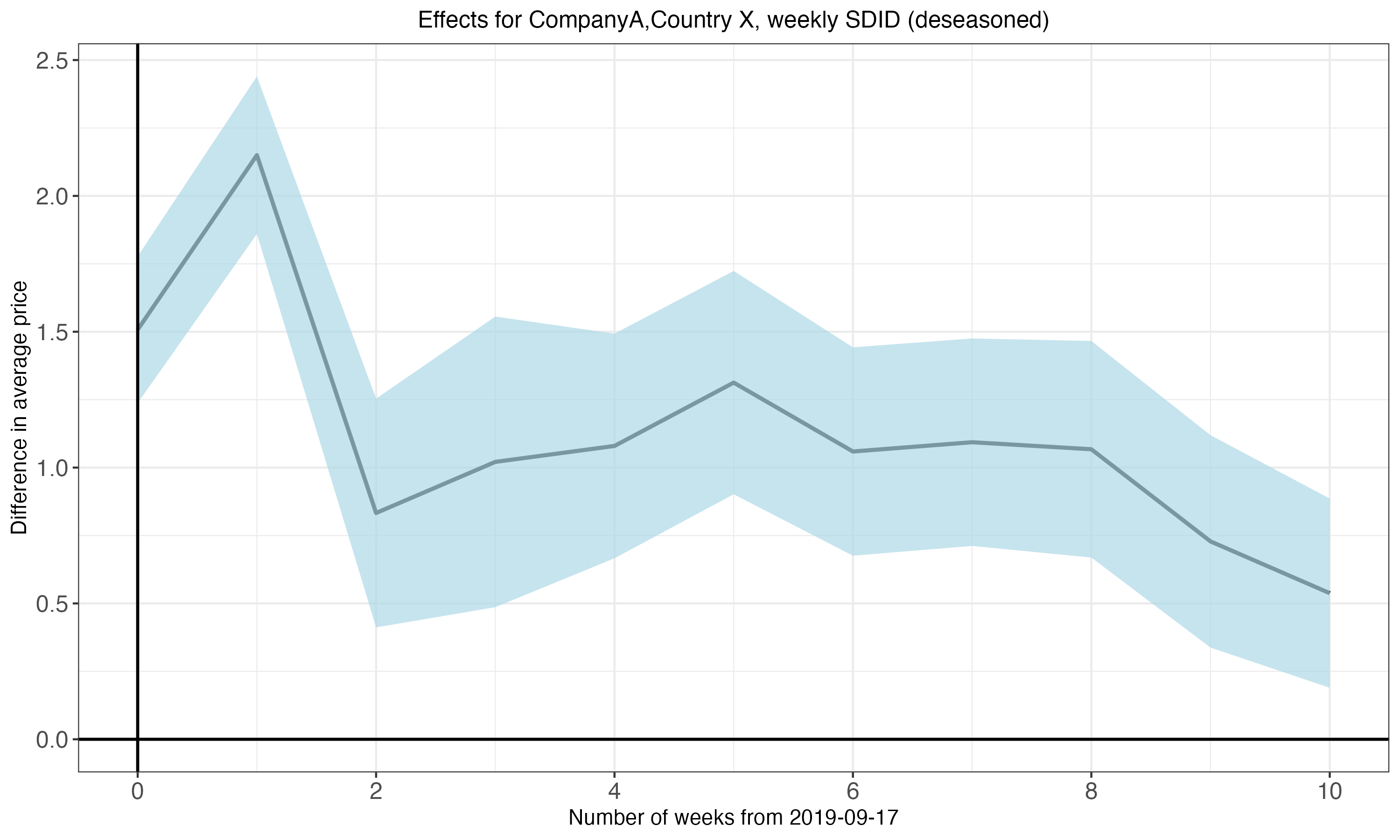} \\
        \includegraphics[width=0.48\textwidth]{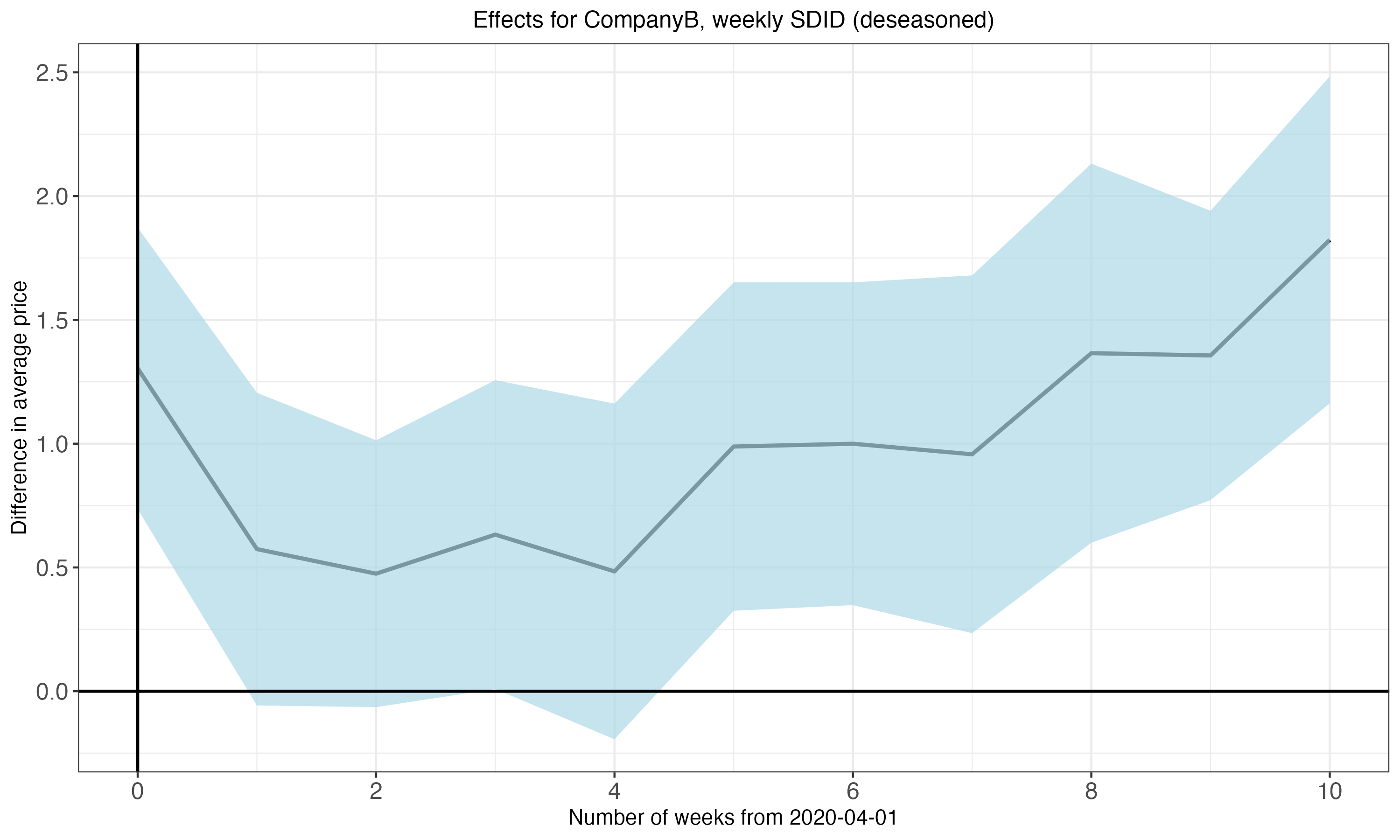}
        \includegraphics[width=0.48\textwidth]{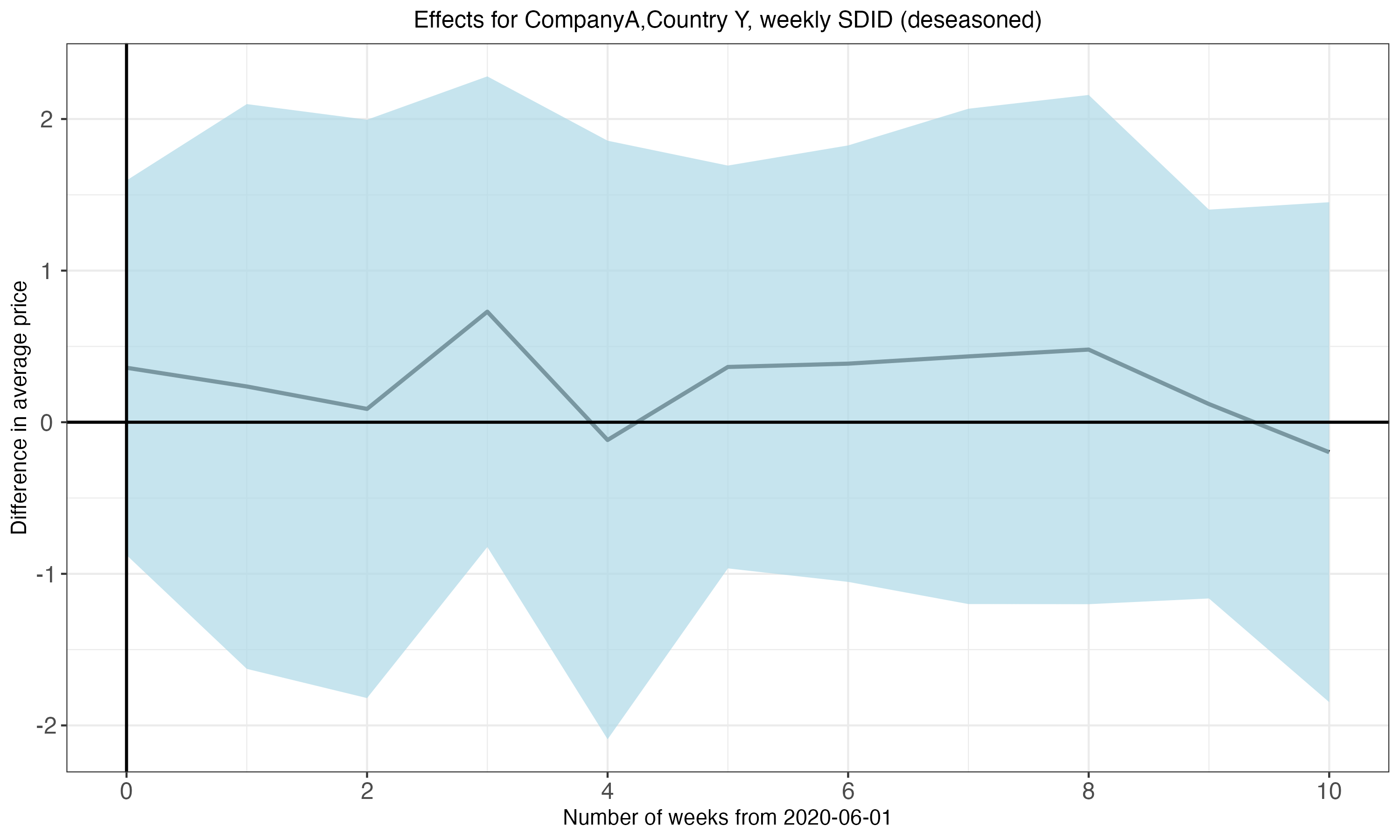}
      \end{center}
    \end{minipage}
  }
  \caption{\newgwR{SDID Estimated Effects of Format Change on Average
    Price (USD/1000) }.  The solid line indicates point estimates of $\tau_{w}$, and the band indicates 95\% confidence intervals.}
  \label{fig:SDID-estimates-rps}
\end{figure}

\newgw{The top right panel shows the format change effects for European Media Company A, Country X. Log average price increases by 0.22, corresponding to a 25\% increase on average price, similar to the DID estimates. Also, in line with DID estimates, the price level decreases to pre-treatment levels around 10 weeks after treatment. }

The estimated increase in log average price for European Media Company B is 0.22 after treatment, corresponding to an average price increase of 25\%. While the lower bound of the confidence interval of average price reaches pre-treatment levels after a few weeks, it starts to increase after that. An extended time horizon shows the average price going back to pre-treatment levels at week 15 \newgwR{(see Figure \ref{fig:SDID-estimates-15w} in Appendix.)}\footnote{\newgwR{We also note that in the extended time horizon, the average price increases above pre-treatment levels for Company A, Country X, which we discuss more below.}}  SDID estimates are less noisy and somewhat differ from DID; in DID  the average price starts to increase after two months and goes back to pre-treatment levels after three months.

\newgw{Finally, the results for European Media Company A, Country Y 
show no immediate change on average prices after the auction format change. 
The  estimated effects are highest in week 8, where log average price increases by 0.52 (68\% increase in average price). However, in this event there is only one treated publisher and confidence intervals are wide and cross zero through out the horizon.}

\newgwR{Figure \ref{fig:SDID-estimates-rps} shows the SDID results in levels rather than logs. The patterns are quite similar across the two specifications, except that for Company A, Country X, in levels the price approaches but does not fully return to its pre-treatment level. In addition, for Company B, the average price gets closer, though not fully, to the pre-treatment level by week 15.}

\newgwR{In the Appendix we also present the results in logs and levels with daily instead of weekly data (Figures \ref{fig:SDID-daily-estimates} and \ref{fig:SDID-daily-estimates-rps}). The results are consistent with the weekly data, with one caveat: in logs, for the Global Company the effect becomes statistically indistinguishable from zero for a few days after day 60.}

\subsection{\newgwR{Re-interpretation of Results}}

\newgwR{In Section \ref{subsec:Pub-Level-Discussion}, we discussed how the difference-in-differences analysis suggests suboptimal bidding, as average prices jump after the format change, consistent with a lack of bid shading when transitioning from SPAs to FPAs, followed by a degree of learning in which prices approach pre-treatment levels.}

\newgwR{The SDID results are broadly consistent with this interpretation, but with some notable caveats. The results for the Global Company are similar to those from the DID analysis (except for a few days in one of the daily specifications, as discussed above), with average prices increasing after the format change and not returning to pre-treatment levels within the horizon we consider. The last format change (Company A, Country Y) does not exhibit a statistically significant price jump after the format change across specifications, which one could interpret as bidders having already learned to bid optimally in FPAs. However, this case involves only one treated publisher, and the resulting estimates are noisy.}

\newgwR{The two middle cases provide evidence of insufficient shading followed by learning, but the effects are more nuanced. In fact, for Company A, Country X, there is an initial price jump that dissipates (or nearly dissipates, depending on the specification) around 10 weeks. In some specifications, average prices increase somewhat after 10 weeks, however; we do not have a clear explanation for this pattern. For Company B, prices do not return to pre-treatment (or close to) levels within 10 weeks across specifications, but do so when considering longer horizons of up to 15 weeks}


\section{Robustness Checks}
\label{sec:robustness}
\subsection{Role of Advertising Campaign Budgets.}

A potential concern with the above interpretation relates to the role
of advertising campaign budgets. \newgw{Advertisers, apart from having an overall ad spending budget, typically set specific campaign budgets, e.g., to target certain users in a specific pre-determined time horizon. (See, e.g.,
\citet{Balseiro_etal15:Budgets} for a theoretical treatment of budget
constraints in display advertising auctions.)}  The concern for our analysis is that
bidders would spend a fixed amount of budget on the treated publishers, and
whatever phenomena take place after the format change are caused by
budget constraints rather than auction format dynamics.

We believe, however, that budget constraints play a limited role, if any, in the
results discussed above. \newgw{First, we show that the auction format change for the most part did not have a significant effect on the volume of impressions sold. Suppose bidders would spend a fixed amount of budget on the treated publishers. Then, if the number of impressions sold on treated publishers decreases after the auction format change, revenue per impression would increase mechanically.} 

\begin{figure}[htbp]
  {
    \begin{minipage}{\textwidth}
      \begin{center}
        \includegraphics[width=0.48\textwidth]{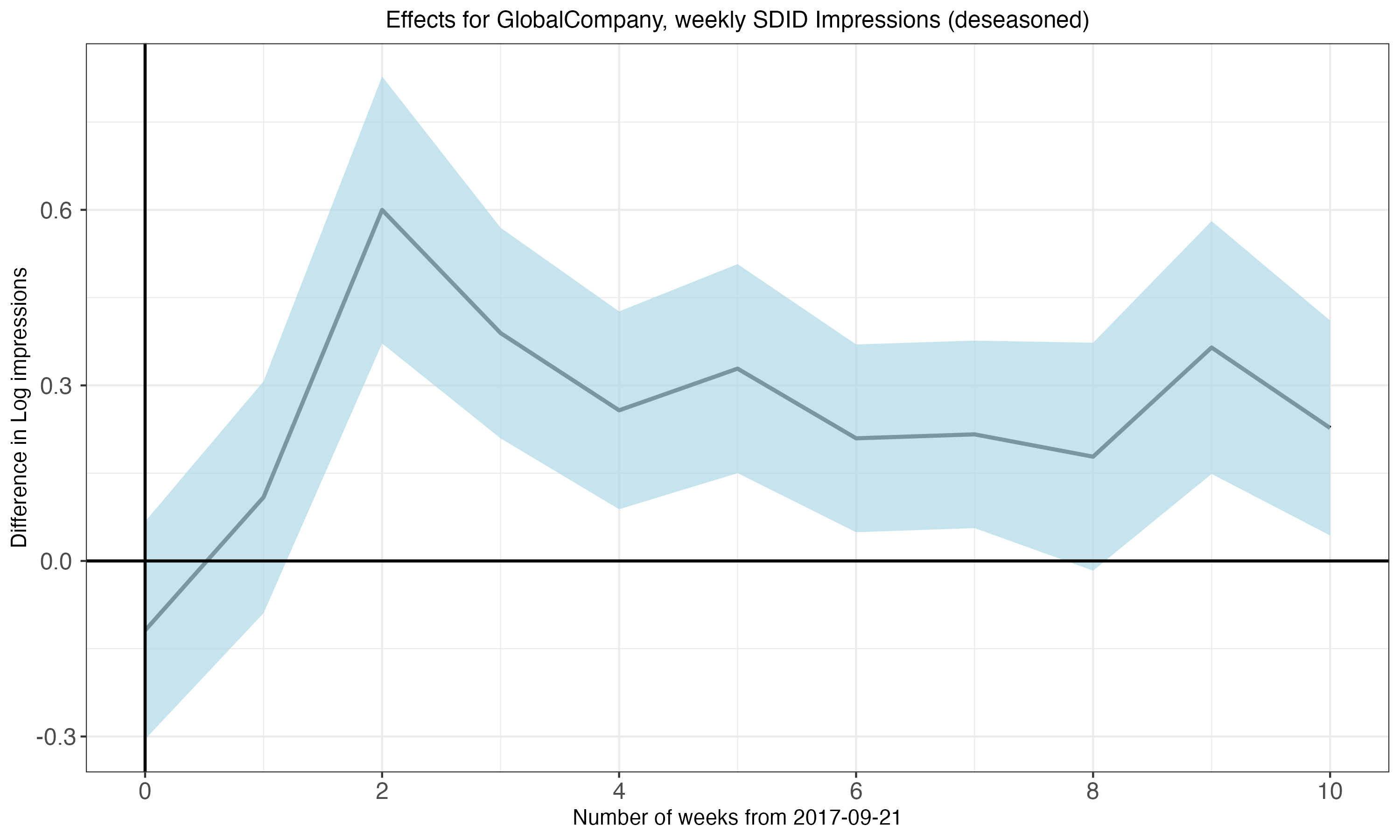}
        \includegraphics[width=0.48\textwidth]{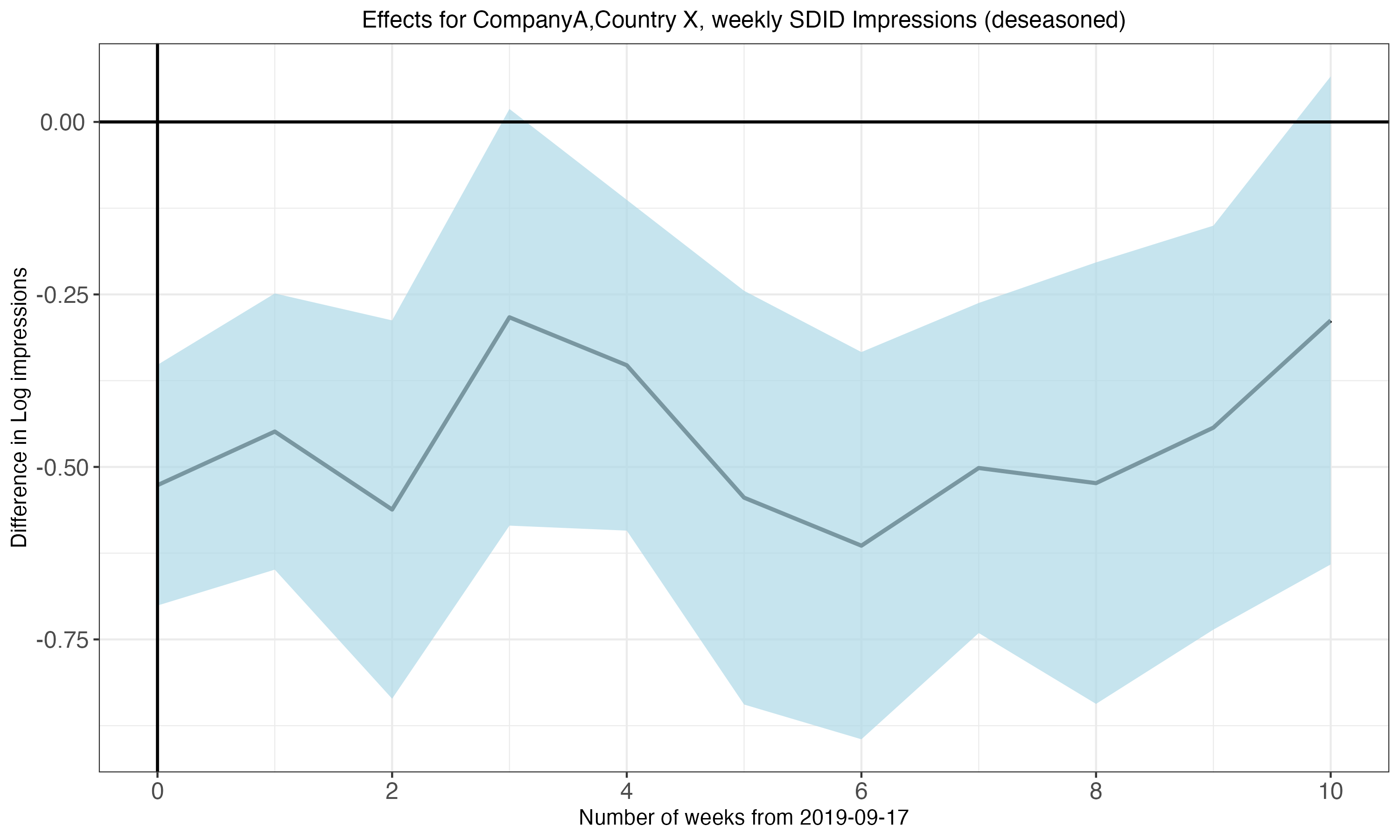} \\
        \includegraphics[width=0.48\textwidth]{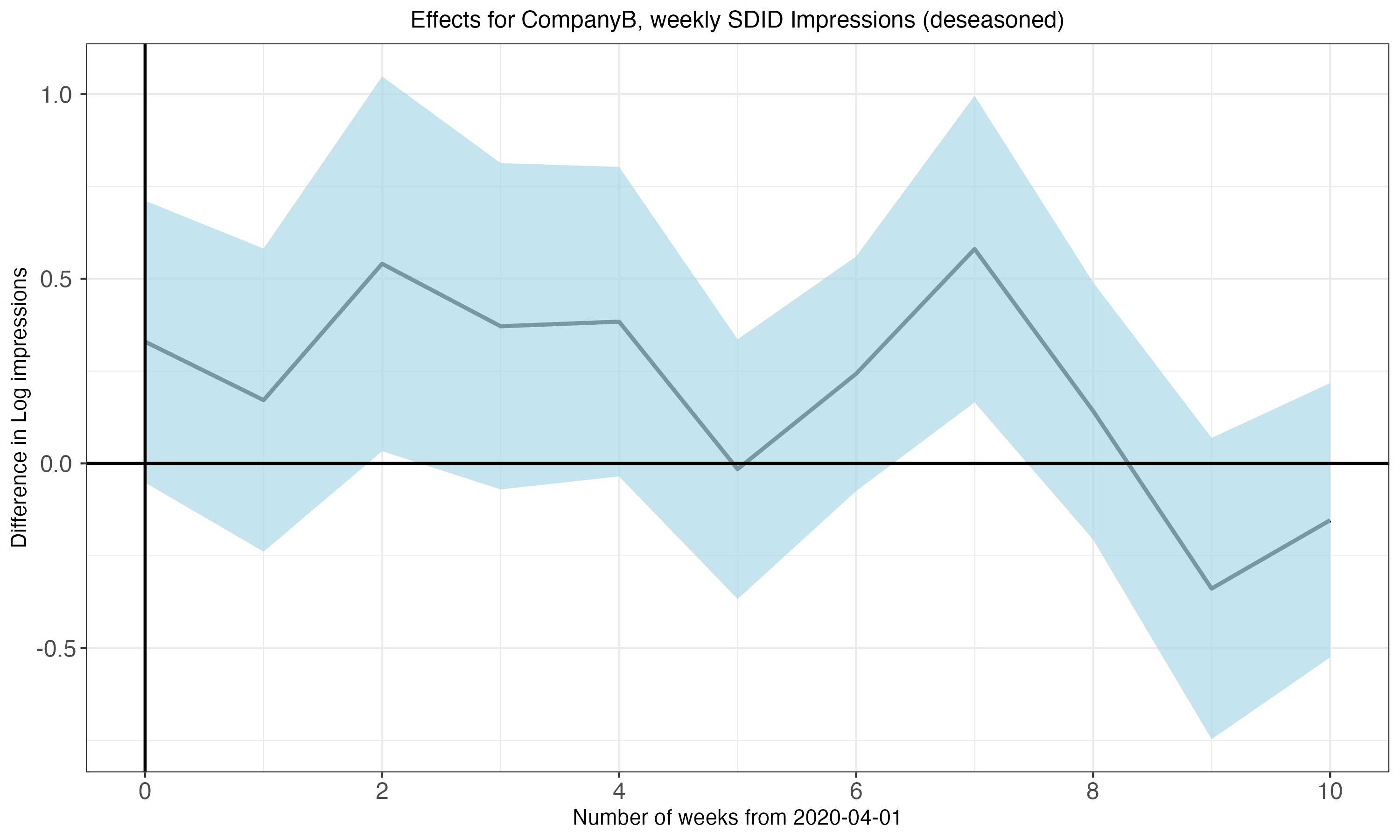}
        \includegraphics[width=0.48\textwidth]{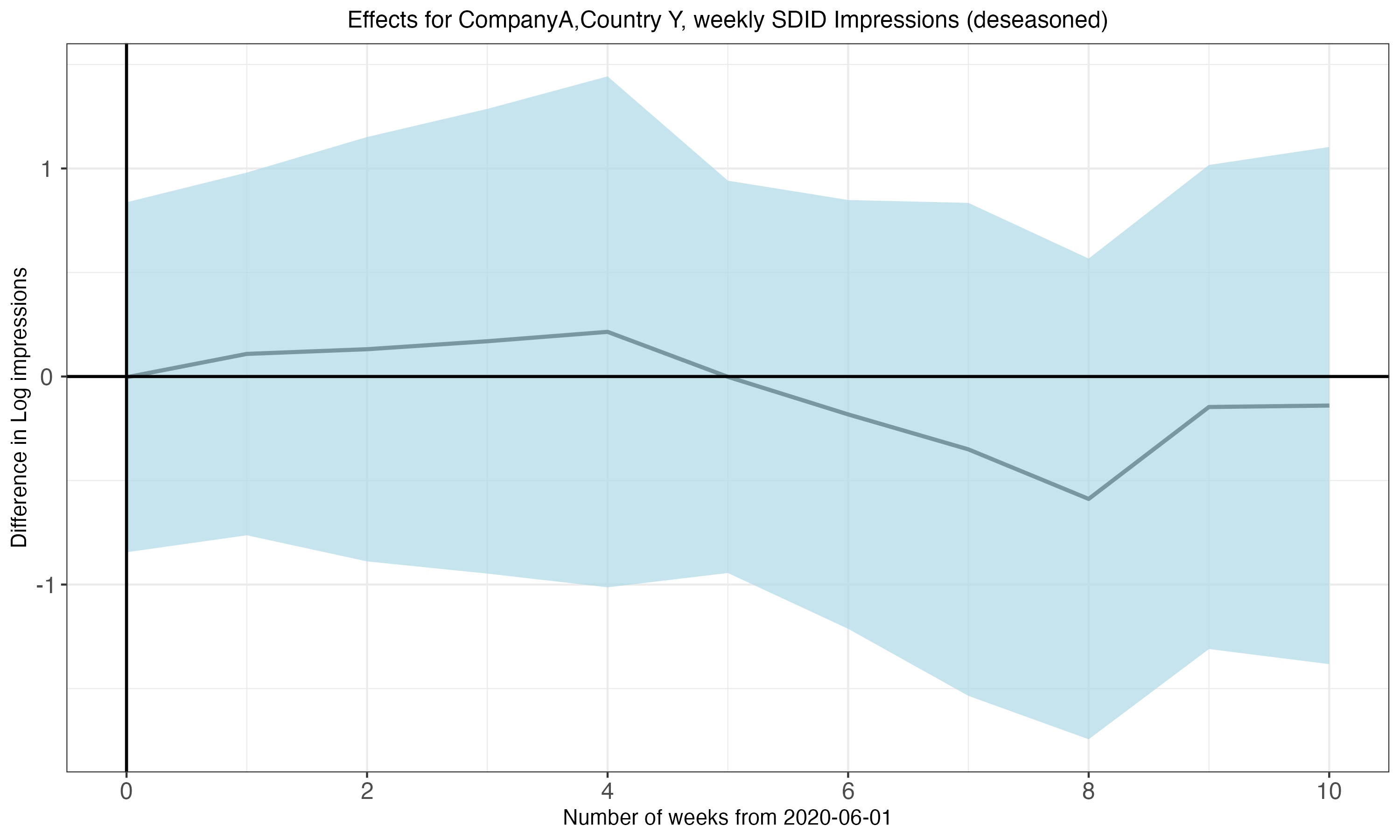}
      \end{center}
    \end{minipage}
  }
  \caption{\newgwR{SDID Estimated Effects of Format Change on Number of Impressions (Logs)}.  The solid line indicates point estimates of $\tau_{w}$, and the band indicates 95\% confidence intervals.}
  \label{fig:SDID_log_imps}
\end{figure}

\newgwR{More specifically, we run the SDID specification from Figure \ref{fig:SDID-estimates} but with log impressions as the dependent variable.\footnote{\newgwR{In the Appendix in Figure \ref{fig:DID_imps}, we also present DID results for impressions for completeness.}}
The results presented in Figure \ref{fig:SDID_log_imps} show that the auction format change did not decrease the volume of impressions on treated publishers, except for Company A, Country X. To complement the analysis, we also run SDID using log total revenue as the dependent variable (see Figure \ref{fig:SDID_revenue} in the Appendix). The results show that not only impressions decline, but total revenue declines as well for Company A, Country X. Because total revenue falls, a fixed-budget explanation cannot be the exclusive interpretation for the increase in average prices, although it may still play a role. Therefore, bid-shading mechanisms remain relevant even in this case.}

\newgw{Furthermore, even if the number of impressions sold would have changed, we show that most bidders/advertisers buy impressions from
multiple publishers and that they do not seem to set fixed budgets for a
particular publisher or group of publishers. To support this claim, we present two pieces of evidence, one using a
cross-section of advertising campaigns and another using the time
series of bidders' spending.}

\begin{figure}[htbp]
  {
    \includegraphics[width=0.6\textwidth]{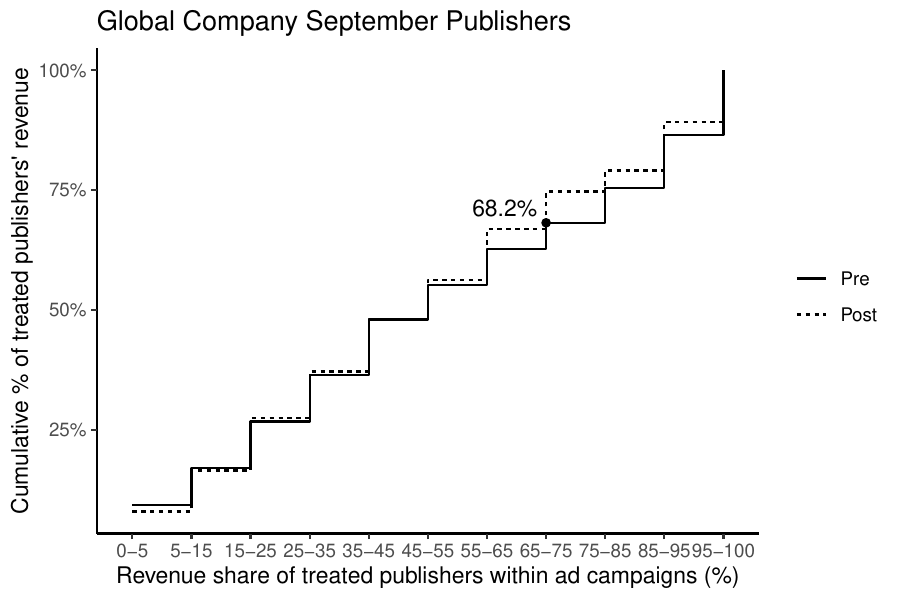}
  }
  \caption{Distribution of Share of Treated Publishers Within Advertising
    Campaign: Global
    Company.  Cumulative percentage of Global Company September
    Publishers' revenue from advertising campaigns that used Xandr's
    DSP service. The horizontal axis represents the share of September
    Publishers within each advertising campaign's spending, rounded to
    nearest multiple of 10\%.  The revenue and share are computed
    separately for 30 days before the format change (``Pre'') and for
    30 days after it
    (``Post'').}
  \label{fig:share-within-campaigns-global}
\end{figure}

First, we take the cross-section of all
advertising campaigns by advertisers that used Xandr's DSP service
including its bidding algorithm and bought impressions from our
treated publishers around the format change dates (30 days before or
after the format change). For each advertising campaign, we compute
the fraction of its spending on treated publishers (i.e., compute the
dollar amount the advertising campaign spent on treated publishers,
divided by the total dollar amount it spent), and round that fraction
to the nearest multiple of 10\%. We then sort those advertising
campaigns in an ascending order of the computed fraction, and plot the
cumulative percentage in those advertising campaigns' total spending
on the treated publishers. Figure
\ref{fig:share-within-campaigns-global} shows such a cumulative
percentage plot for advertising campaigns that were bought from Global
Company September Publishers, separately for the 30-day period before
the format change and the 30-day period after it.

We observe
that the share of September Publishers varies considerably across
campaigns. For instance, the solid point on the plot indicates that,
out of September Publishers' revenue from advertising campaigns using
Xandr's DSP service during the 30-day pre-period, 68.2\% comes from
advertising campaigns that spent less than 75\% on September
Publishers (i.e., spent more than 25\% on other publishers). \newgwR{ The figure also shows that following the auction format change, the share of revenue from advertising campaigns allocating less than 75\% of their spend to September Publishers increased by several percentage points.}\footnote{\newgwR{Consistent with our previous results, this shift suggests that advertisers redistributed spending across a broader set of publishers after the switch. One possible interpretation is that bidders adjusted their strategies in response to higher prices under FPAs, spreading budgets more evenly across publishers to avoid paying higher prices.}} Figure
\ref{fig:share-within-campaigns-Europe} shows plots for the other
three groups of treated publishers, with similar observations. These
figures indicate that advertisers/bidders buy from a diverse set of
publishers and not just from treated publishers.

\begin{figure}[htbp]
  {
    \includegraphics[width=0.8\textwidth]{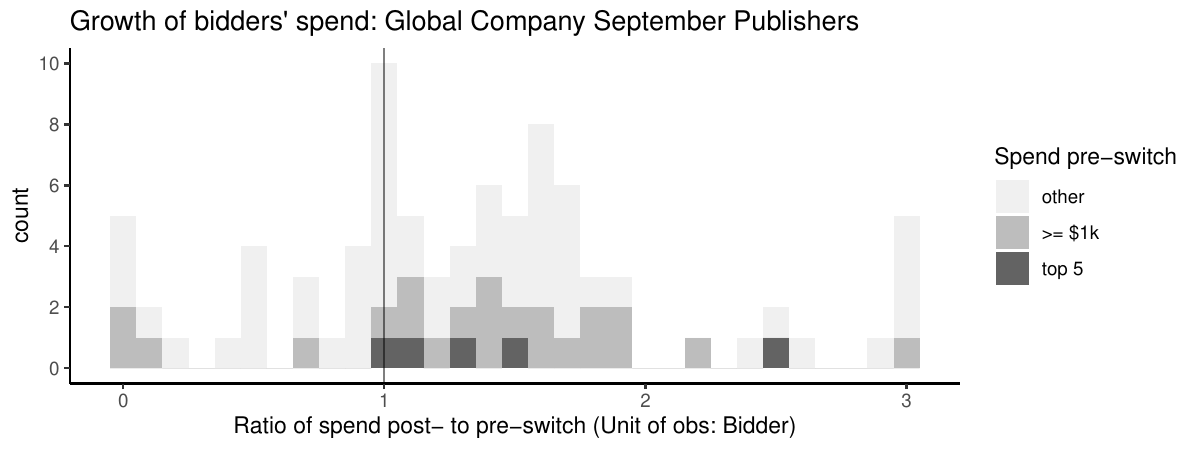}
  }
  \caption{Growth Rates of Bidders' Spending on Treated Publishers: Global
    Company.  Histogram of growth rates of bidders' spending on Global Company
    September Publishers from 7 days before the auction format change
    to 7 days after it, color-coded by the importance of each bidder
    to the September Publishers' revenue during the 7-day period
    before the change.}
  \label{fig:growth-bidder-global}
\end{figure}

Second, the time series of bidders' spending on treated publishers
exhibit quite a bit of temporal variations after the switch to FPAs,
and these variations show diverse patterns across bidders (we are
using all bidders here, in contrast to the cross-sectional evidence in
the previous paragraph). For each bidder, we compute the growth rate
of the bidder's spending as the ratio of the bidder's spending on the
treated publishers during the 7-day period after the format change to
that bidders's spending on the treated publishers during the 7-day
period before the format change.  Figure \ref{fig:growth-bidder-global}
shows the histogram of such growth rates for bidders buying from
September Publishers (the unit of observation is the bidder). Bidders
are color-coded by their importance to the September Publishers'
revenue, i.e., according to whether (i) the bidders are the top 5
bidders in terms of spending on September Publishers, (ii) they
otherwise spend at least 1,000 USD in the 7 days before the format
change, and (iii) they spend less than 1,000 USD in the 7 days before
the format change. The growth rates show an important variation from
0 to above 3 (where the horizontal axis is capped) and, importantly,
they differ from 1 in many cases.  Figure
\ref{fig:growth-bidder-Europe} shows plots for the other three groups
of treated publishers; we again see substantial variation in the
growth rates of spending across bidders. These facts suggest that the
bidders do not have a fixed budget for treated publishers.

\subsection{Other Robustness Checks.} \label{sec:robust}

We first rerun the \newgw{DID} study regressions on the European Media
Companies using alternative definitions for the control group.  First,
we use publishers of Company A other than in Country X as the control
group to estimate the effects for the publishers of Company A in
Country X. In other words, we exclude publishers of Companies B and C
that we included in the control group for the main
specification. Publishers of Company A are arguably more similar to
each other than to publishers of Companies B and C, and therefore a
treatment--control comparison without Company A may potentially be
more appropriate. The results
are shown in the top panel of Figure \ref{fig:event-alt-control-grp},
and are similar to the results for the main specification in Figure
\ref{fig:event-estimates}. 

Second, to estimate the effects for the
publishers of Company B and the publisher of Company A in Country Y,
we use publishers of Company A other than in Countries X and Y as the
control group. In the main regression specification, we include
publishers of Company C as part of the control group. This might raise
concerns of confounding through equilibrium effects, as Company C
targets internet users in the same geographical region (Country Y) as
the treatment group. Excluding publishers of Company C from the
control group mitigates such concerns. The results are shown in the
bottom two panels of Figure \ref{fig:event-alt-control-grp}. Again, the
results are similar to those for the main specification.

Next, we rerun the \newgw{DID} regressions specified in equation
\eqref{eq:1step-reg} by replacing the outcome variable (left-hand-side
variable) with $\log y_{pt}$, log of average price, \newgw{to consider multiplicative treatment effects}. Figure
\ref{fig:event-multiplicative} shows the estimates. Apart from showing
some pretrends---which is the reason why we prefer $y_{pt}$ to
$\log y_{pt}$ as the main specification---the basic observation stays
the same, i.e., there is (i) a significant jump in the average price
immediately after the auction format change, and (ii) a decline in the
increase within approximately 60 days (the publishers of Company A in
Country X) or 30 days (the publishers of Company B and the publisher
of Company A in Country Y) after the change.

To further investigate whether seasonality adjustments are affecting the
estimates, we also estimate the \newgw{DID} regressions in two
alternative ways. In the first method, we estimate the regression in
two steps: we first remove the seasonality of $y_{pt}$ by regressing
$y_{pt}$ on dummy variables, separately for each $p$, and obtain a
``deseasonalized'' time series $\widetilde{y}_{pt}$ for each $p$ \newgw{(similar to the SDID two-stage approach implemented for weekly data)}. We
then run the regression as in \eqref{eq:1step-reg}, except
that $y_{pt}$ is replaced with $\widetilde{y}_{pt}$ and the seasonal
fixed effects ($\gamma_{p,\mathrm{dow}(t)}$,
$\gamma_{p,\mathrm{dom}(t)}$, $\gamma_{p,\mathrm{month}(t)}$, and
$\gamma_{p,\mathrm{eoq}(t)}$) are removed; see Appendix
\ref{subsec:alt-seasonality} for details. In the second method, we
estimate \eqref{eq:1step-reg} without any seasonal fixed
effects. Figures \ref{fig:two-step} and \ref{fig:no-seasonality} show
estimates of $\beta_k$ for (i) and (ii), respectively. Again, the
estimates show patterns similar to Figure \ref{fig:event-estimates},
although the estimates under Figure \ref{fig:no-seasonality} exhibit
more fluctuations because of day-of-week effects. These results
indicate that the seasonality adjustments in the main regression do
not drive our main results.

Finally, as a falsification test, we run the \newgw{DID} regressions by
picking hypothetical dates for the auction format change that are one
year before the actual dates. Figure \ref{fig:event-falsification}
shows the results.  The estimates are no longer statistically
significant in three out of the four pairs. For the remaining pair
(the publishers of Company A in Country Y vs.~their controls), the
estimated effects of the hypothetical auction format change are
negative.

\section{Bidder Heterogeneity}\label{sec:bidder-hetero}

To investigate further the relation between the effects of the auction
format change on publishers' revenue and bidders' behavior, we
estimate how the effects of the format change on spending differ
across different types of bidders. For that purpose, we aim to classify
bidders into different levels of sophistication as defined below, and
then estimate the following regression equation:
\begin{equation}
  y_{pbt} = \alpha_{pb} + \sum_{\underline{k}\leq k\leq\overline{k},\,k\neq-1}\beta_{bk}D_{pb}\cdot\boldsymbol{1}(K_{t}=k)
  + \gamma_{t}
  + \gamma_{pb,\mathrm{dow}(t)} + \gamma_{pb,\mathrm{dom}(t)} + \gamma_{pb,\mathrm{month}(t)} + \gamma_{pb,\mathrm{eoq}(t)}
  + \varepsilon_{pbt}.
  \label{eq:hte-regress}
\end{equation}
Here, $b$ indicates the type of bidders. The difference with the main
regression specification (\ref{eq:1step-reg}) is the additional index
$b$: (i) the outcome variable $y_{pbt}$ is now the average spending
per impression by all bidders of type $b$ for each publisher--day
pair, (ii) the treatment effects $\beta_{bk}$ are estimated separately
for each bidder type $b$, and (iii) publisher fixed effects $\alpha_{pb}$ and
seasonal fixed effects $\gamma_{pb,\mathrm{dow}(t)}$,
$\gamma_{pb,\mathrm{dom}(t)}$, $\gamma_{pb,\mathrm{month}(t)}$,
$\gamma_{pb,\mathrm{eoq}(t)}$ are made bidder-type-publisher specific.

We first conjecture that larger bidders are more sophisticated and
shade their bids more aggressively than smaller bidders once the
format switches to FPAs. To see this, we classify the bidders into
three types, ``large,'' ``medium,'' and ``small,'' as follows. We
calculate each bidder's spending on the treated publishers in the
30-day period immediately before the format change, and compute each
bidder's share within the total revenue of the treated publishers
during that period. For bidders of the Global Company, we classify a
bidder as ``large'' if the share is above 10\%, ``medium'' if the
share is above 1\%, and ``small'' if the share is below 1\%.\footnote{\newgw{The thresholds were set to create a relatively `balanced' split of bidders. 
In particular, the 10\% threshold and 1\% yield 3 Large bidders, 6 Medium bidders and 86 Small bidders. 
Furthermore, there was a huge distance in the share between the 3rd bidder (14.6\%) and the 4th bidder (4.3\%), so it was natural to set the boundary between Large and Medium there. As for the division between Medium and Small, the 1\% threshold left a meaningful number of bidders in the Medium bucket.}} \newgw{That is, a ``large'' bidder has an important spend relative to all the spend {\em by all bidders in our data set on treated publishers} (in the 30-day period before the format change). We note that we do not have access to spend of bidders outside the publishers in our data set, so we cannot categorize them in terms of overall size.}

Figure \ref{fig:hte-by-size-global} shows the estimated $\beta_{bk}$ for
$b\in \{\text{large}, \text{medium}, \text{small}\}$. However,
contrary to our initial hypothesis, we do not see a monotonic pattern:
the effect on spending is the highest among ``small'' bidders and the
lowest among ``medium'' bidders, while ``large'' bidders' responses
were in between the other two types.

We next use an alternative classification of bidders that is arguably more directly
related to bidder sophistication.  Some advertisers use Xandr's DSP
service, including its bidding algorithm, and Xandr assigns a single
bidder ID to such advertisers: we refer to them as the
``AppNexus/Xandr bidder.'' Since the AppNexus/Xandr bidder uses the
bidding algorithm of Xandr, which also coordinated the format change from
SPA to FPA, one can expect this bidder to be more sophisticated than other
bidders in changing its bidding algorithm.

\begin{figure}[htbp]
  {
    \includegraphics[height=0.3\linewidth]{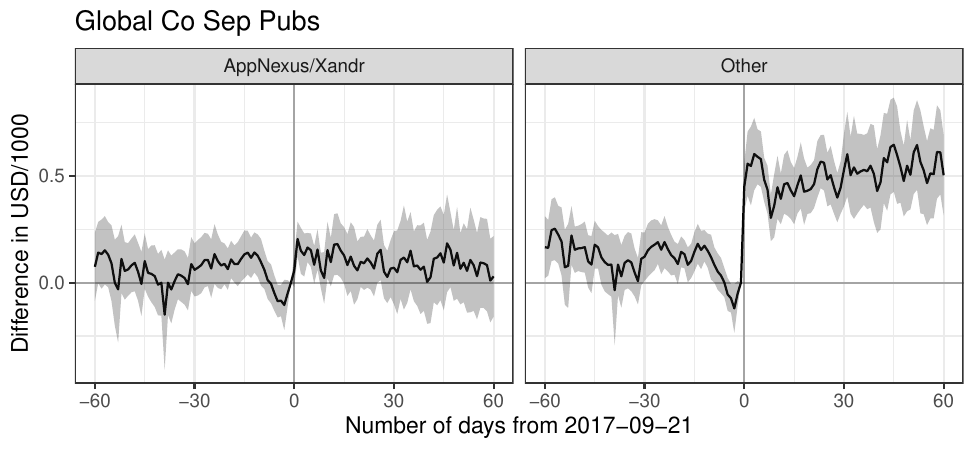}
  }
  \caption{AppNexus/Xandr Bidder and Non-AppNexus/Xandr Bidders: Global
    Company.  Effects of auction format changes on spending per sold impression
    by bidders on Global Company September Publishers, separately for
    the AppNexus/Xandr bidder and non-AppNexus/Xandr bidders.}
  \label{fig:hte-console-global}
\end{figure}

Figure \ref{fig:hte-console-global} shows the estimated effects when
Global Company September Publishers changed to FPAs.\footnote{\newgw{For Global Company publishers, AppNexus/Xandr bidder purchases
38.08\% of total impressions in the 30 day period prior to their auction format change date.}} We observe that
spending by non-AppNexus/Xandr bidders jumped immediately
($\beta_{b1}=0.56/1000$ USD, or 84\% of the average price for the
30-day period before the format change) and that increase persisted
for 60 days, while spending by the AppNexus/Xandr bidder increased
only moderately ($\beta_{b1}=0.20/1000$ USD, or 39\% of the average
price for the 30-day period before the format change) and became
statistically insignificant after 6 days. These results suggest that
the AppNexus/Xandr bidder was able to adjust to the new environment of
FPA more quickly than other bidders due to the former's sophisticated
bidding algorithm incorporating bid shading, suggesting the increase
in revenue is the result of suboptimal bid shading by naive bidders.
Figure \ref{fig:hte-console-others} shows estimates for other
publishers, and we observe a pattern similar to that of the Global
Company September Publishers.\footnote{The $\beta_{b1}$'s for the
  publishers of Company A in Country X are 2.02/1000 USD for
  non-AppNexus/Xandr bidders (42\% of the average price for the 30-day
  period before the format change) and 1.16/1000 USD for the
  AppNexus/Xandr bidder (27\%). For the publishers of Company B, they
  are 1.25/1000 USD (41\%) for non-AppNexus/Xandr bidders and
  2.98/1000 USD (68\%) for the AppNexus/Xandr bidder. For the
  publisher of Company A in Country Y, they are 0.92/1000 USD for
  non-AppNexus/Xandr bidders (67\%) and 0.76/1000 USD (37\%) for the
  AppNexus/Xandr bidder.  The impact for the AppNexus/Xandr bidder
  becomes statistically insignificant seven days after the format
  change.}  The exception is Company B, which sees a larger increase
in spending by the AppNexus/Xandr bidder.  For Company B, unlike other
publishers, the AppNexus/Xandr bidder represents only a small fraction
of impressions and revenue (3.5\% of impressions and 5\% of revenue
for the 30-day period before the format change). It even reduced the
number of impressions it bought from publishers of Company B by 80\%,
when we compare the number for 30 days before and after the format
change. \newgw{We interpret this as an extreme case of compositional
change due to more aggressive bid shading of the AppNexus/Xandr bidder so that  the AppNexus/Xandr bidder may buy
only impressions with high willingness to pay after the format
change.}

 
\section{Conclusion}\label{sec:conclusion}

Using the data of internet display advertising auctions, we have
analyzed the impacts of auction format change from second-price
auctions (SPAs) to first-price auctions (FPAs). By estimating \newgw{DID and SDID}
regressions, we find that the average price jumps up immediately
after the format change from SPAs to FPAs. \newgwR{Depending on the specification, we also find that the increase attenuates over time. This pattern is consistent with initially insufficient bid shading by some bidders, followed by gradual adjustment as bidders learn to shade their bids.} Our
heterogeneity analysis reveals that the AppNexus/Xandr bidder---who
used a more sophisticated bidding algorithm---shaded their bids more
aggressively than non-AppNexus/Xandr bidders once the format changed
to FPAs, supporting our argument that suboptimal bid shading caused
the transitory increase in price.

We have also collected a more granular bid-level dataset for Company B, which would allow us to develop a structural model of bidding behavior. Such data would enable us to simulate counterfactual bids for each bidder under rational bid shading and compare them with observed bidding behavior.

\clearpage

\appendix

\section{SDID}

\begin{figure}[htbp]
  {
    \begin{minipage}{\textwidth}
      \begin{center}
        \includegraphics[width=0.48\textwidth]{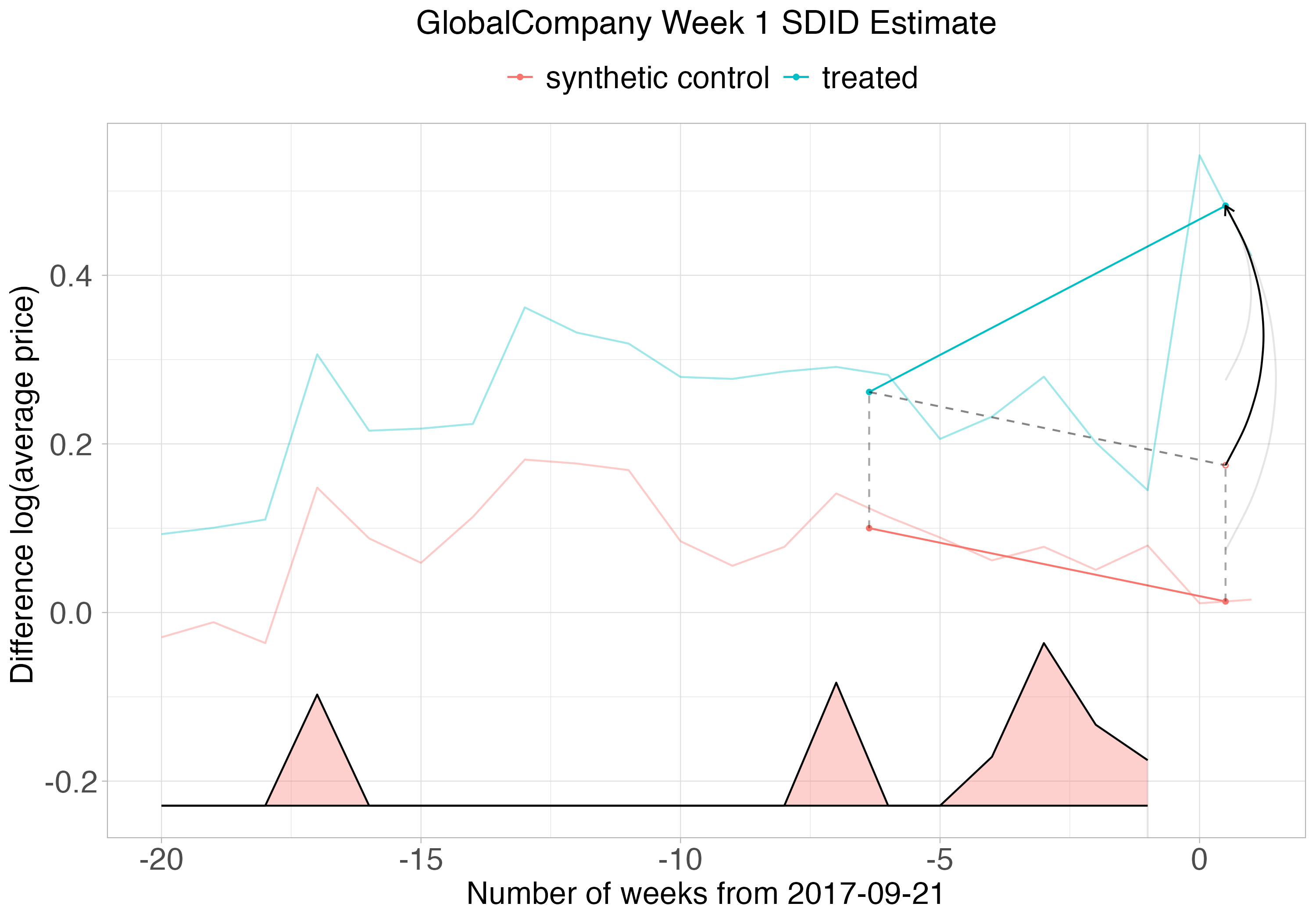}
        \includegraphics[width=0.48\textwidth]{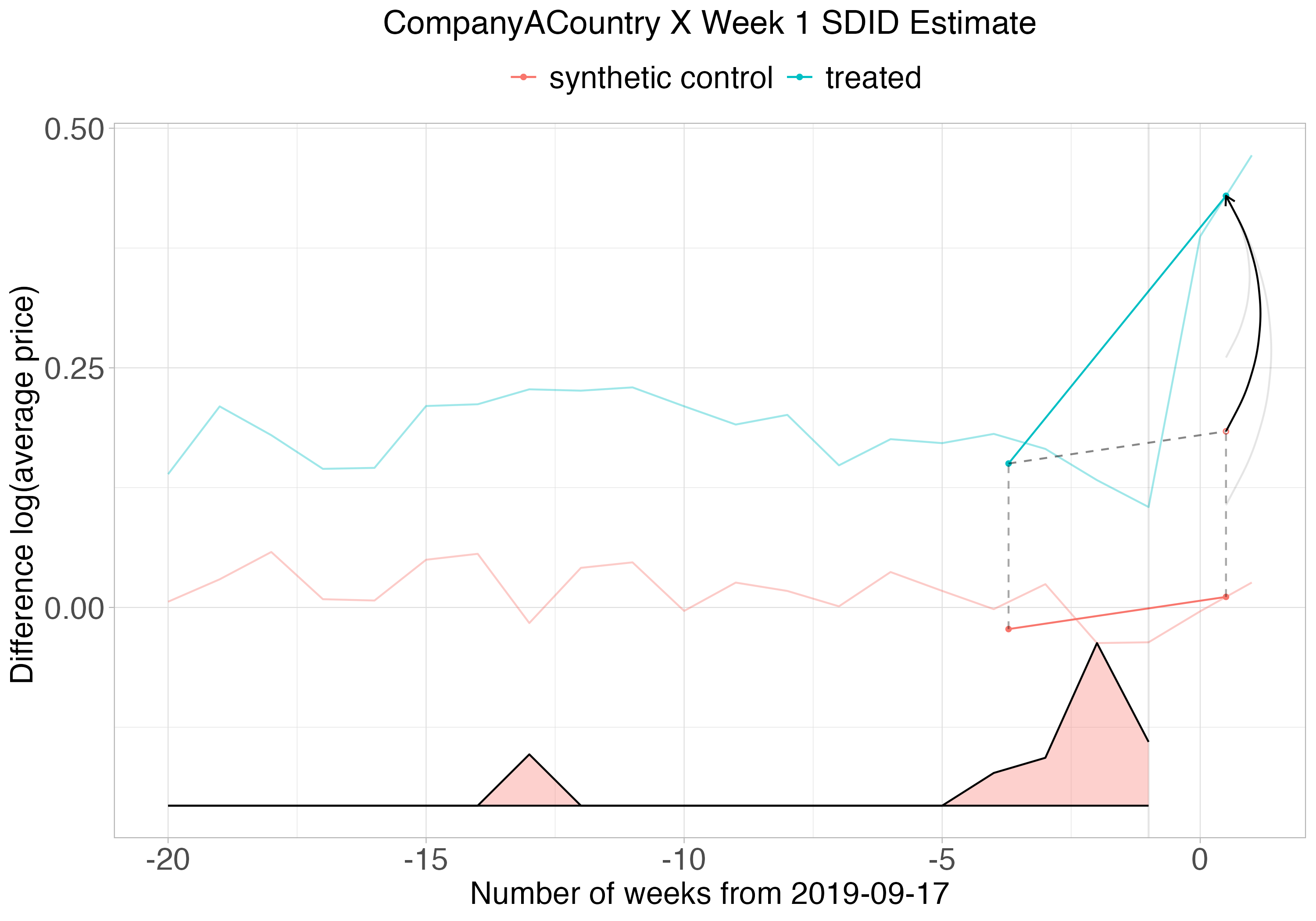} \\
        \includegraphics[width=0.48\textwidth]{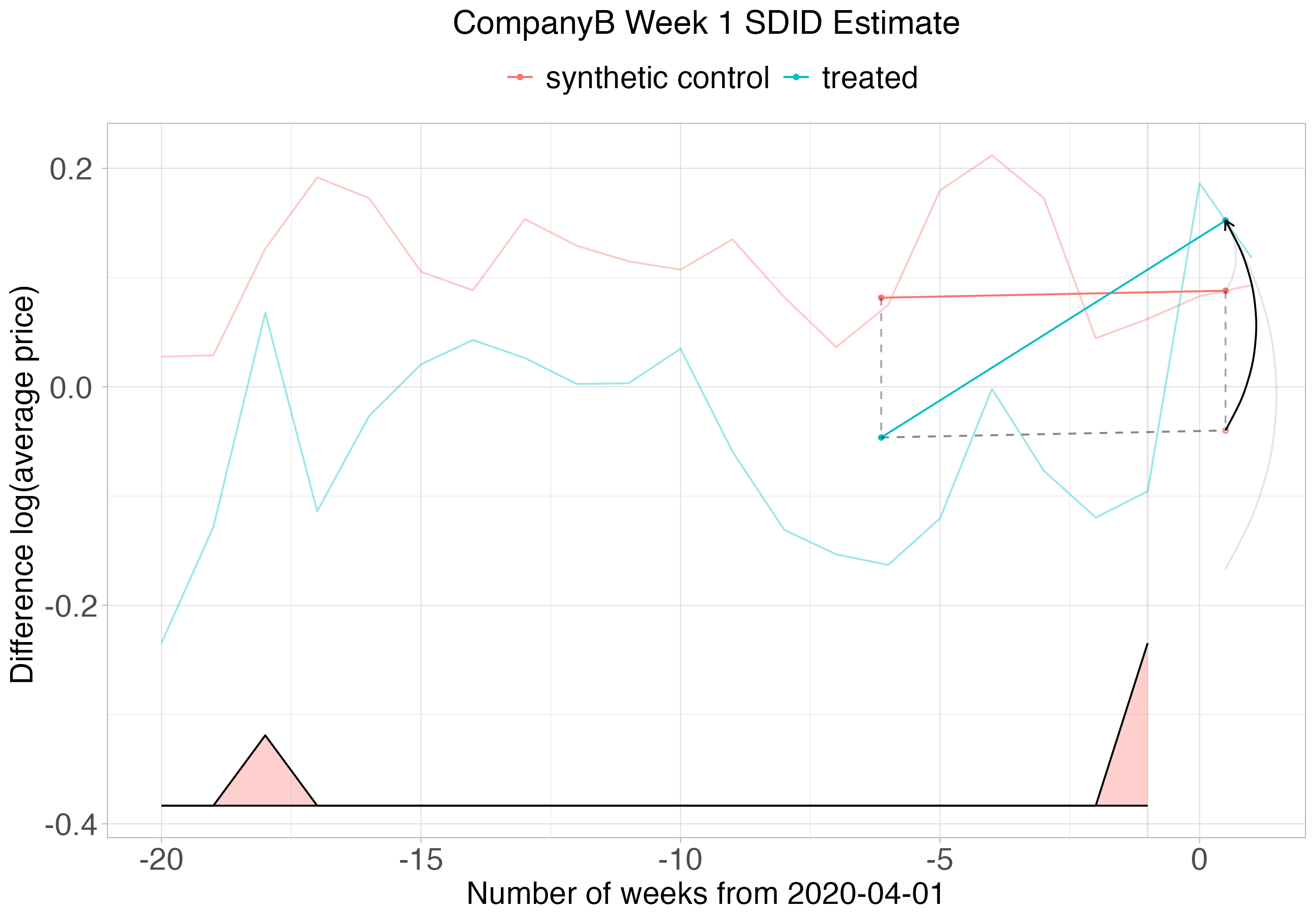}
        \includegraphics[width=0.48\textwidth]{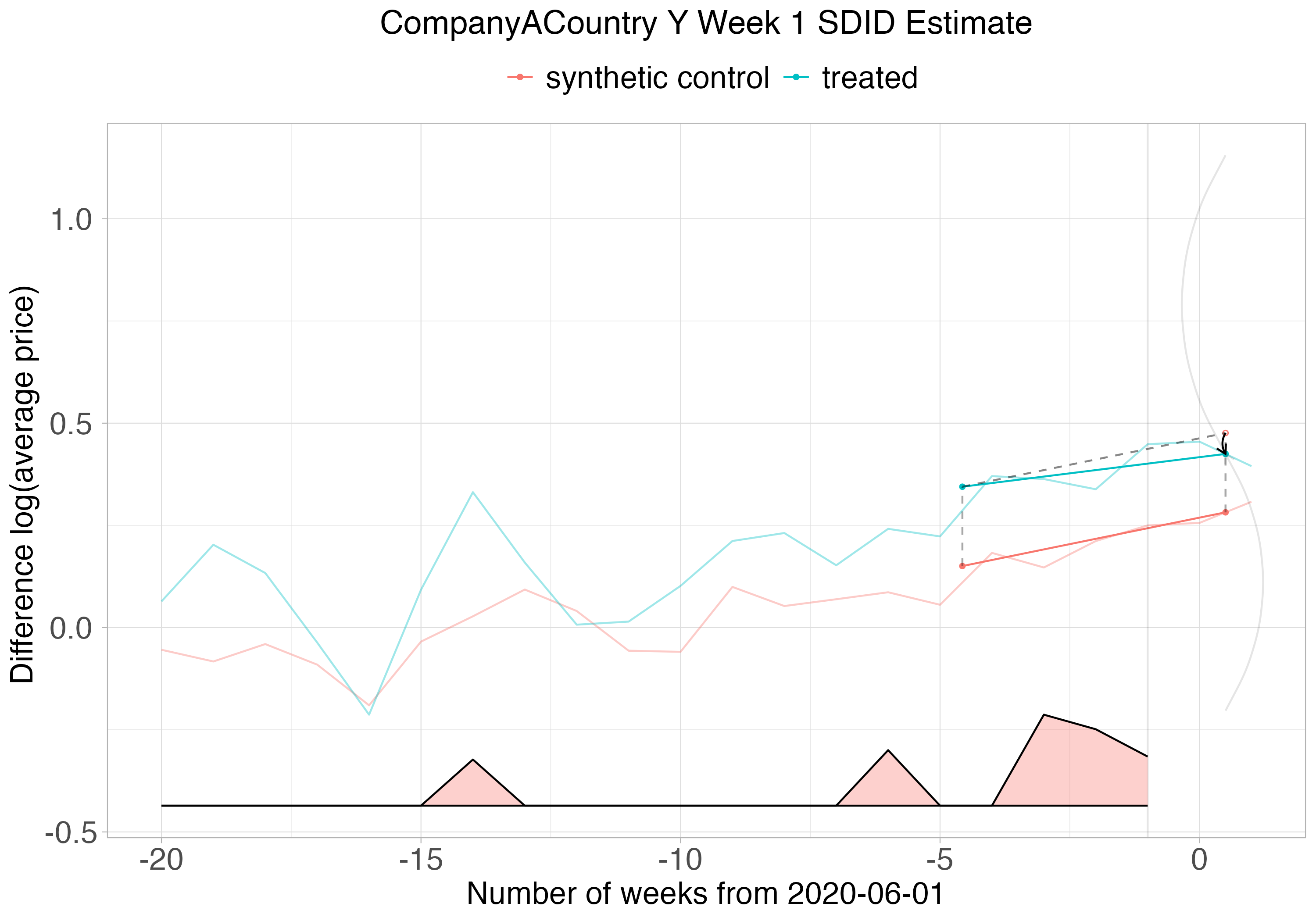}
      \end{center}
    \end{minipage}
  }
  \caption{SDID Control and Treatment Trajectories for Week 1 Estimates.  The shaded regions on the bottom of the plot show how pre-treatment time periods are weighted. 
}
  \label{fig:parallel-trends}
\end{figure}

\begin{figure}[htbp]
  {
    \begin{minipage}{\textwidth}
      \begin{center}
        \includegraphics[width=0.48\textwidth]{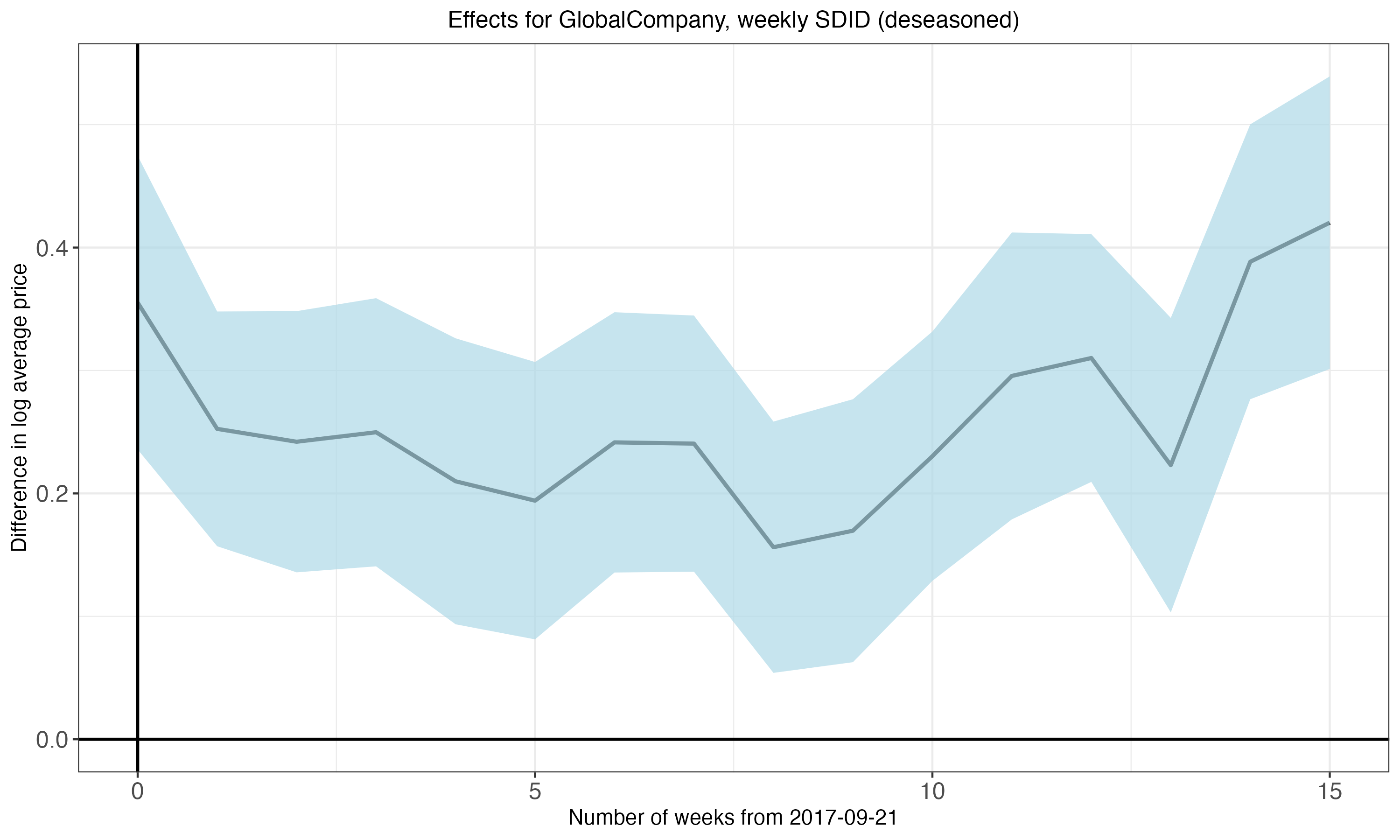}
        \includegraphics[width=0.48\textwidth]{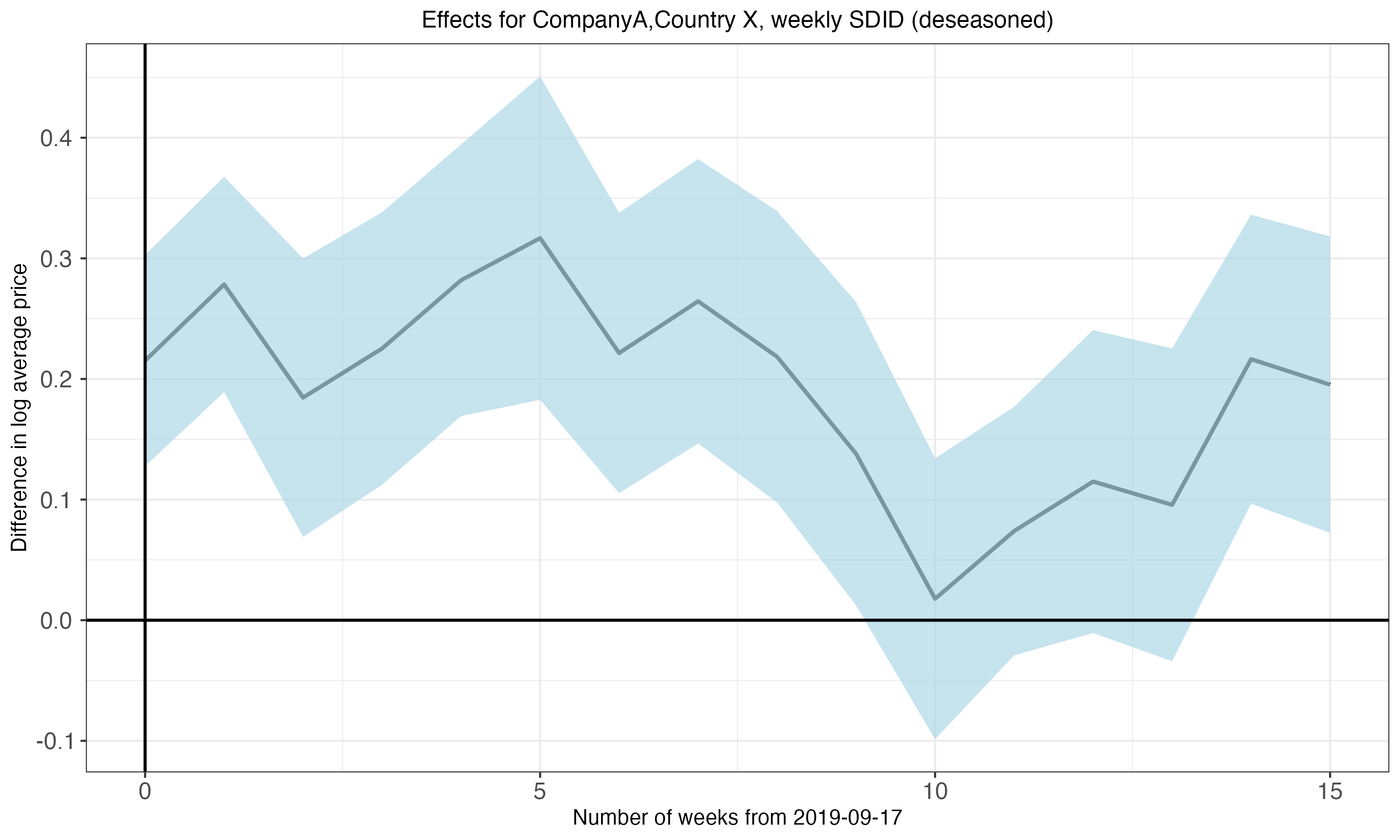} \\
        \includegraphics[width=0.48\textwidth]{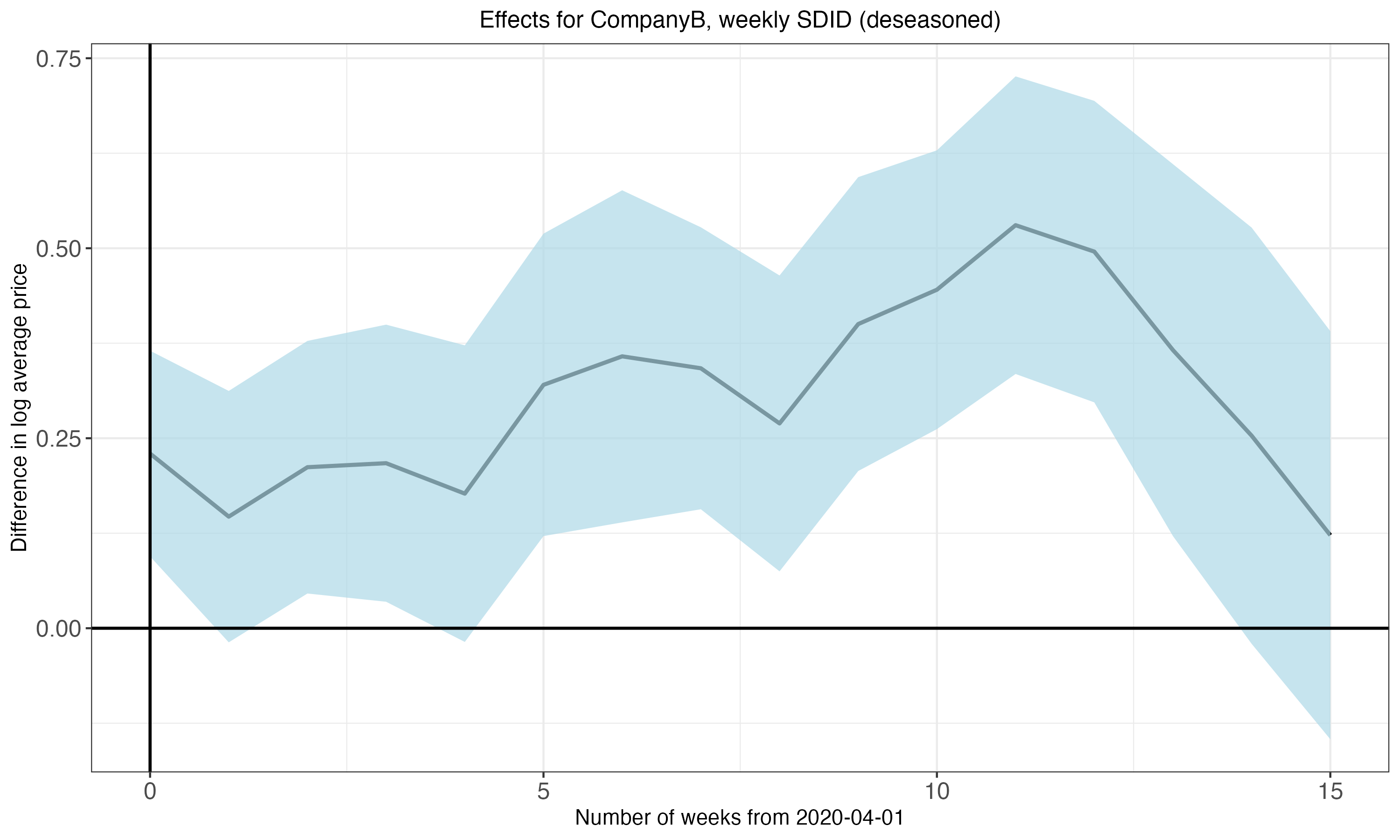}
        \includegraphics[width=0.48\textwidth]{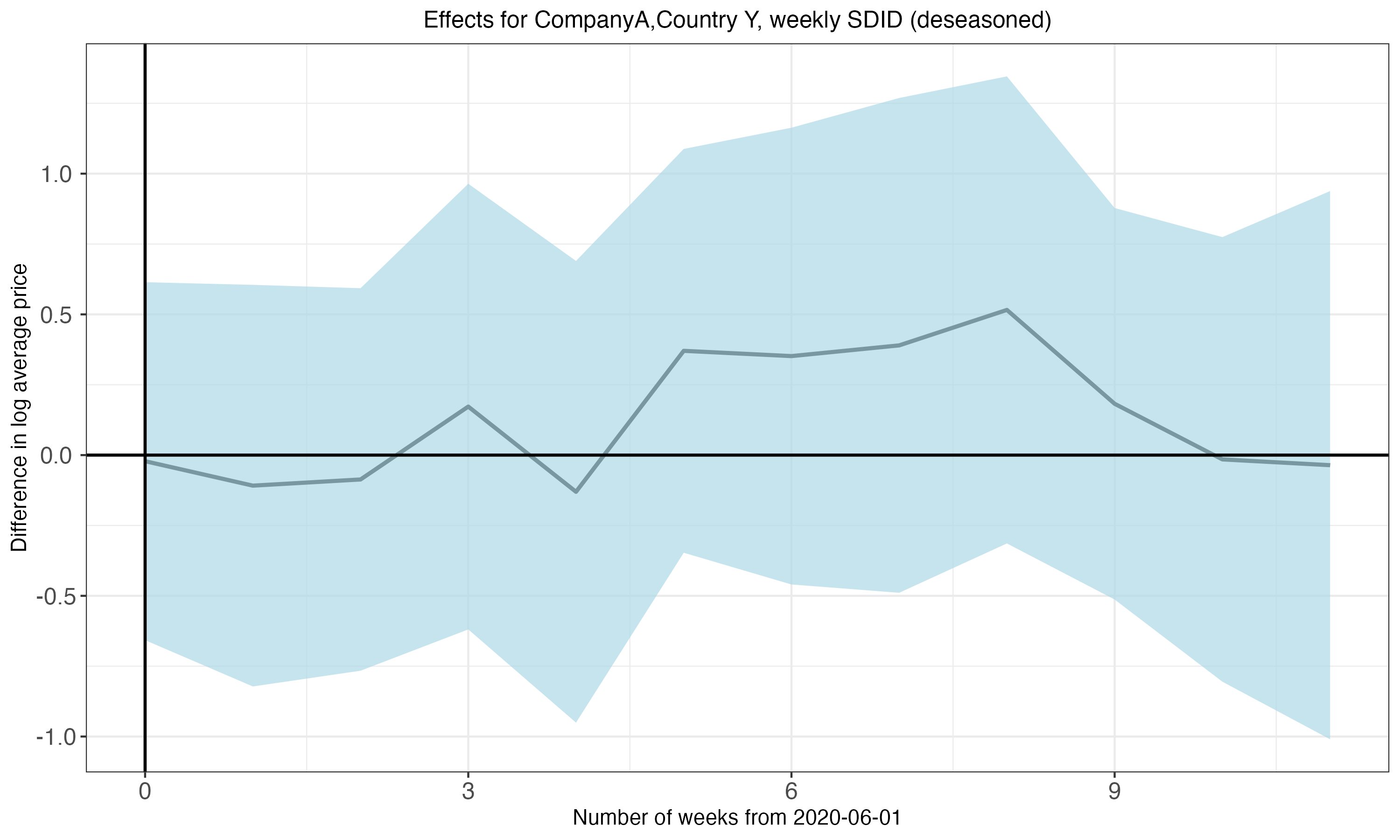}
      \end{center}
    \end{minipage}
  }
  \caption{\newgwR{SDID Estimated Effects of Format Change on Average
    Price - 15 weeks (Logs)}.  The solid line indicates point estimates of $\tau_{w}$, and the band indicates 95\% confidence intervals. Company A, Country X finishes at week 11 because of lack if further data.}
  \label{fig:SDID-estimates-15w}
\end{figure}

\begin{figure}[htbp]
  {
    \begin{minipage}{\textwidth}
      \begin{center}
        \includegraphics[width=0.48\textwidth]{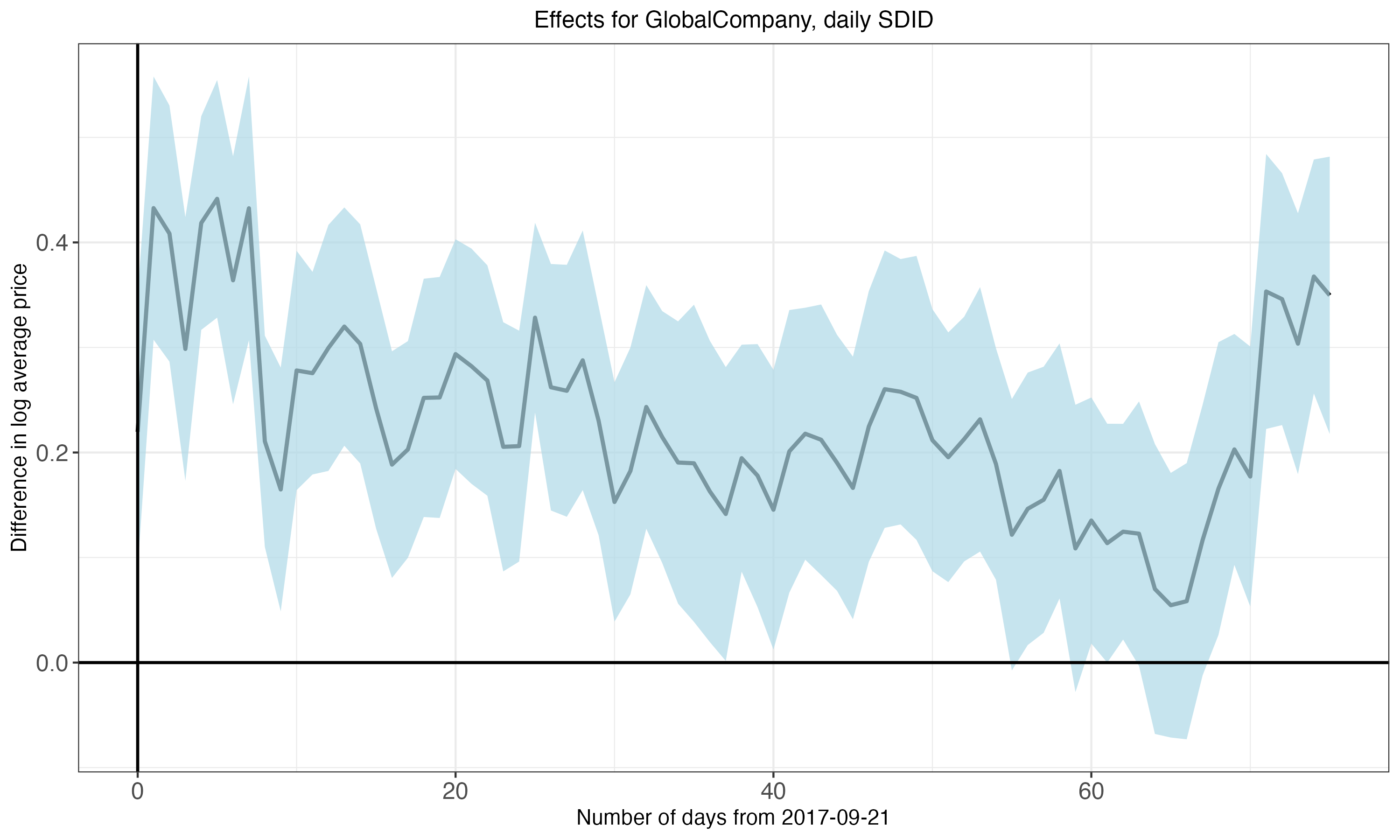}
        \includegraphics[width=0.48\textwidth]{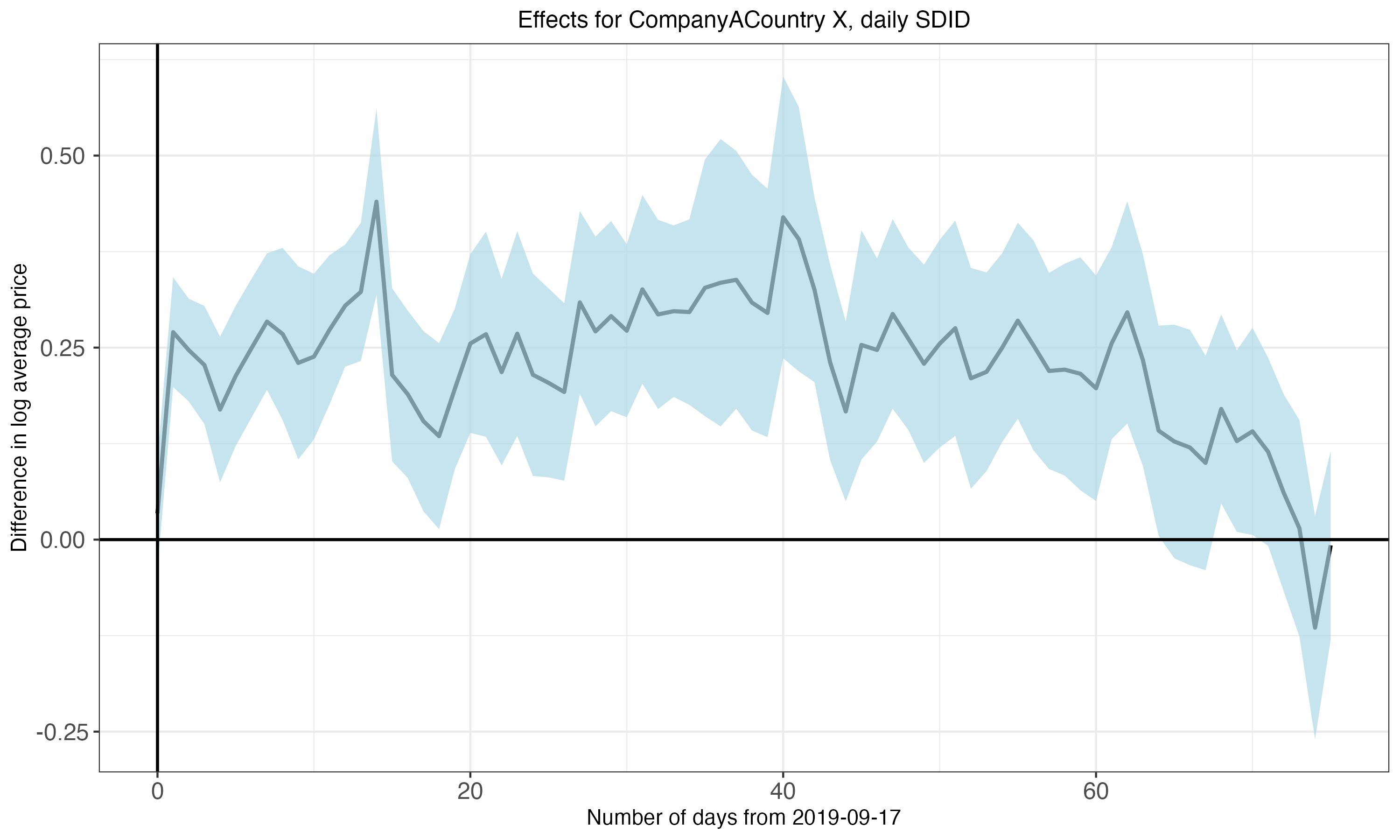} \\
        \includegraphics[width=0.48\textwidth]{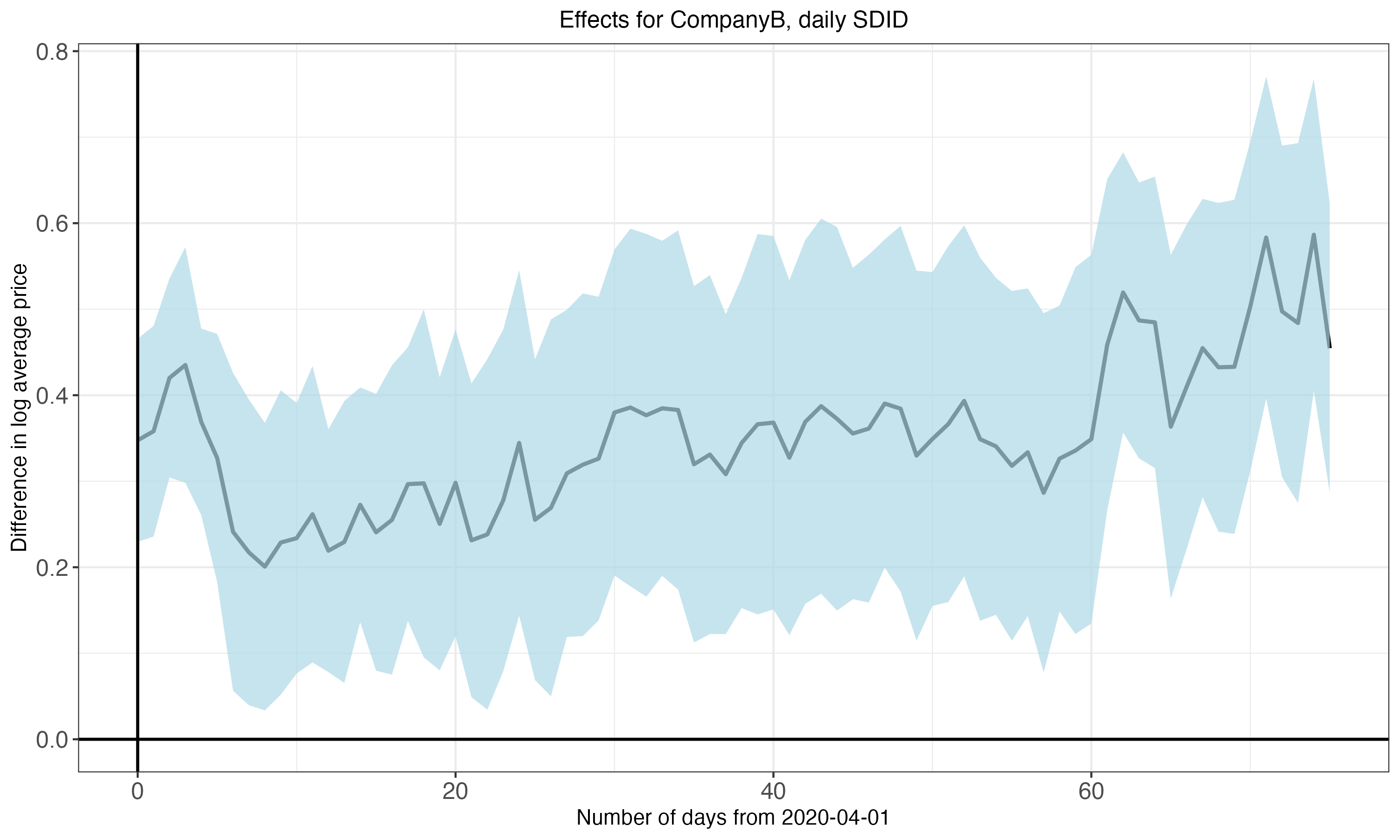}
        \includegraphics[width=0.48\textwidth]{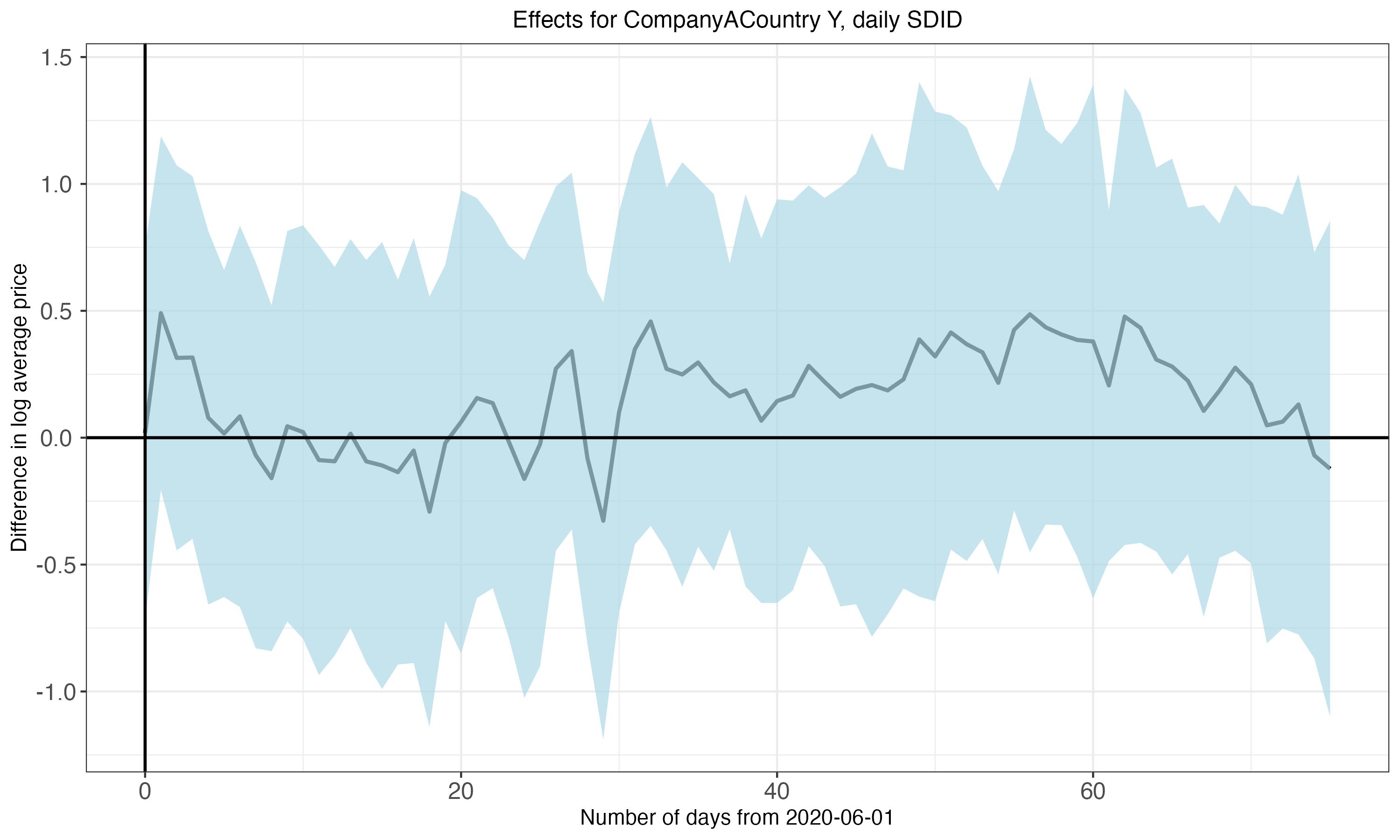}
      \end{center}
    \end{minipage}
  }
  \caption{\newgwR{SDID Estimated Effects of Format Change on Average
    Price with Daily Data (Logs)}.  The solid line indicates point estimates of $\tau_{w}$, and the band indicates 95\% confidence intervals.}
  \label{fig:SDID-daily-estimates}
\end{figure}

\begin{figure}[htbp]
  {
    \begin{minipage}{\textwidth}
      \begin{center}
        \includegraphics[width=0.48\textwidth]{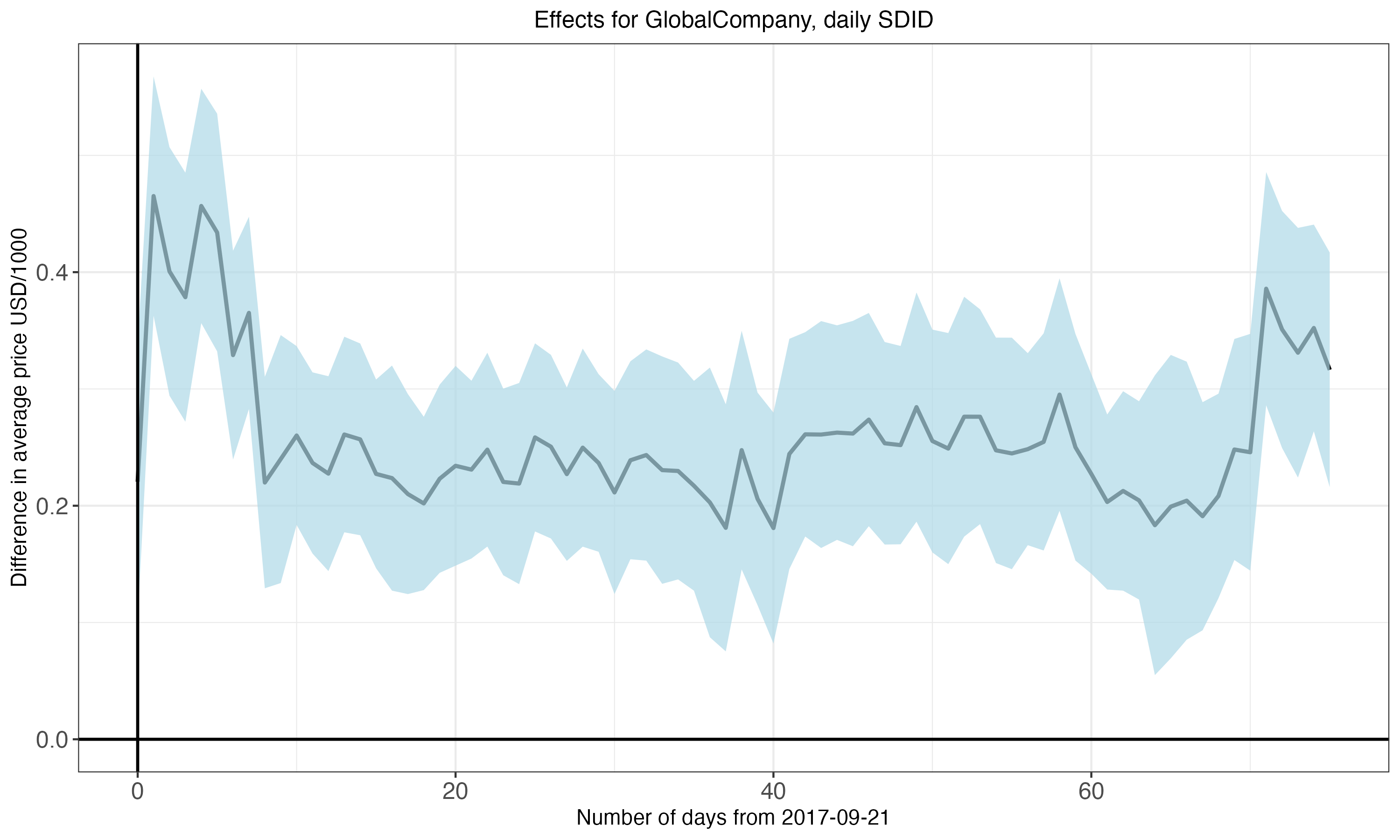}
        \includegraphics[width=0.48\textwidth]{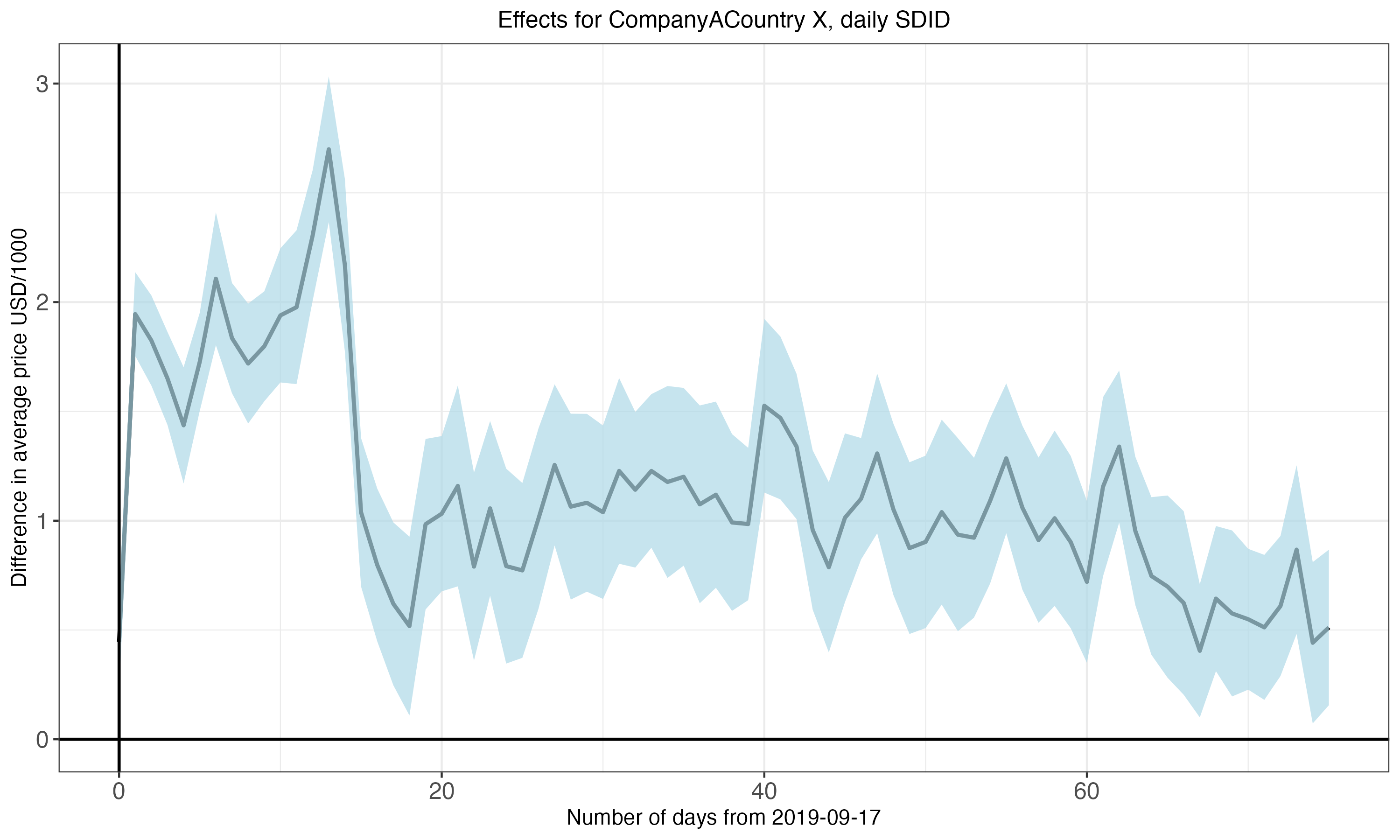} \\
        \includegraphics[width=0.48\textwidth]{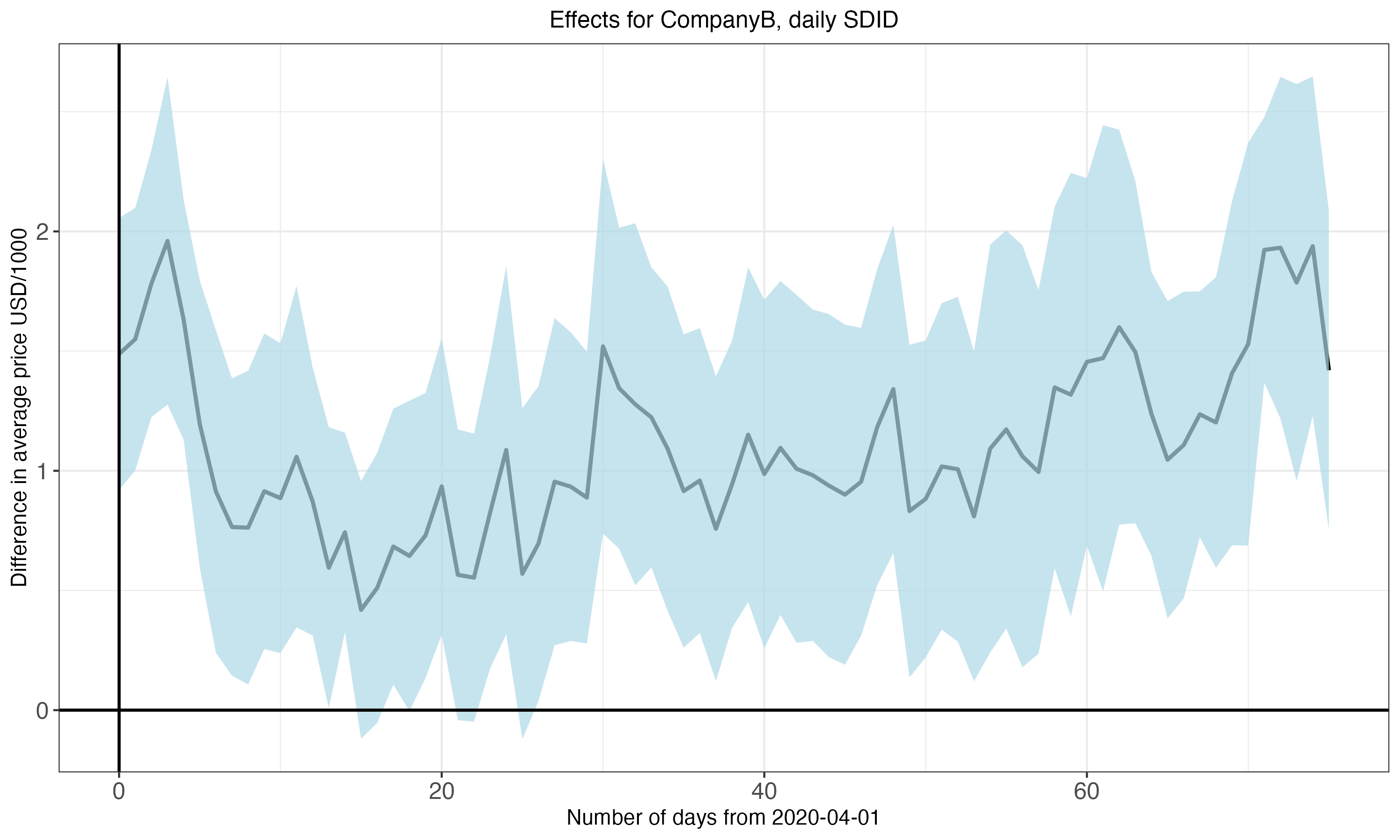}
        \includegraphics[width=0.48\textwidth]{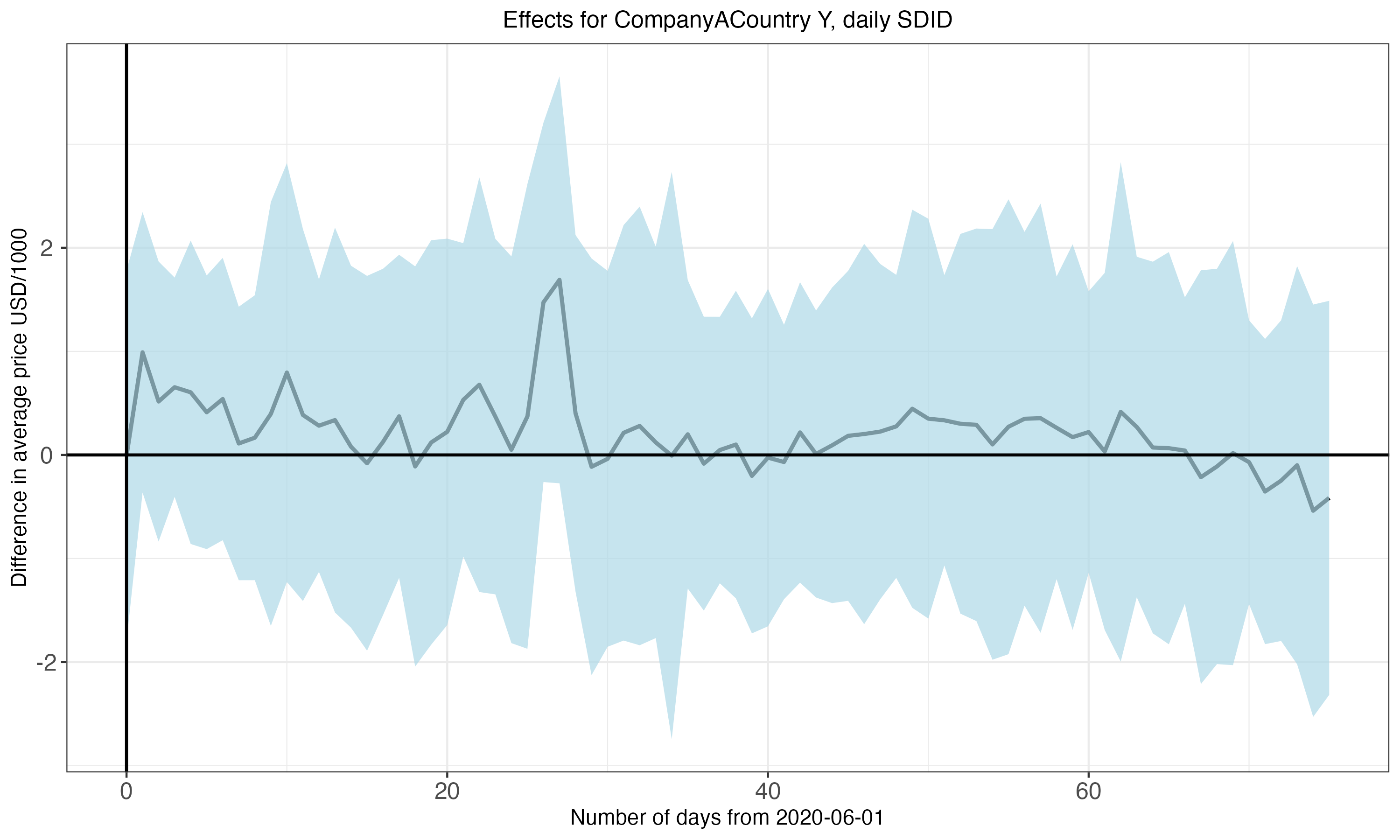}
      \end{center}
    \end{minipage}
  }
  \caption{\newgwR{SDID Estimated Effects of Format Change on Average
    Price with Daily Data }.  The solid line indicates point estimates of $\tau_{w}$, and the band indicates 95\% confidence intervals. }
  \label{fig:SDID-daily-estimates-rps}
\end{figure}
 
\clearpage
\section{Supplementary Figures on Ad Campaign Budgets}

\begin{figure}[htbp]
  {
    \begin{minipage}{\textwidth}
      \begin{center}
        \includegraphics[width=0.48\textwidth]{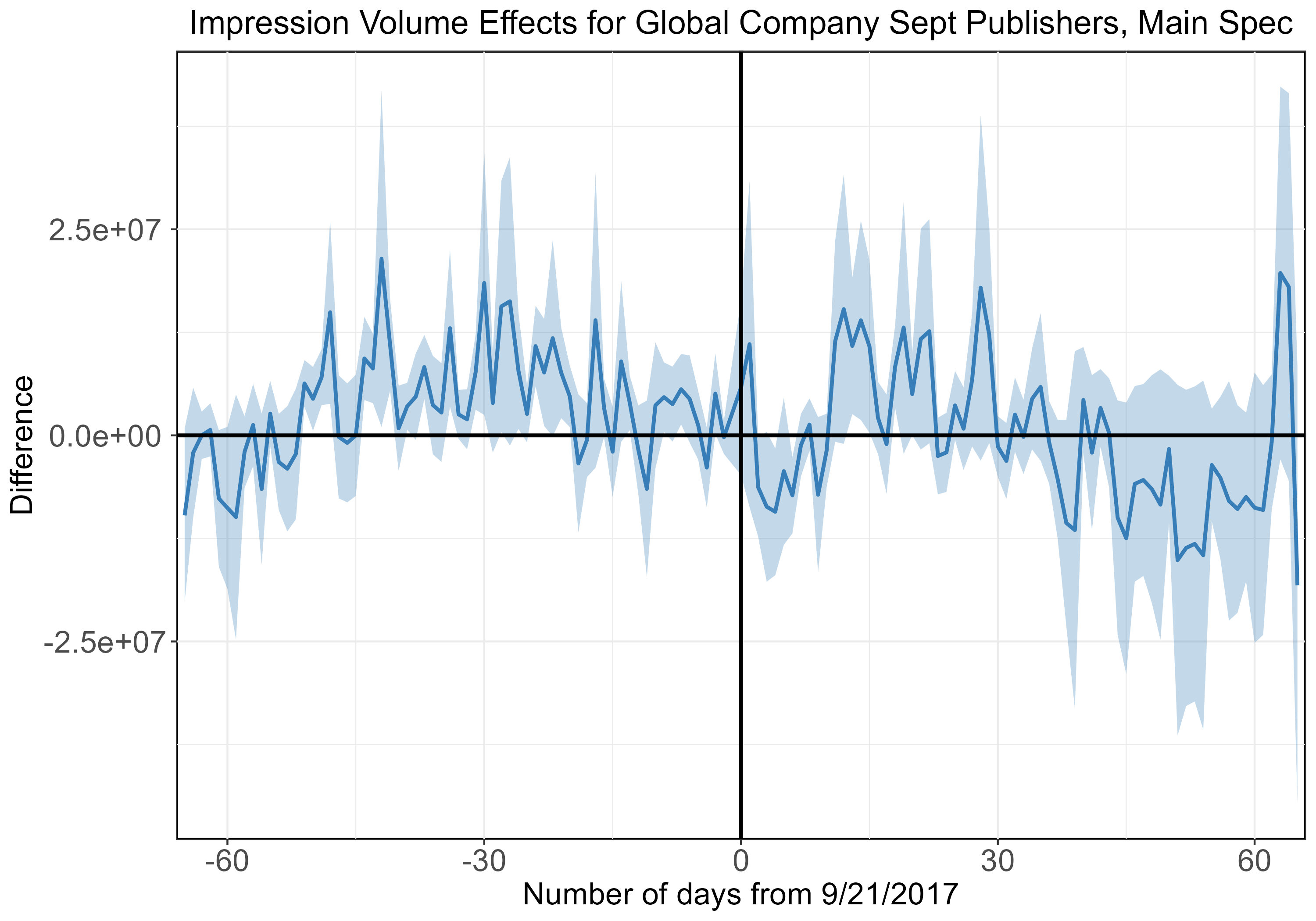}
        \includegraphics[width=0.48\textwidth]{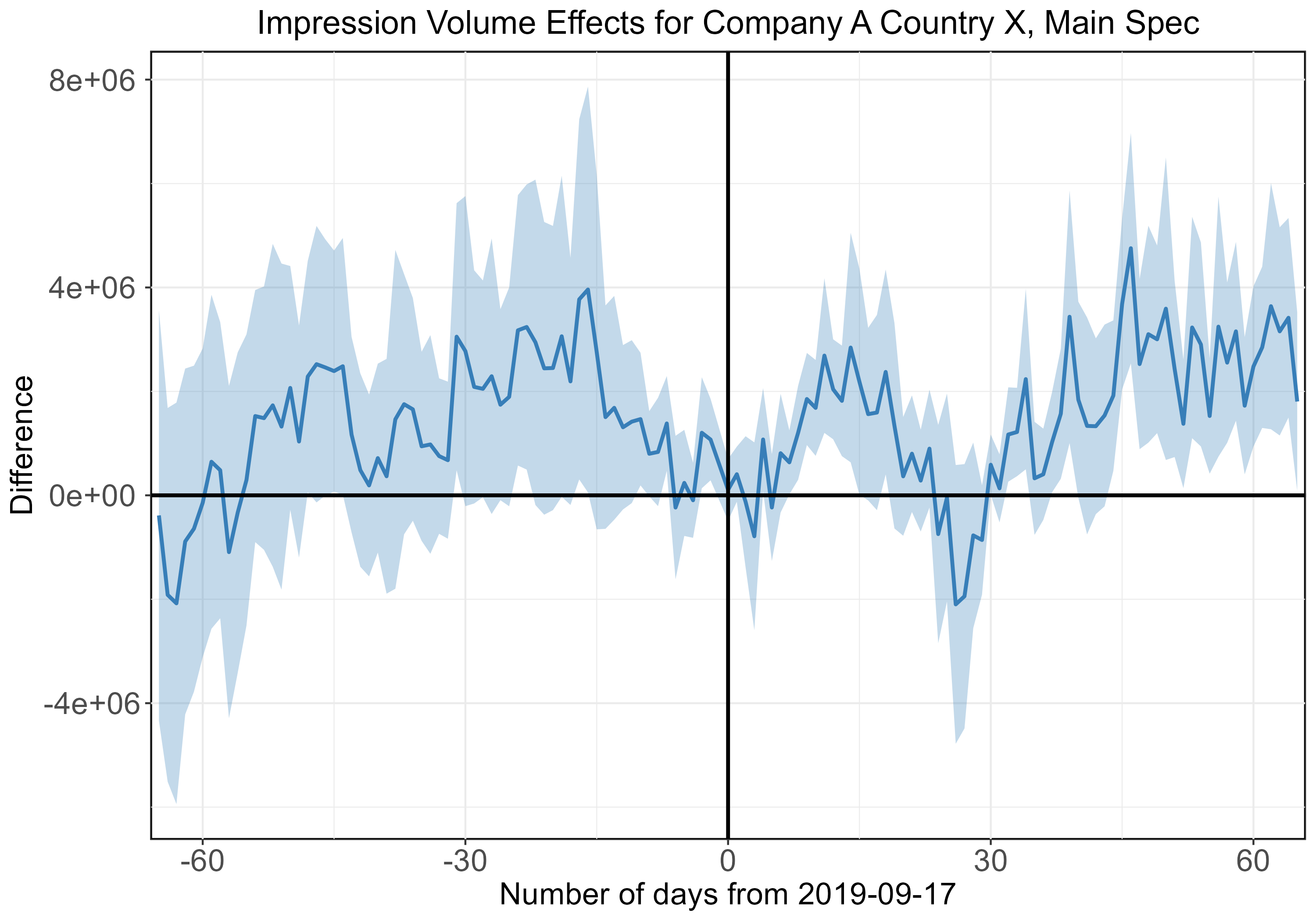} \\
        \includegraphics[width=0.48\textwidth]{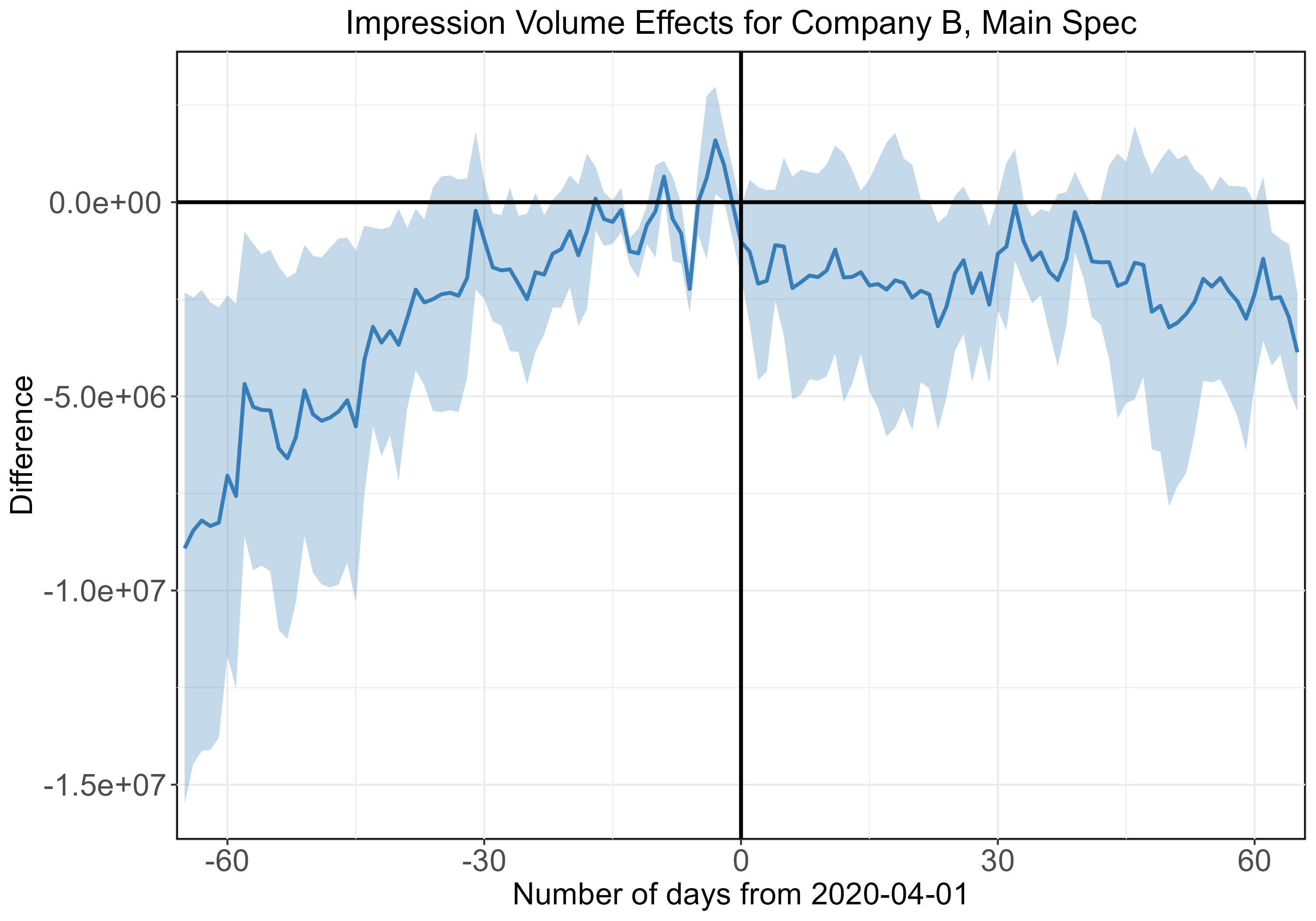}
        \includegraphics[width=0.48\textwidth]{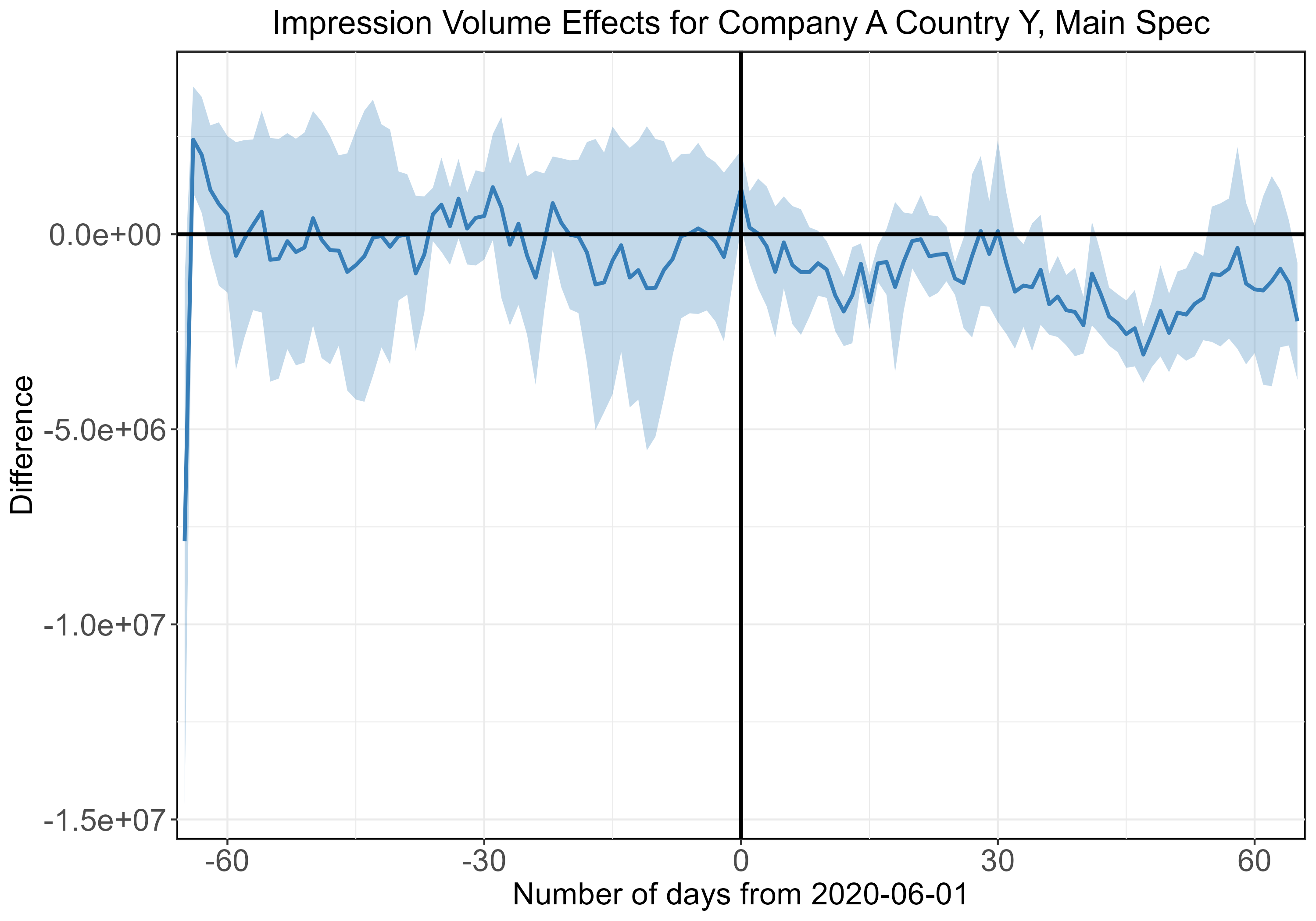}
      \end{center}
    \end{minipage}
  }
  \caption{Estimated Effects on Volume of Impressions Sold.  The solid line indicates point estimates, and the band indicates 95\% confidence intervals.}
\label{fig:DID_imps}
\end{figure}

\begin{figure}[htbp]
  {
    \includegraphics[width=0.7\textwidth]{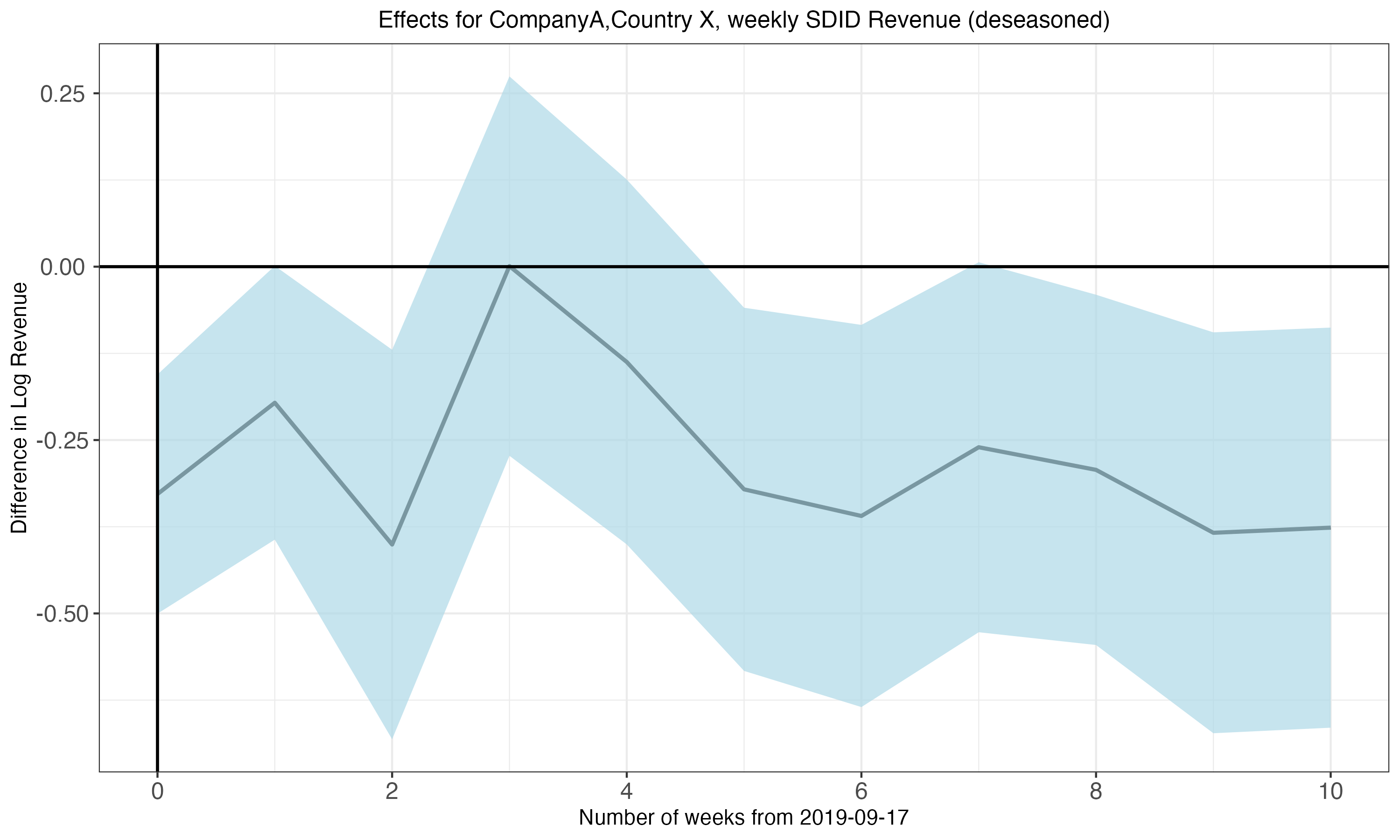}
  }
  \caption{\newgwR{SDID Estimated Effects of Format Change on Total Revenue (Logs)}.  The solid line indicates point estimates of $\tau_{w}$, and the band indicates 95\% confidence intervals.}
  \label{fig:SDID_revenue}
\end{figure}

\begin{figure}[htbp]
  {
    \begin{minipage}{\textwidth}
      \begin{center}
        \includegraphics[width=0.48\textwidth]{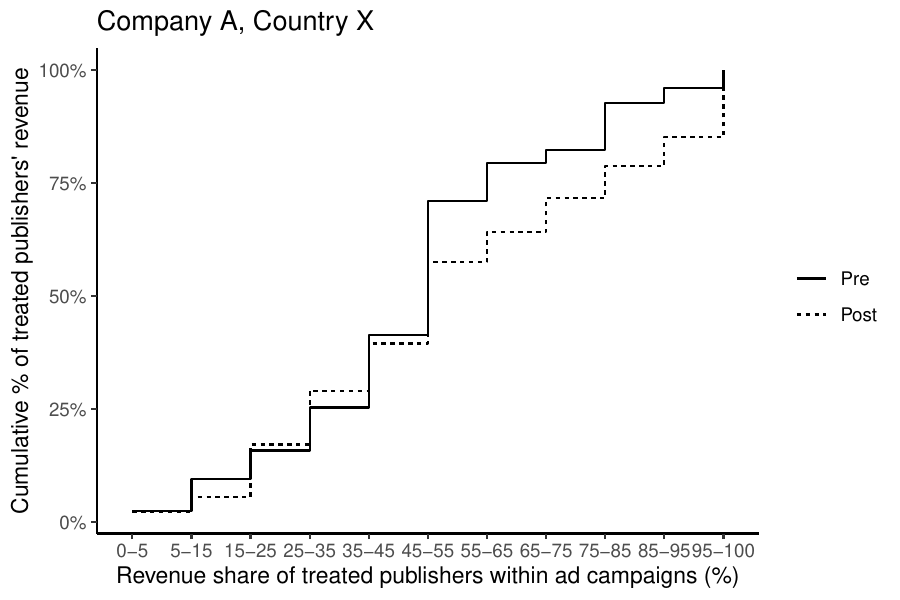}
        \hfill
        \includegraphics[width=0.48\textwidth]{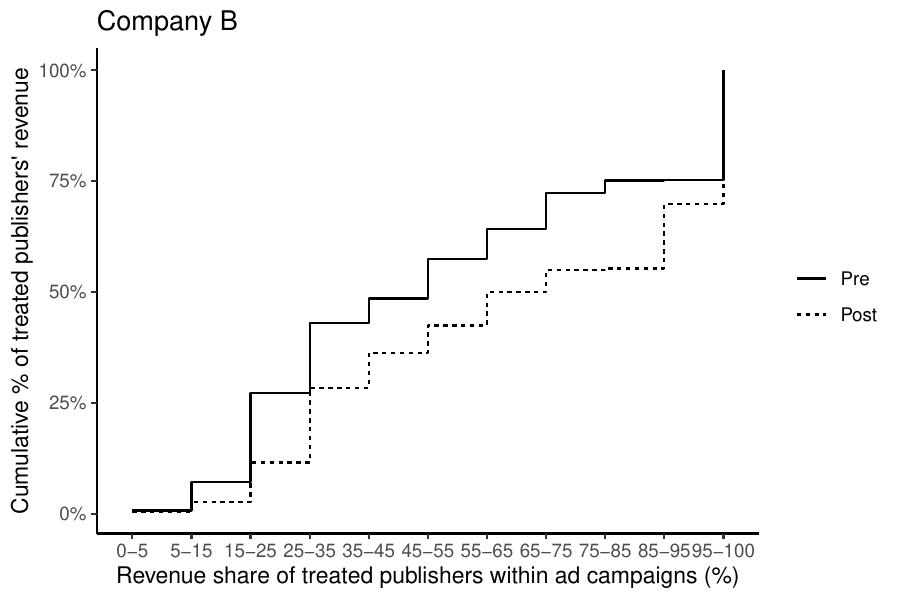} \\
        \includegraphics[width=0.48\textwidth]{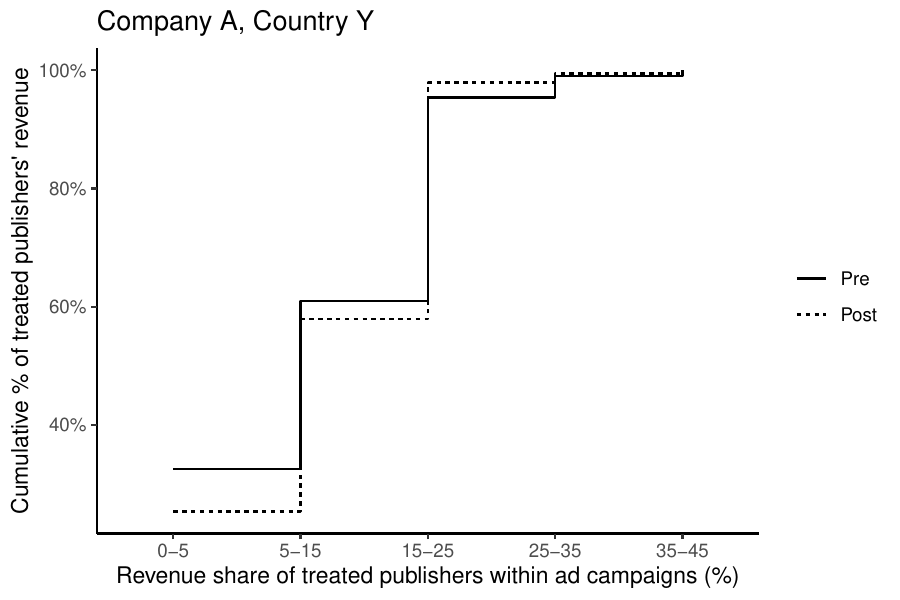}
      \end{center}
    \end{minipage}
  }
  \caption{Distribution of Share of Treatment Publishers Within Advertising
    Campaign: European Media
    Companies.  Cumulative percentage of treated publishers' revenue from ad
    campaigns that used Xandr's DSP service. The horizontal axis
    represents the share of treated publishers within each ad
    campaign's spending, rounded to the nearest multiple of 10\%.  The
    revenue and share are computed separately for 30 days before the
    format change (``Pre'') and for 30 days after it (``Post'').}
  \label{fig:share-within-campaigns-Europe}
\end{figure}

\begin{figure}[htbp]
  {
    \begin{minipage}{\textwidth}
      \begin{center}
        \includegraphics[height=0.3\linewidth]{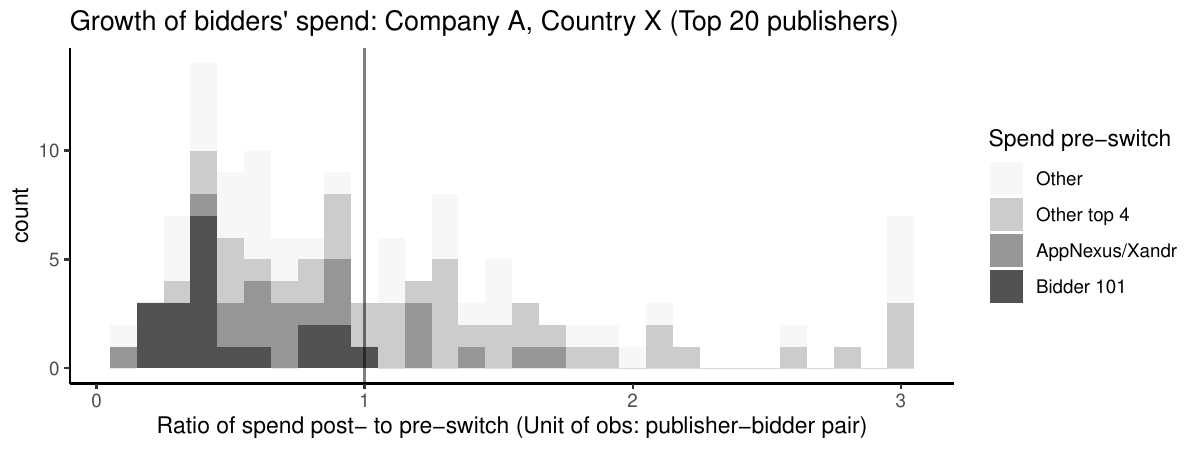} \\
        \includegraphics[height=0.3\linewidth]{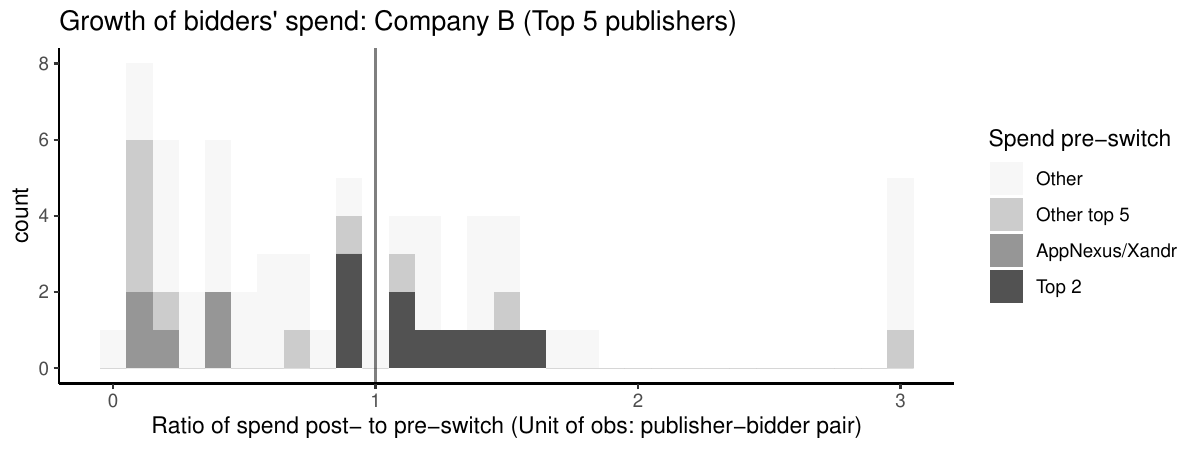} \\
        \includegraphics[height=0.3\linewidth]{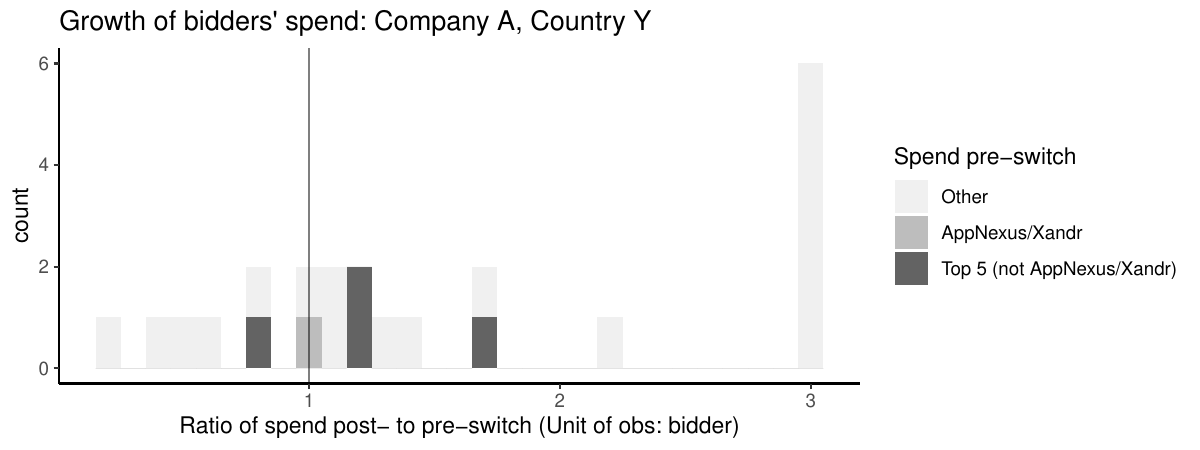}
      \end{center}
    \end{minipage}
  }
  \caption{Growth Rates of Bidders' Spending on Treated Publishers: European
    Media Companies.  Histogram of growth rates of bidders' spending on treated
    publishers of European Media Companies from 7 days before format
    change to 7 days after, color-coded by the importance of each
    bidder in treated publishers' revenue during the 7-day period
    before change (``AppNexus/Xandr'' indicates the AppNexus/Xandr
    bidder as explained in Section \ref{sec:bidder-hetero}).}
  \label{fig:growth-bidder-Europe}
\end{figure}

\clearpage

\section{Other Robustness Checks}

\subsection{Alternative Control Groups within European Media Companies}\label{subsec:alt-control-grp}

\begin{figure}[htbp]
  {
    \begin{minipage}{\textwidth}
      \begin{center}
        \includegraphics[width=0.48\textwidth]{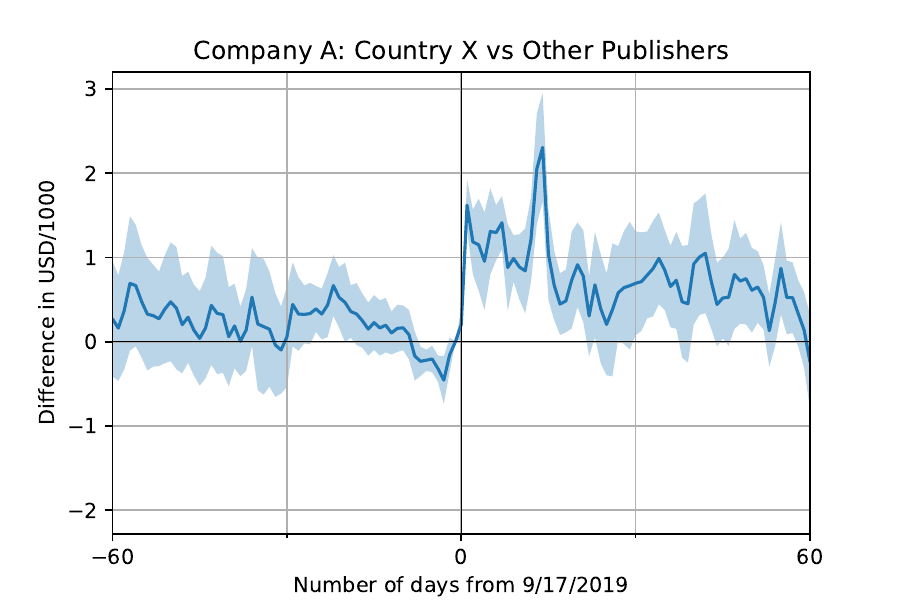} \\
        \includegraphics[width=0.48\textwidth]{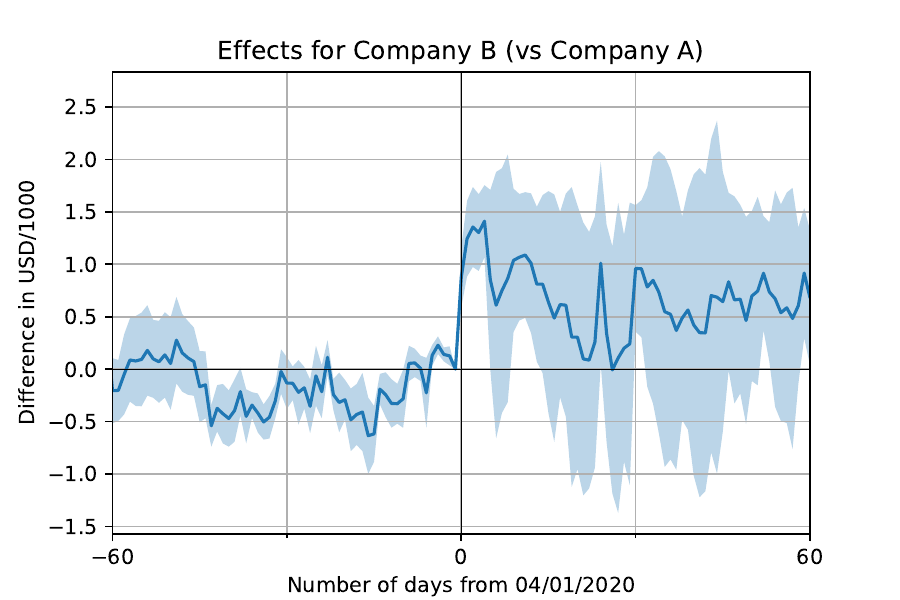}
        \includegraphics[width=0.48\textwidth]{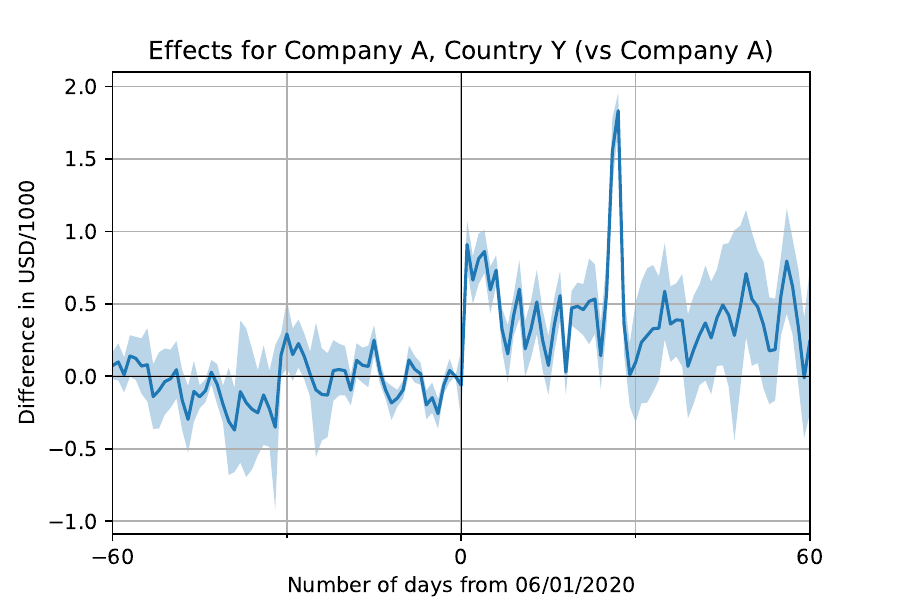}
      \end{center}
    \end{minipage}
  }
  \caption{Estimated Effects of Format Change Using Alternative Control
    Groups.  The solid line indicates point estimates of $\beta_{k}$, and the
    band indicates 95\% confidence intervals. Top: publishers of
    Company A in Country X (treatment group) vs.~publishers of Company
    A other than in Countries X and Y. Bottom left: publishers of
    Company B (treatment group) vs.~publishers of Company A other than
    in Countries X and Y. Bottom right: the publisher of Company A in
    Country Y vs.~publishers of Company A other than in Countries X
    and Y.  }
  \label{fig:event-alt-control-grp}
\end{figure}

\clearpage
\subsection{Log Average Price as Outcome Variable} \label{sec:log}

\begin{figure}[htbp]
  {
    \begin{minipage}{\textwidth}
      \begin{center}
        \includegraphics[width=0.48\textwidth]{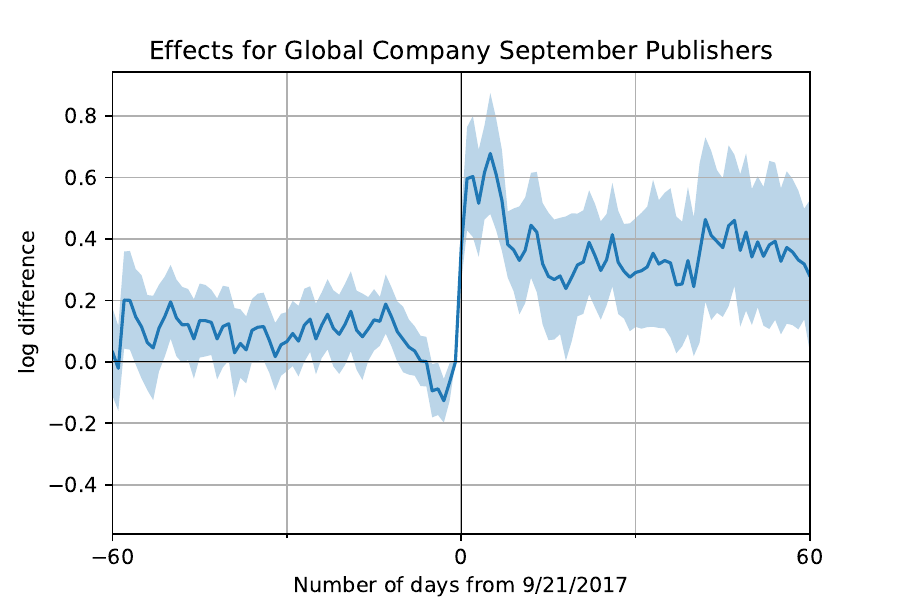}
        \includegraphics[width=0.48\textwidth]{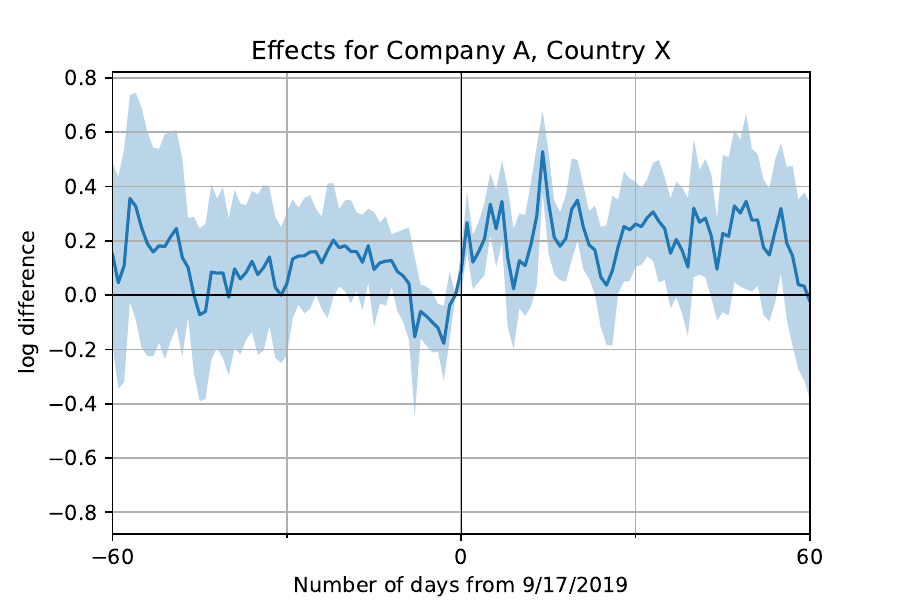} \\
        \includegraphics[width=0.48\textwidth]{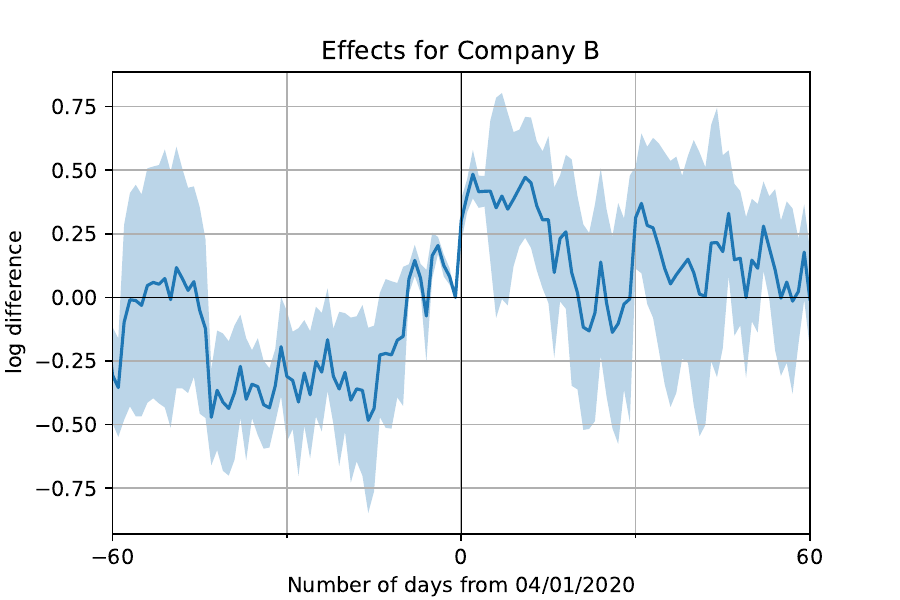}
        \includegraphics[width=0.48\textwidth]{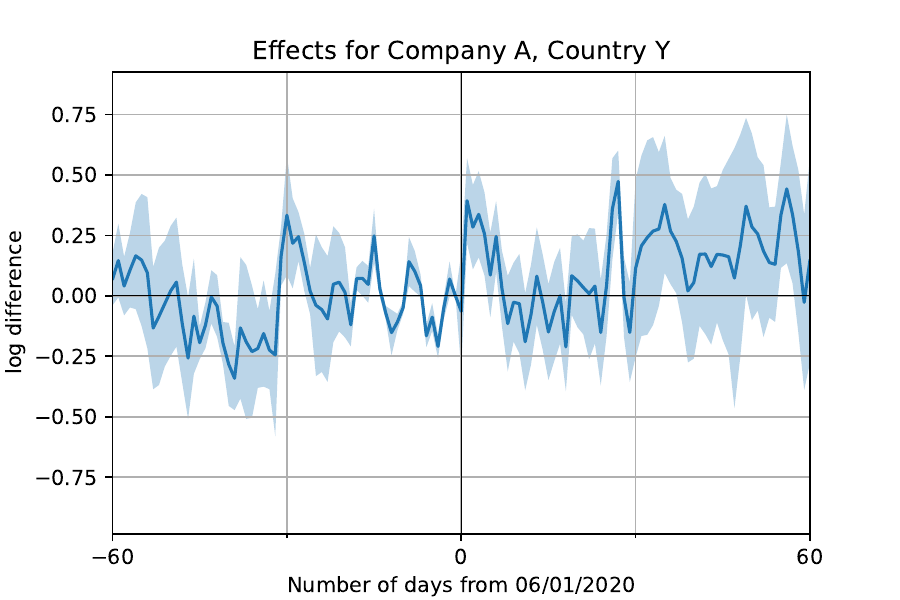}
      \end{center}
    \end{minipage}
  }
  \caption{Estimated Effects of Format Change on Log Average
    Price.  The solid line indicates point estimates of $\beta_{k}$, and the
    band indicates 95\% confidence intervals. The model is estimated
    by replacing the LHS variable in equation \eqref{eq:1step-reg}
    with $\log y_{pt}$.}
  \label{fig:event-multiplicative}
\end{figure}

\clearpage
\subsection{Alternative Seasonality Adjustments}\label{subsec:alt-seasonality}

\begin{figure}[htbp]
  {
    \begin{minipage}{\textwidth}
      \begin{center}
        \includegraphics[width=0.48\textwidth]{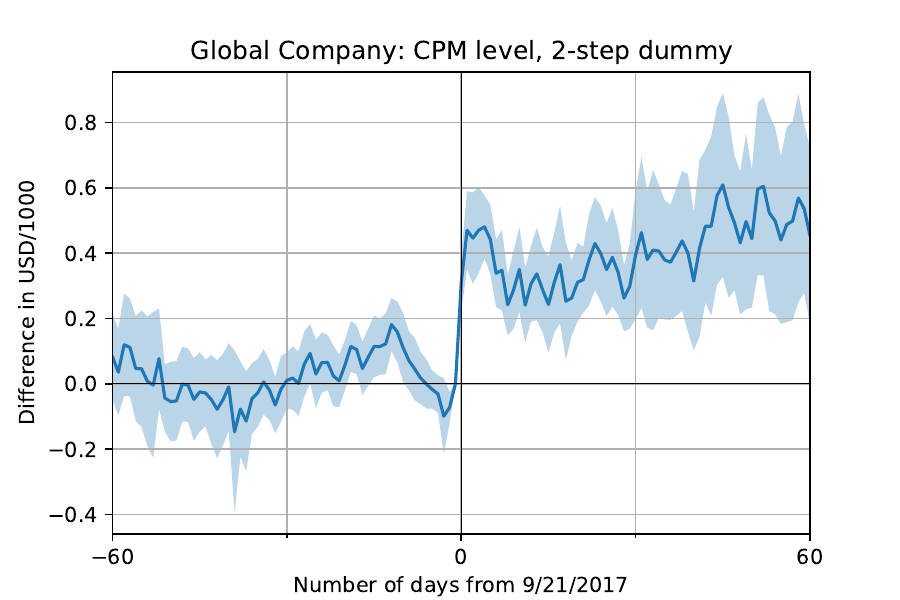}
        \includegraphics[width=0.48\textwidth]{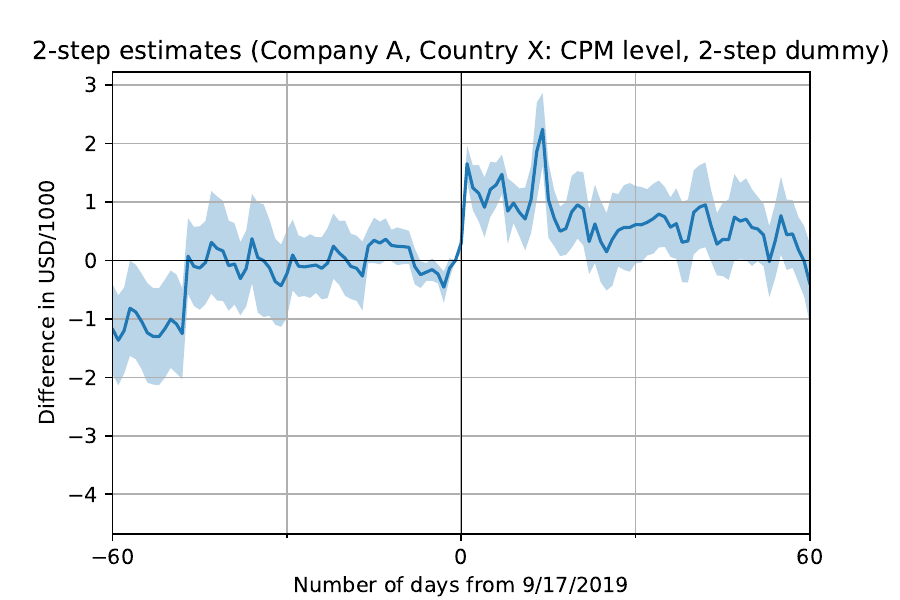}
      \end{center}
    \end{minipage}
  }
  \caption{Estimates of $\beta_{k}$ under ``Two-Step'' Method.}
  \label{fig:two-step}
\end{figure}

Figure \ref{fig:two-step} uses the estimates under the following
``two-step'' method:

\begin{enumerate}
\item We first regress, for each publisher $p$'s time series
  $\{y_{pt}\}$,
  \[ y_{pt} = \gamma_{p,\mathrm{dow}(t)} + \gamma_{p,\mathrm{dom}(t)} +
    \gamma_{p,\mathrm{month}(t)} + \gamma_{p,\mathrm{eoq}(t)}+\delta_{pt}, \]
  using the data before the format change. We weight the observations
  by the number of impressions.
\item We compute the fitted values of the previous regression
  $\widehat{y}_{pt}$.
  \item The residuals of the regression $y_{pt}-\widehat{y}_{pt}$
subtract the seasonal component from $y_{pt}$ and
  gives the deseasonalized time series $\widetilde{y}_{pt}$.
\item Regress $\widetilde{y}_{pt}$ as follows:
  \[ \widetilde{y}_{pt} = \alpha_{p} + \sum_{\underline{k}\leq
      k\leq\overline{k},\,k\neq-1}\beta_{k}D_{p}\cdot\boldsymbol{1}(K_{t}=k)
    + \gamma_{t}+\widetilde{\epsilon}_{pt} \ .\]
\end{enumerate}

\begin{figure}[htbp]
  {
    \begin{minipage}{\textwidth}
      \begin{center}
        \includegraphics[width=0.48\textwidth]{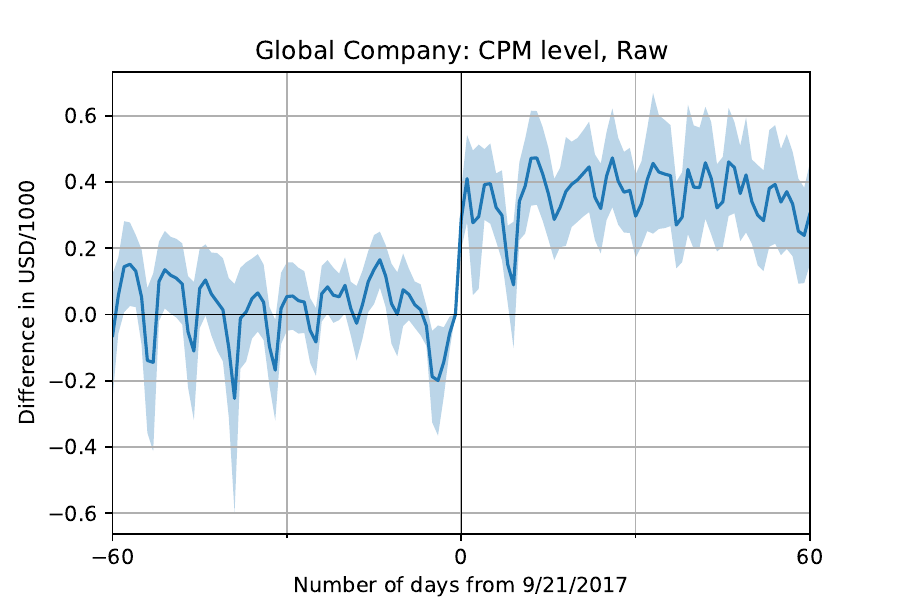}
        \includegraphics[width=0.48\textwidth]{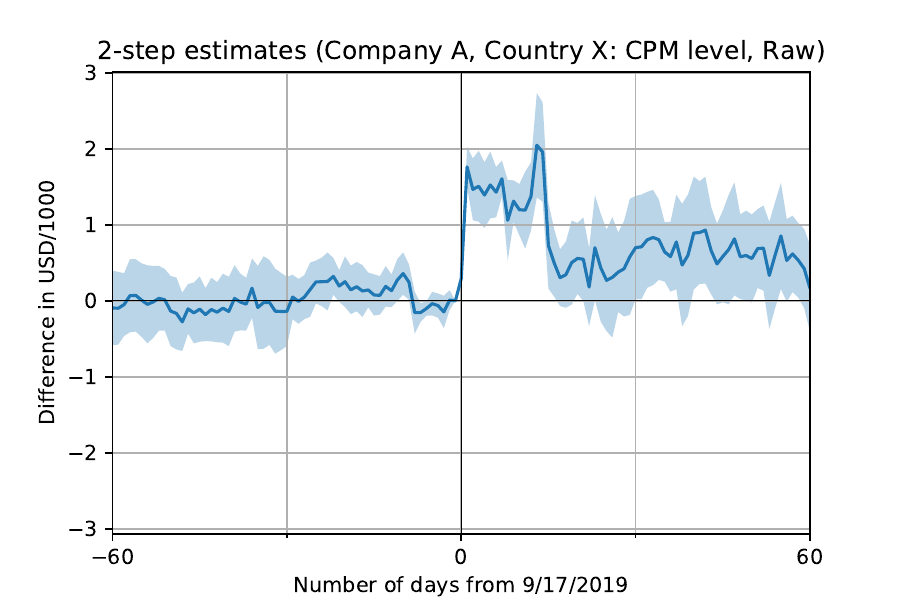}
      \end{center}
    \end{minipage}
  }
  \caption{Estimates of $\beta_{k}$ When No Seasonality Adjustments Are Made.}
  \label{fig:no-seasonality}
\end{figure}

Figure \ref{fig:no-seasonality} estimates the main regression
\eqref{eq:1step-reg} but without correcting with any seasonal fixed effects.

\clearpage
\subsection{Falsification Test with Hypothetical Event Dates}

\begin{figure}[htbp]
  {
    \begin{minipage}{\textwidth}
      \begin{center}
        \includegraphics[width=0.48\textwidth]{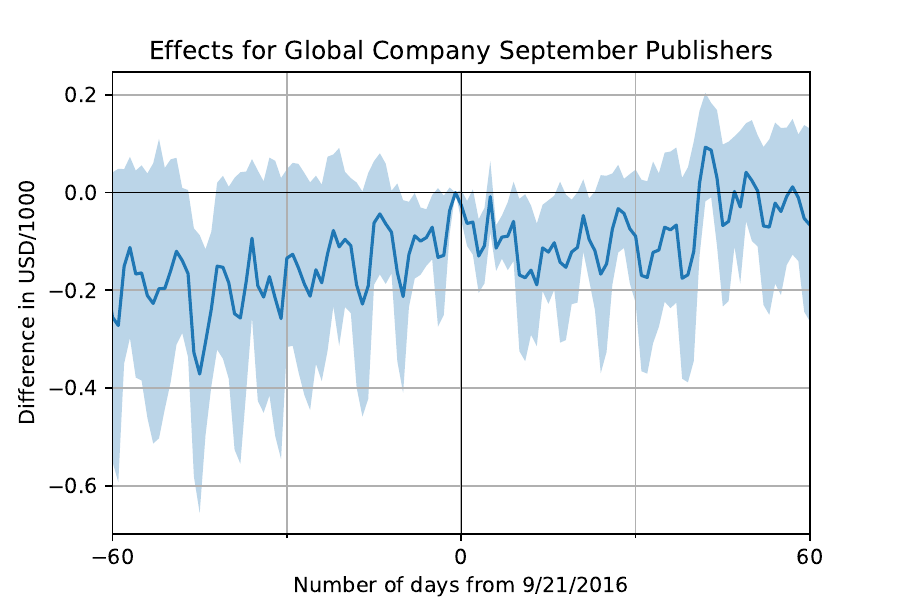}
        \includegraphics[width=0.48\textwidth]{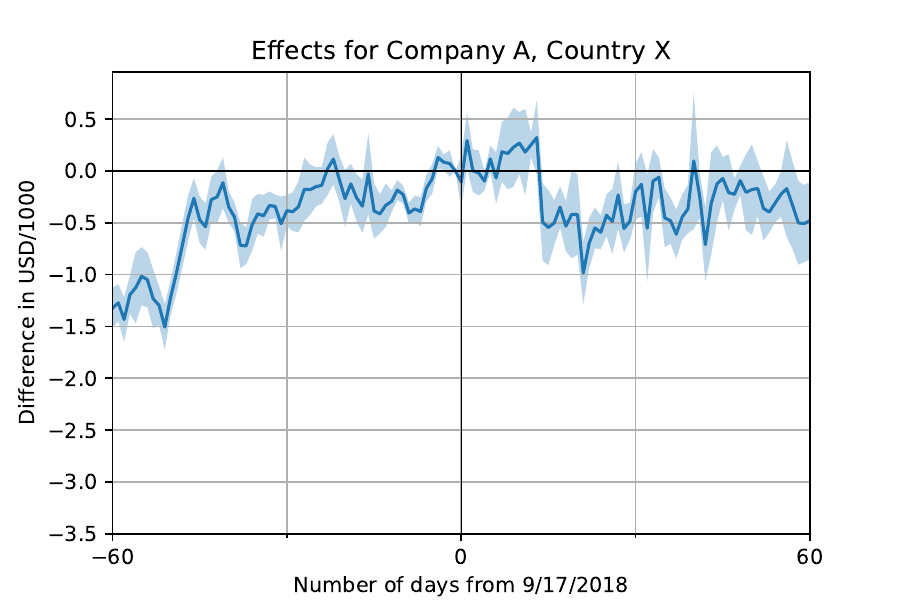} \\
        \includegraphics[width=0.48\textwidth]{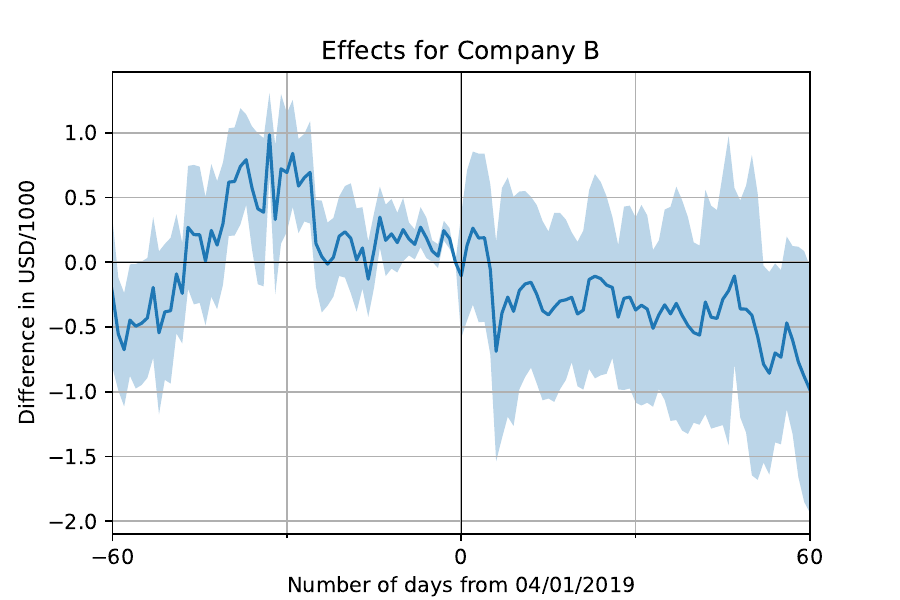}
        \includegraphics[width=0.48\textwidth]{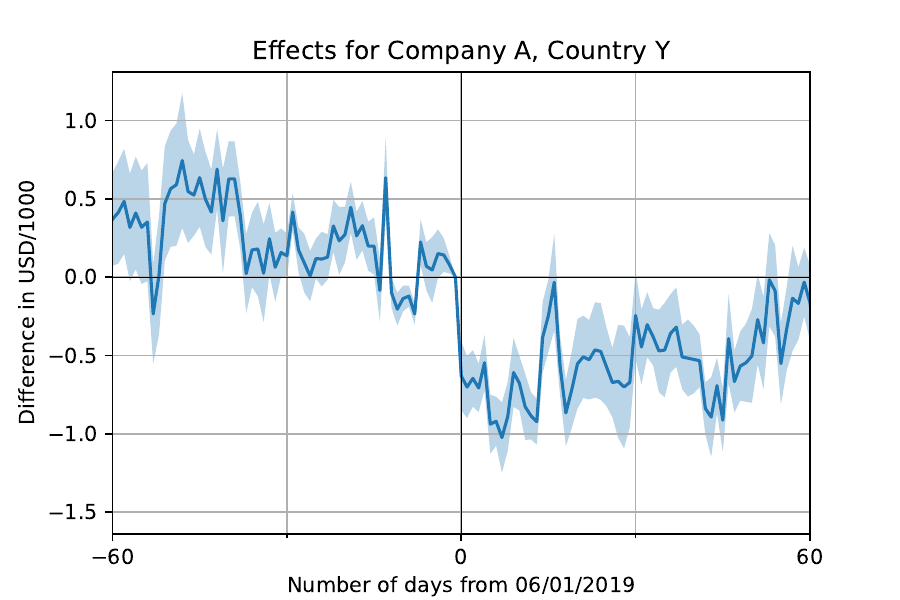}
      \end{center}
    \end{minipage}
  }
  \caption{Estimated Effects of Hypothetical Format Change on Average
    Price.  Hypothetical change dates are set as one year before the actual
    format changes. }
  \label{fig:event-falsification}
\end{figure}

\clearpage
\section{Bidder Heterogeneity}

\begin{figure}[htbp]
  {
    \includegraphics[width=0.7\textwidth]{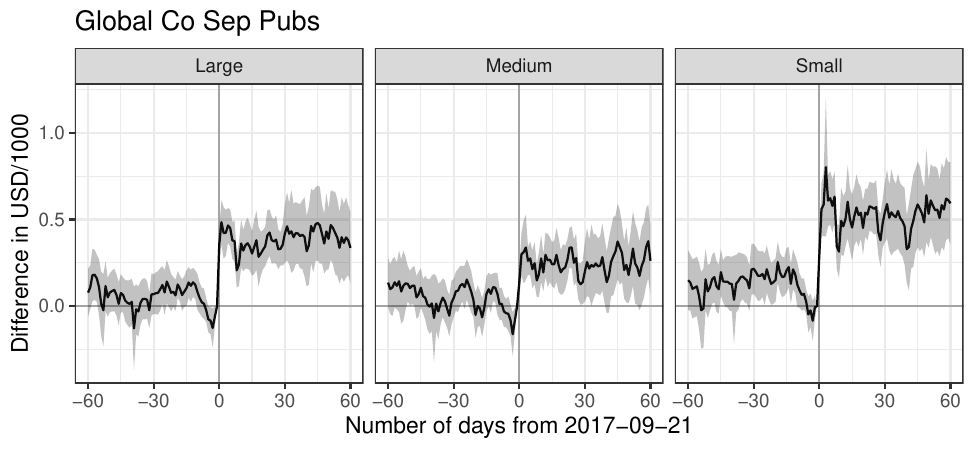}
  }
  \caption{Estimated Increase in Bidders' Spending by Bidder Size: Global
    Company.  Effects of auction format changes on spending per sold impression
    by bidders on Global Company September Publishers, separately by
    bidder size.  Bidders are classified as ``large'' if their share
    of September Publishers' aggregate revenue during the 30 days
    before the format change date (September 21, 2017) is above 10\%,
    ``medium'' if the share is above 1\%, and ``small'' if the share
    is below 1\%. }
  \label{fig:hte-by-size-global}
\end{figure}

\begin{figure}[htbp]
  {
    \begin{minipage}{\textwidth}
      \begin{center}
        \includegraphics[height=0.3\linewidth]{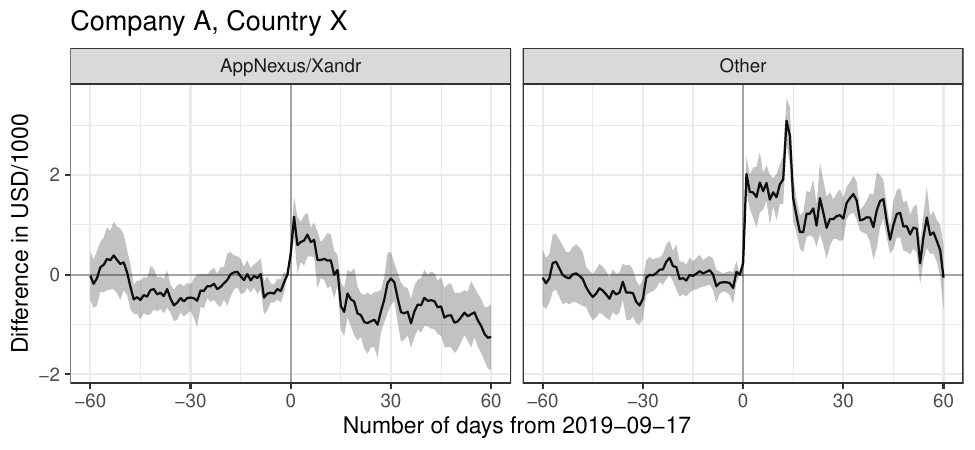} \\
        \includegraphics[height=0.3\linewidth]{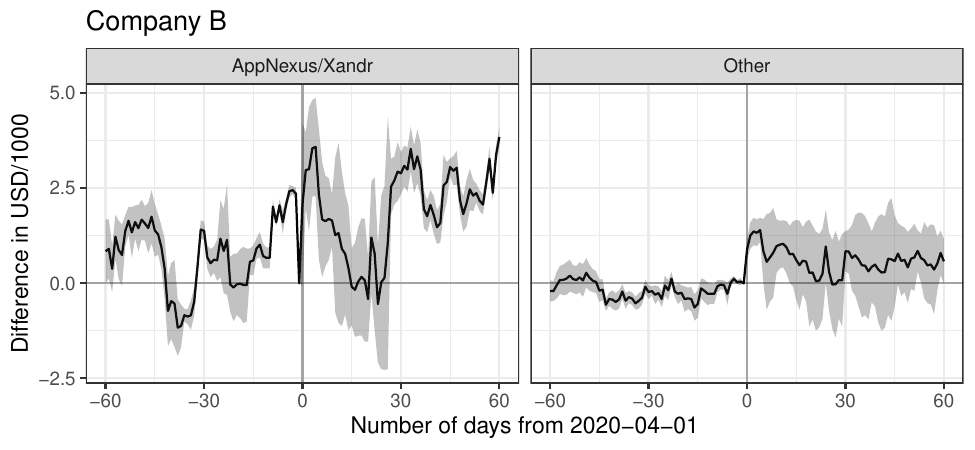} \\
        \includegraphics[height=0.3\linewidth]{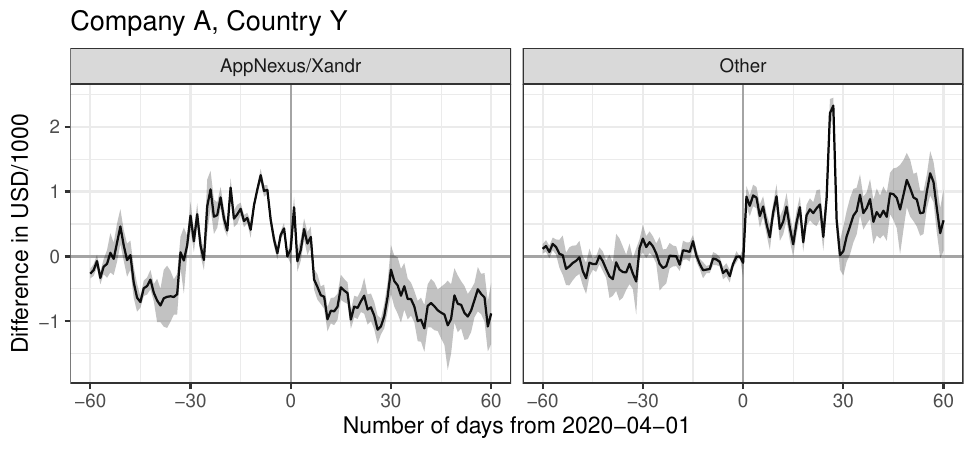}
      \end{center}
    \end{minipage}
  }
  \caption{AppNexus/Xandr Bidder and Non-AppNexus/Xandr Bidders: European
    Media Companies.  Effects of auction format changes on spending per sold impression
    by bidders on treated publishers, separately for the
    AppNexus/Xandr bidder and for other bidders.}
  \label{fig:hte-console-others}
\end{figure}




\clearpage

\bibliographystyle{informs2014} 
\bibliography{biblist.bib} 

\end{document}